\documentclass[fleqn,10pt]{wlscirep}
\usepackage[utf8]{inputenc}
\usepackage[T1]{fontenc}
\usepackage{algorithm}
\usepackage{algpseudocode}
\usepackage{graphicx}
\usepackage{hyperref}
\usepackage{xcolor,colortbl}
\usepackage{todonotes}
\usepackage{xr}
\usepackage{cleveref}

\DeclareMathOperator*{\argmax}{argmax}

\newcommand*{\argmaxl}{\argmax\limits}

\makeatletter
\newcommand*{\addFileDependency}[1]{
\typeout{(#1)}

\@addtofilelist{#1}
\IfFileExists{#1}{}{\typeout{No file #1.}}
}\makeatother

\newcommand*{\myexternaldocument}[1]{%
\externaldocument[S-]{#1}%
\addFileDependency{#1.tex}%
\addFileDependency{#1.aux}%
}

\myexternaldocument{si}

\title{HDBind: Encoding of Molecular Structure with Hyperdimensional Binary Representations}

\author[1,2,*]{Derek Jones}
\author[3]{Xiaohua Zhang}
\author[3]{Brian J. Bennion}
\author[1]{Sumukh Pinge}
\author[1]{Weihong Xu}
\author[1]{Jaeyoung Kang}
\author[1]{Behnam Khaleghi}
\author[1]{Niema Moshiri}
\author[2]{Jonathan E. Allen}
\author[1]{Tajana S. Rosing}

\affil[1]{University of California - San Diego, Department of Computer Science and Engineering, La Jolla, CA}
\affil[2]{Lawrence Livermore National Laboratory, Global Security Computing Applications Division, Livermore, CA}
\affil[3]{Lawrence Livermore National Laboratory, Biosciences and Biotechnology Division, Livermore, CA}
\affil[*]{wdjones@ucsd.edu, allen99@llnl.gov, tajana@ucsd.edu}


\begin{abstract} 
Traditional methods for identifying ``hit'' molecules from a large collection of potential drug-like candidates rely on biophysical theory to compute approximations to the Gibbs free energy of the binding interaction between the drug and its protein target. These approaches have a significant limitation in that they require exceptional computing capabilities for even relatively small collections of molecules. Increasingly large and complex state-of-the-art deep learning approaches have gained popularity with the promise to improve the productivity of drug design, notorious for its numerous failures. However, as deep learning models increase in their size and complexity, their acceleration at the hardware level becomes more challenging. 
Hyperdimensional Computing (HDC) has recently gained attention in the computer hardware community due to its algorithmic simplicity relative to deep learning approaches. 
The HDC learning paradigm, which represents data with high-dimension binary vectors, allows the use of low-precision binary vector arithmetic to create models of the data that can be learned without the need for the gradient-based optimization required in many conventional machine learning and deep learning methods. This algorithmic simplicity allows for acceleration in hardware that has been previously demonstrated in a range of application areas (computer vision, bioinformatics, mass spectrometery, remote sensing, edge devices, etc.).  
To the best of our knowledge, our work is the first to consider HDC for the task of fast and efficient screening of modern drug-like compound libraries. We also propose the first HDC graph-based encoding methods for molecular data, demonstrating consistent and substantial improvement over previous work. We compare our approaches to alternative approaches on the well-studied MoleculeNet dataset and the recently proposed LIT-PCBA dataset derived from high quality PubChem assays. We demonstrate our methods on multiple target hardware platforms, including Graphics Processing Units (GPUs) and Field Programmable Gate Arrays (FPGAs), showing at least an order of magnitude improvement in energy efficiency versus even our smallest neural network baseline model with a single hidden layer. Our work thus motivates further investigation into molecular representation learning to develop ultra-efficient pre-screening tools. We make our code publicly available at \url{https://github.com/LLNL/hdbind}.
\end{abstract}

\begin{document}

\flushbottom
\maketitle

\thispagestyle{empty}

\section*{Introduction}\label{sec:intro}
The modern drug discovery process consists of multiple sequential steps that progress from an initial large collection of candidates, sampled from the estimated $10^{60}-10^{100}$ possible drug-like small molecule structures, to a smaller targeted set of \textit{hit} or lead  compounds with potential activity with protein targets of interest\cite{Schneider2020-wz}. These candidates are filtered according to their likelihood of success based on a scoring function that uses either physics-based modeling \cite{Yu2023-nt} or, increasingly, properties inferred directly from data using machine learning\cite{Volkov2022-vc, Jones2021-al, Minnich2020-iz}. The results of the \textit{virtual screen} are then used to identify molecular \textit{leads} for more rigorous---and expensive---experimental validation\cite{Schneider2020-wz}.
Public catalogs of drug-like molecules have grown to comprise tens of billions of possibilities\cite{Grygorenko2020-nw}, while the number of available protein structures has simultaneously grown with the introduction of AI-enabled 3D structure prediction tools, resulting in over 200 million publicly available predicted structures\cite{alphafold_database, Baek2021-wq}. Even the exhaustive interrogation of the approximately 20,000 human proteins poses a considerable computational challenge. While increasingly complex deep learning architectures are demonstrating state-of-the-art (SOA) results on a wide range of molecular property prediction tasks \cite{Ross2022-ce, Wang2022-hg, Wu2018-or}, it is becoming increasingly clear that energy efficiency will become a greater priority over time as these models are, repeatedly, trained and deployed on increasingly vast human protein-drug interactome\cite{Schwartz2020-bl}.

Hyper-dimensional computing (HDC) is an emerging paradigm of lightweight machine learning that leverages the orthogonality of vectors in high dimensional space coupled with simple arithmetic operations for learning that are, comparatively to SOA deep learning architectures, simple to implement in hardware, and thus primed to take advantage of emerging hardware acceleration breakthroughs\cite{Plate1995-gz, Kanerva2009-sz, Thomas2021-uo, Karunaratne2019-zu, Ge2020-dp, Rahimi2019-vb, Burrello2019-qf, Rasanen2016-ac, Mitrokhin2019-op,Salamat2020-ld, Kanerva2000-uw, Kang2022-dac, Kang2022-pm, Pinge2023-wn, Kazemi2022-zw}. 
HDC has been demonstrated as a versatile and efficient approach for a growing variety of application domains including proteomics\cite{Xu2023-iq}, molecular property prediction\cite{Ma2021-xu}, medical image classification,  visual scene understanding\cite{Abhijith2021-fi, Rahimi2017-ox}, and biosignal classification\cite{Rahimi2019-vb, Burrello2019-qf}.
HDC requires the specification of an \textit{encoding} method to transform the original input data representation into a high-dimensional vector space as \textit{hypervectors}\cite{Kanerva2009-sz, Thomas2021-uo}. 
Then, given a similarity metric defined on the high-dimensional space, commonly chosen as the cosine similarity, which is sensitive only to the relative orientation, similar hypervectors can then be aggregated in order to build higher-level class prototype representations that form the \textit{associative memory} of the model\cite{Kanerva2009-sz, Thomas2021-uo}.
Inference then simply requires computing the similarity between a query hypervector and the elements of the associative memory\cite{Kanerva2009-sz, Thomas2021-uo}.
Despite the potential of HDC to provide a lightweight and energy efficient method for classification in the context of screening protein-ligand interactions, to the best of our knowledge, there has only been a single previously reported study of HDC on a molecular machine learning task in general~\cite{Ma2021-xu}, but this work does not consider the problem of protein-drug interactions. Our work is the first to use HDC to accelerate protein-drug interactions, in combination with a range of molecular representations including the well-studied Extended Connectivity FingerPrint (ECFP)\cite{Rogers2010-xp} as well as representations extracted from a state of the art Large Language Model (LLM) and self-supervised graph pretraining algorithms\cite{Wang2022-hg, Ross2022-ce}. 
Our work considers a panel of 6 molecular property prediction tasks derived from MoleculeNet\cite{Wu2018-or}. We show improved performance compared to the SOA HDC-based approach MoleHD\cite{Ma2022-wu} as well as baseline traditional ML methods. Our results additionally show improvement in some cases over the finetuned LLM, MolFormer-XL\cite{Ross2022-ce}.   
We  consider the LIT-PCBA binding interaction dataset, which collects experimental data across 15 protein targets selected from high-confidence PubChem Bioassay data\cite{Tran-Nguyen2020-xq}. Our work is thus the first HDC-based study of a real-world collection of molecular activity data beyond the benchmark datasets that have been considered until now, demonstrating a compelling use case for HDC in a challenging real-world application with a fair comparison to traditional physics-based molecular docking and baseline Multi-layer Perceptron (MLP) that is typically trained on top of a given molecular vector representation for a downstream task.

\section*{Materials and Methods}
\subsection*{Hyperdimensional Computing (HDC)}
Hyperdimensional Computing (HDC) is an emerging paradigm for building lightweight and error-robust models for classification and clustering\cite{Kanerva2009-sz, Thomas2021-uo}. HDC leverages the properties of high-dimensional vector spaces. With increasingly large dimension size $D: D \in \mathbb{Z}$, the distance between any pair of randomly selected vectors converges towards the expected distance between all vectors\cite{Thomas2021-uo, Kainen1997-cf}. 
 Thus, nearly all vectors are unrelated and can be considered as \textit{quasi-orthogonal}; it is then possible to attribute unique vectors to semantically meaningful properties of the dataset $X: x \in \mathbb{R}^{n}$ (i.e. element type, number of bonds, etc.)\cite{Thomas2021-uo, Yu2022-no}. 
An \textit{encoding} function $\phi(x)$ is specified to produce the representations in the high-dimension space $H: h \in \mathbb{R}^D$ from the samples of the dataset.
The encoding function $\phi$ may incorporate prior knowledge about the mapping between the ambient data dimension and the high dimensional space or may be a parameterized function such as a neural network that is learned from the data\cite{Xu2023-zg, Ma2022-wu, Thomas2021-uo}. 
Simple arithmetic operations can be used to reason with the the high-dimensional vectors $h$.
The \textit{binding} operator $\otimes: H \times H \rightarrow H$ is used to create ordered tuples of points in $H$. We define $\otimes$ as the hadamard or element-wise product, which is associative and commutative:
\begin{equation}
 \otimes(a,b) = \sum_i a_i b_i   
\end{equation}
The bundling operator $\oplus: H \times H \rightarrow H$ allows for the composition of information from disparate sources into a single representation\cite{Thomas2021-uo}. We define $\oplus$ as the element-wise sum, which is associative and commutative:
\begin{equation}
 \oplus(a,b) = \sum_i a_i + b_i   
\end{equation}
Lastly, permutation $\Pi$ is used to, efficiently, incorporate positional information into the representation $h$\cite{Ma2021-xu, Gupta2020-cp, Thomas2021-uo}.

\subsection*{Learning in HDC}
\label{sec:learning_in_hdc}
\begin{algorithm}[H]
        \begin{algorithmic}
        \Require
        \Statex Encoded training hypervectors, $H$
        \Statex Training labels, $Y$
        \Statex Batch size, $B$
        \Procedure{BuildAM}{$H$, $Y$, $B$}
        \For{$(h_b, y_b) \in$ GenerateBatches($H$, $Y$; $B$)}
            \For{$k \gets 0$ to $K$}
                \State $h_k = h_b[y_b == k]$ \Comment{collect all hypervectors in class $k$ for batch $b$}
                \State $\mathbb{A}[k] += \sum_{i=1}^{B} h_{k,i}$ \Comment{sum the hypervectors $h_k$ and add to the associative memory}
            \EndFor
        \EndFor
        \State \Return $\mathbb{A}$ \Comment{return initialized associative memory}
        \EndProcedure
        \end{algorithmic}
    \caption{HDC AM: Build Associative Memory Module $\mathbb{A}$}
    \label{alg:hdc_am}
    \end{algorithm}

\begin{algorithm}[H]
        \begin{algorithmic}
        \Require
        \Statex Associative memory module, $\mathbb{A}$
        \Statex Encoded training hypervectors, $H: h \in H$
        \Statex Training labels, $Y: y\in Y$
        \Statex Batch Size $B$
        \Procedure{UpdateAM}{$\mathbb{A}$, $H$, $Y$, $B$}
        \For{$(h_b, y_b) \in$ GenerateBatches($H$, $Y$; $B$)}

            \State $s_b = \rho(\mathbb{A}, h_b)$ \Comment{Compute pairwise cosine similarity in parallel for the batch and the associative mem.}
            \State $\hat{y_b} = \argmaxl_{k}s_b[:,k]$ \Comment{Get the indices of the most similar AM element as the predicted class}
            \State $e$ = $\hat{y_b} \neq y_b$ \Comment{compute a binary mask using the model's errors}
            \State $M$ =($h_b[e, :]$, $y_b[e, :]$)
            \For{($h$, $y$) $\in$ $M$} \Comment{for each mistake}
                \State $\mathbb{A}[y] \mathrel{{+}{=}} h$ \Comment{Add the mistake to the correct a.m. entry}
                \State $\mathbb{A}[1 - y] \mathrel{{-}{=}} h$ \Comment{Subract the mistake from the incorrect a.m. entry} 
            \EndFor
        \EndFor
        \State \Return $\mathbb{A}$ \Comment{return updated associative memory}
        \EndProcedure
        \end{algorithmic}
    \caption{HDC Retrain: Update Associative Memory Module $\mathbb{A}$}
    \label{alg:hdc_retrain}
    \end{algorithm}

\begin{algorithm}[H]
        \begin{algorithmic}
        \Require
        \Statex Associative memory module, $\mathbb{A}$
        \Statex Encoded testing hypervectors, $H: h \in H$
        \Statex Batch Size $B$
        \Procedure{PredictAM}{$\mathbb{A}$,$H$, $B$}
        \State $\hat{y} = \{\}$
        \For{$h_b \in$ GenerateBatches($H$; $B$)}

            \State $s_b = \rho(\mathbb{A}, h_b)$ \Comment{Compute pairwise cosine similarity in parallel for the batch and the associative mem.}
            \State $\hat{y_b} = \argmaxl_{k}s_b[:,k]$ \Comment{Get the indices of the most similar AM element as the predicted class}
            \State
            \State $\hat{y} = \hat{y} \cup \hat{y}_b$ \Comment{append the batch predictions}

        \EndFor

        \State \Return $\hat{y}$ \Comment{return predictions}
        \EndProcedure
        \end{algorithmic}
    \caption{HDC Test}
    \label{alg:hdc_test}
    \end{algorithm}

HDC supports the development of lightweight classification models without the need of numerical optimization approaches such as stochastic gradient descent (SGD) or more sophisticated alternatives typically used to train deep neural networks\cite{Rumelhart1986-ul, Kingma2014-ti, You2019-an}. Learning in HDC for a set of $K$ classes proceeds by the construction of prototypes $h_k: k \in K $ for each class $k$:
\begin{equation}\label{eq:hdc_training}
    h_k = \bigoplus_{i\vert y_i = k} \phi(x_i)
\end{equation}
where $x_i$ is the $i^{\text{th}}$ sample from the dataset $X$ and $y_i$ is the respective class label. The initial epoch of training consists of building the \textit{associative memory} $\mathbb{A}$ of the model by applying $\phi(x)$ to the input dataset $X$ producing a representative prototype vector $h_k$ for each class with a single pass over the training set (Algorithm \ref{alg:hdc_am}). To perform inference on a query hypervector $h_q$, we simply compute:
\begin{equation}
\label{eq:hdc_inference}
    \hat{y} = \underset{k \in K}{\text{argmax }} \rho (h_k, h_q) = \underset{k \in K}{\text{argmax }} \rho (h_k, \phi(x_q))
\end{equation}
where $\rho$ denotes a user-specified similarity metric and $x_q$ is the query data point. \cite{Thomas2021-uo}. In our work we implement $\rho$ as the cosine similarity:
\begin{equation}
    \rho(h_k, h_q) = \rho(h_k, \phi(x_q)) = \cos(\theta) = \frac{h_k \cdot \phi(x_q)}{\vert \vert h_k \vert \vert~ \vert \vert \phi(x_q) \vert \vert}
\end{equation}

After constructing $\mathbb{A}$ with single-pass learning, it can be further refined with a \textit{re-training} phase (Algorithm \ref{alg:hdc_retrain}). This phase tests the model predictions on the training set then updates $\mathbb{A}$ accordingly. This operation functions to increase the distance from the incorrect class prototype(s) while decreasing the distance to the correct class prototype. For testing the learned $\mathbb{A}$, we simply compare the hypervectors of the test set with $\mathbb{A}$ using the user-defined similarity metric $\rho$ and select the index of the most similar prototype to represent the predicted class (Algorithm \ref{alg:hdc_test}).

\subsection*{Encoding Molecular Data for HDC}
\label{sec:hdc_molecule_encoders}

Small drug-like molecules are often described using the ``simplified molecular-input line-entry ststem'' (i.e. SMILES) which encodes the structure as an ASCII string\cite{Weininger1988-mh}. The SMILES string itself describes a depth-first traversal of the 2D molecular graph structure. The ECFP representation considers the graph representation of the molecule and is widely used in computational chemistry for tasks such as similarity search in chemical libraries as well as a feature for ML models. ECFP is based on the Morgan algorithm \cite{Morgan1965-zp}, which was originally proposed to solve the molecular isomorphism problem and is widely used for chemical similarity analysis as well as general purpose representations for machine learning.  
The ECFP algorithm makes changes to MorganFP that improve efficiency, such as a user-defined iteration limit, a cache to store intermediate atom identifiers between iterations, and a hashing scheme to record the resulting representations \cite{Rogers2010-xp}. 
Thus, ECFP effectively uses a bottom-up approach to collect progressively larger molecular substructures that are guaranteed to coherently preserve the graph structure as any entry in the ECFP corresponds to a valid subgraph of the input molecular graph whereas a randomly selected substring of a SMILES may not correspond to a valid subgraph or even a valid SMILES string\cite{Rogers2010-xp}. 
ECFP allows for a user to specify the number of bits (i.e. vector length) $n\in\mathbb{Z}^{+}$ in a representation, commonly chosen as 1,024 or 2,048 \cite{Minnich2020-iz, Wu2018-or}. 
Further, a maximum radius size $r\in\mathbb{Z}^{+}$ (i.e., number of edges (bonds) from a root node (atom)) for collecting substructure-graphs is specified to constrain the search for substructure information.
Thus each binary value in the ECFP representations indicates the presence or lack thereof for a chemical substructure.

\subsubsection*{Random Projection FingerPrint encoding (RPFP)}\label{sec:encoding_rp}

\begin{figure}
    \centering

        \includegraphics[width=.75\textwidth]{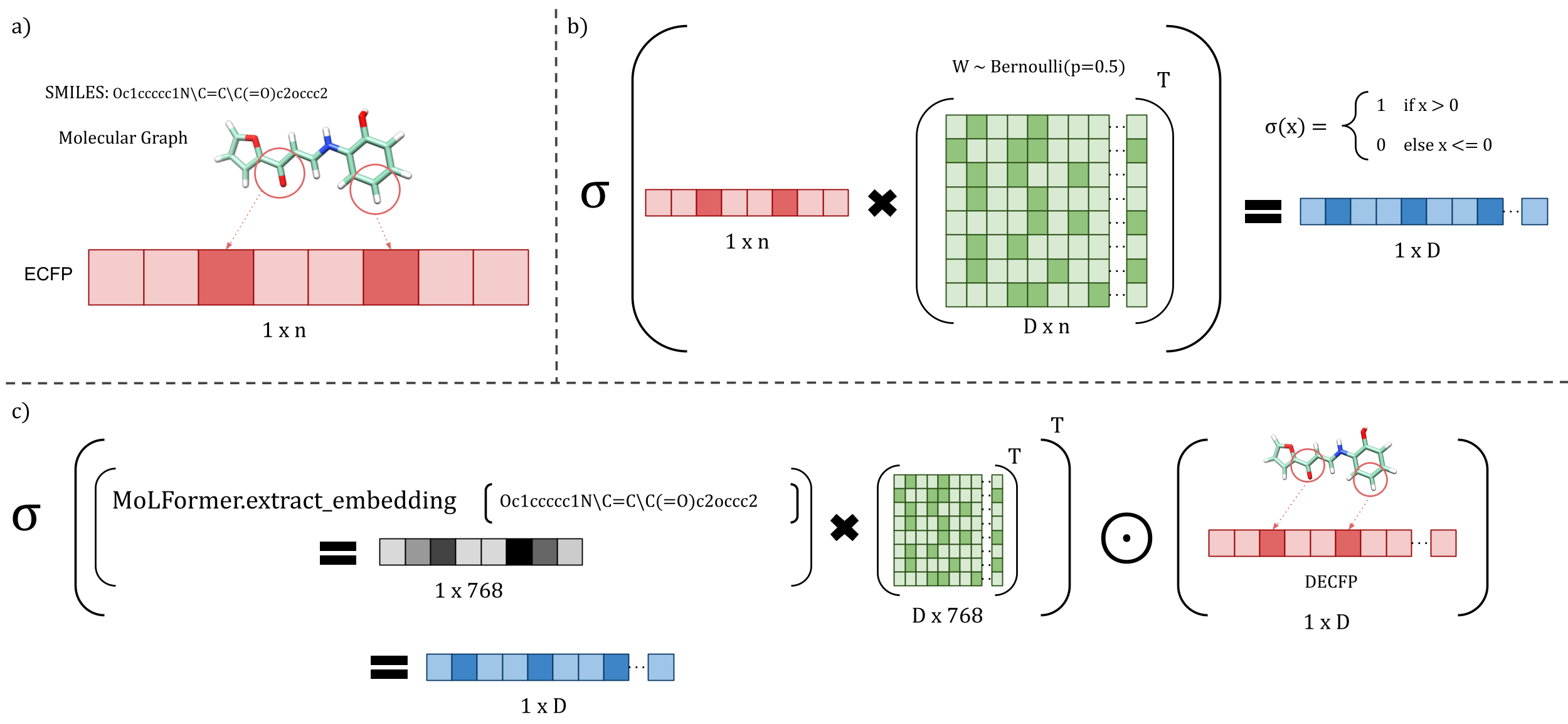}
    \caption{Description of (a) the ECFP representation, (b) the Random Projection FingerPrint (RPFP), and (c) the HDB-Combo encoding. The HDB-Combo encoding uses the Hadamard (i.e., element-wise) product of the ECFP hypervector (DECFP) with the random projection of the MoLFormer representation, embedding the features learned from the large-scale self-supervised pretraining along with the coherent graph substructure information provided by the ECFP algorithm.}
    \label{fig:hdb_summary_fig}
\end{figure}

Random Projection (RP) provides a simple method for dimensionality reduction \cite{Dasgupta2000-hb, Bingham2001-rk}. 
RP can also be considered as the basis of an encoding method to produce high-dimensional embeddings $h$ that preserve the relative distances of the input data\cite{Thomas2021-uo, Gupta2020-cp}:
\begin{align}
    z = xW^{\top} \\
    h = \sigma(z)
\end{align}
where $W \in \mathbb{R}^{D \times n }$ is a matrix whose rows are randomly sampled from the surface of the unit sphere \cite{Thomas2021-uo}. 
The quantization operator $\sigma(z)$ is defined as:
\begin{align}\label{eq:sign}
    \sigma(z) =
    \begin{cases}
        -1 & \text{where } z \leq 0\\
        1 & \text{where } z > 0
    \end{cases}
\end{align}

\subsubsection*{Direct ECFP Encoding (DECFP)}
The direct ECFP encoding (DECFP) approach simply uses the rdkit\cite{Landrum2021-zy} function \texttt{GetMorganFingerprintAsBitVect} to compute fingerprints for each molecule, which given their sparse binary properties satisfy our definition of hypervectors. The nBits parameter is adjusted to equal $D$ corresponding to the hypervector dimension. This can be described as:

\begin{align}
z = \text{E}_{n,r}(s)
\end{align}
\begin{align}
h = \sigma(z)
\end{align}

where $s$ denotes the SMILES string corresponding to a particular sample.
As no matrix multiplications are required, the entire encoding process is carried out on the CPU.

\subsubsection*{Large-scale Self-supervised Representations}
\begin{figure}
    \centering
    \includegraphics[width=\textwidth]{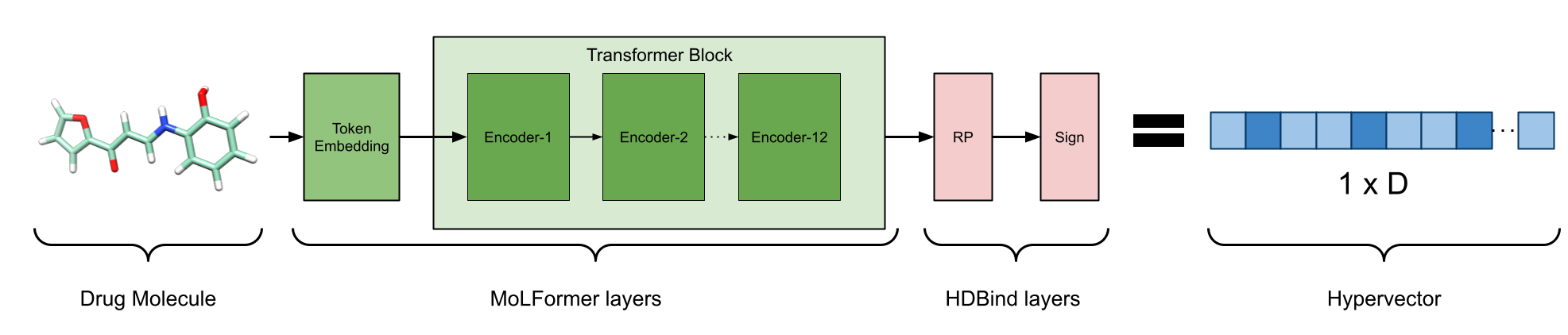}
    \caption{HDB-MoLFormer architecture description.}
    \label{fig:hdb-molformer}
\end{figure}

Data-driven molecular representation learning has caught much attention in recent years in tandem with the rise of deep learning\cite{Duvenaud2015-kl,Chithrananda2020-el, Ross2022-ce, Wang2022-hg}. We investigate the SOA approach, MoLFormer\cite{Ross2022-ce}, as the basis of the molecular representation we consider. MoLFormer uses the masked language model framework\cite{Liu2019-lz, Devlin2019-vh} and thus employs self-supervision to learn to predict missing tokens from within a SMILES sequence\cite{Ross2022-ce}.
An alternative pretraining paradigm instead uses the molecular graph representation along with graph-centric augmentations (atom masking, bond deletion, subgraph removal) and self-supervised contrastive learning objectives\cite{Wang2022-hg, Sun2019-fa}. The SOA approach MolCLR\cite{Wang2022-hg} is considered in our work.
Previous work has considered the use of neural networks for the basis of an HDC embedding \cite{Ma2022-wu}, however we are the first to our knowledge to consider a model obtained from an extensive training run on large collections of publicly available molecular data\cite{Ross2022-ce}. Similarly to the ECFP encoding, we use the random projection approach described previously to realize the HDC embeddings as HDB-MoLFormer (Fig. \ref{fig:hdb-molformer}) and HDB-MolCLR. This is a similar strategy to previous work which uses a deep convolutional neural network as a feature extractor to generate input representations for the random projection layer\cite{Dutta2022-kl}. 

Effectively, the HDB-MoLFormer and HDB-MolCLR strategies may be considered as a neural network of $L$ layers where the initial $0 \leq l \leq L - 1: l \leq L$ layers are trained using a gradient-based optimization scheme with a self-supervised (pre-)training objective. The $L^\text{th}$ layer in this network then uses a randomly sampled linear projection layer (bias omitted) with a sign activation function (eq. \ref{eq:sign}) to truncate the input values to be in the binary space $\{-1, 1\}$. The outputs and their labels are collected to form the associative memory of the model which are subsequently used for HDC training and inference (Algorithm \ref{alg:hdc_am}, \ref{alg:hdc_retrain}, and \ref{alg:hdc_test}).

\subsection*{Ranking compounds with HDC}
To rank compounds for the HDC methods, we use the confidence estimation equation as described in MoleHD\cite{Ma2021-xu}. For a binary classifier, the range of similarity differences between the positive and negative classes are transformed linearly to the interval [0,1].:

\begin{equation}
    \eta = \frac{1}{2} + \frac{\rho(h_{q}, h_{1}) - \rho(h_q, h_0)}{4} 
\end{equation}

where $h_0$ and $h_1$ are respectively the negative and positive class prototype hypervectors contained in the model associative memory $\mathbb{A}$ and $h_q$ is the query hypervector. Intuitively, if $h_q$ is equally similar to both $h_0$ and $h_1$, $\eta = \frac{1}{2}$. If $h_q$ is more similar to $h_1$, $\eta > 0.5$, otherwise if $h_q$ is more similar to $h_0$, then $\eta < 0.5$.

\subsection*{Metrics}

To facilitate comparison with previous work on MoleculeNet\cite{Wu2018-or}, we use the receiver operating characteristic - area under the curve (ROC-AUC) to measure performance of different models. The ROC-AUC metric compares the true positive rate (TPR) and false positive rate (FPR) of a classifier at various thresholds of a models score to identify a positive class. The area under the curve produced by the various thresholds is measured with respect to a perfect classifier (TPR=1, FPR=0 for all thresholds).

It is common in the high-throughput screening literature to measure performance in terms of a scoring function in terms of the to encounter the enrichment factor (EF) metric\cite{Bender2005-vc, Gentile2020-tj}, which attempts to measure how well a screening method may be able to improve the density of actives in a large database of molecular candidates. The EF metric is typically defined in terms of the hit rate for a sample compared to the background hit rate of the full database. As modern databases may reach billions, a tractable sample is chosen for ruther validation, such as the top 1\% of compounds as ranked by the outputs of some scoring function. Let $a_s, a_b$ represent the number of actives and $n_s, n_b$ the size of the sample and database respectively. Then let $p_s=a_s/n_s$ be the probability of selecting an active from a sample of ranked compounds (i.e. sample hit rate) and $p_b=a_b/n_b$ be the probability of selecting an active compound from the database (i.e. background hit rate) of the database. The enrichment factor (EF) is then calculated as the ratio between the two quantities: 
\begin{align}\label{eq:enrich_factor}
    \text{EF-}{x\%} = \frac{p_s}{p_b} = \left( \frac{a_s}{n_s} \right) / \left( \frac{a_b}{n_b} \right) = \frac{a_s}{a_b} \cdot \frac{n_b}{n_s}
\end{align}
where $x=n_s/n_b$ is the fraction of top ranked molecules sampled from the database (e.g. $x=1\%$). This measurement of enrichment however is subject to the limitation of its sensitivity to the proportion of the active to inactive compounds in the test set, which is typically highly skewed in binding activity datasets\cite{Mysinger2012-hn, Tran-Nguyen2020-xq}. Several works have proposed an alternative metric which instead uses a fixed false positive rate to measure the enrichment factor\cite{Jain2008-nr, Cleves2020-ye, Tran-Nguyen2021-jw}. This approach removes the limitation of being dependent on the active to inactive ratio. To facilitate direct comparison to previously published methods\cite{Tran-Nguyen2021-jw}, we report this definition of \textit{roc-enrichment} using a false positive rate of $x\%$ as $\text{ER}$-$x\%$:

\begin{align}\label{eq:roc-enrich_factor}
    \text{ER-}{x\%} = \text{ROC-Curve}(\text{FPR-}x\%) \times 100 = \text{TPR}_{\text{FPR-}x\%} \times 100
\end{align}

where TPR is the true positive rate given by the ROC-Curve at the false positive rate of $x\%$ ($\text{FPR-}x\%$). We use $x=1\%$ to compare with previous work\cite{Tran-Nguyen2021-jw}, however when considering large databases it may be more tractable to consider smaller sample sizes (i.e. $x=.1\%$, $.2\%$, and $.5\%$).

\subsection*{Training Details}\label{sec:training_details}
All methods presented are trained on the Lassen high-performance computing cluster at Lawrence Livermore National Laboratory. Coarse-grained parallelism was achieved for each dataset by randomly sampling a task and running independently on each node of a given allocation. Each node is equipped with an IBM Power 9 CPU, 256GB of main memory, and 4x Nvidia V100 GPUs. Our experiments only consider a single GPU for all methods. All HDC methods share the same training and testing algorithms (Algorithm \ref{alg:hdc_am}, \ref{alg:hdc_retrain}, and \ref{alg:hdc_test}), with the only difference being the encoding algorithms used to produce the high-dimensional vector representations. A batch size of 128 was used for training all HDC models considered to enable fair comparison between different hypervector dimension sizes $D$ and GPU memory usage. All MLP models are optimized using \texttt{Ray.Tune} hyperparameter optimization library\cite{liaw2018tune}. We use the AsynchronousHyperBand scheduler with default parameters to sample 50 configurations. The best model, according to the minimum validation loss, is selected to train on the full dataset and evaluated on the test set for performance metrics.

\subsection*{Energy Analysis}
To estimate energy usage, we use the following equation:

\begin{equation}
    E = \bar{P} \times t
\end{equation}

where $E$ is the energy usage (Joules), $\bar{P}$ is the average power output (Watts) of the processor (CPU, GPU, or FPGA) over the course of the program execution, and $t$ is the execution time or latency of the program. 
To collect power measurements for CPU and GPU we use the the variorum power and performance measurement tool\cite{variorum}. We collect all performance measurements, not including the FPGA, on the Lassen HPC cluster using a single Nvidia V100 GPU.

\section*{Results}\label{sec:results}
\subsection*{Molecular Property Classification on MoleculeNet}
\subsubsection*{Previous Work on Supervised Learning Approaches}
The MoleculeNet benchmark is a common performance benchmark for machine learning methods across a variety of regression and classification tasks. We consider a series of 6 classification tasks to compare with recently published SOA methods\cite{Ross2022-ce, Wang2022-hg}. N-gram\cite{Liu2018-cn}, GeomGCL\cite{Li2021-od}, MolCLR\cite{Wang2022-hg}, and MolFormer-XL\cite{Ross2022-ce} represent self-supervised methods with SOA results as reported previously\cite{Ross2022-ce}. 
MolCLR\cite{Wang2022-hg} is a molecular graph pretraining method composed of atom masking, bond deletion, and subgraph removal graph augmentations whose encoded representations are used as input to the normalized temperature-scaled cross-entropy (NT-Xent) contrastive loss\cite{Chen2020-ub}. MolCLR is trained on approximately 10 million SMILES strings collected from the PubChem database\cite{Kim2023-vo}. MoLFormer-XL\cite{Ross2022-ce} is another recently proposed self-supervised pretraining method that is instead built using the masked language model framework\cite{Liu2019-lz, Devlin2019-vh} and further expands the training set considered by MolCLR\cite{Wang2022-hg} by two orders of magnitude, training on over 1 billion SMILES from PubChem\cite{Kim2023-vo}. The pre-trained MoLFormer and MolCLR models are then fine-tuned on the target MoleculeNet classification tasks by training an MLP on top of the output layers of the pre-trained networks using a supervised loss (e.g. cross-entropy or negative log-likelihood). Representative baseline supervised machine learning methods are collected from previously published methods\cite{Ross2022-ce, Lu2019-fr, Yang2019-nv} except for our own implementation of the MLP.

\subsubsection*{HDC Methods on MoleculeNet}
To our knowledge, MoleHD\cite{Ma2022-wu} is the only known previously published HDC approach for molecular property prediction in general. MoleHD uses an encoding of the SMILES string that is built upon the byte-pair encoding algorithm that accounts for atoms as cohesive structures and is trained using ChEMBL\cite{Li2021-pf, Zdrazil2024-xc}. MoleHD collects the unique tokens collected by the SmilesPair Encoding algorithm\cite{Li2021-pf} and maps these tokens to unique, quasi-orthogonal vectors of high dimension (e.g. 10,000). MoleHD also considers $n$-gram encoding methods, however the SPE method appears to produce the best overall method which we base our implementation on and our comparison. Results for all of the discussed models are compared to our proposed HDBind (HDB) approaches that consider two state-of-the-art self-supervised pretraining frameworks, MoLFormer\cite{Wang2022-hg} and MolCLR\cite{Ross2022-ce} and the well studied Extended Connectivity Fingerprint (ECFP)\cite{Rogers2010-xp} which incorporates substructure information derived directly from the molecular graph and its atom types and connectivity. Our hypothesis is that the explicit graph representation considered by the ECFP algorithm\cite{Rogers2010-xp} provides coherent substructure information (i.e. each ECFP bit corresponds to a valid molecular subgraph) that is crucial to identify in molecular property classification tasks\cite{Liu2019-lz}. Further, our hypothesis for large scale pretraining methods is that the random projection will preserve the structure of the original data in a randomly selected high dimensional space, with low required precision, allowing for extremely large vectors to be stored. Previous work has demonstrated the utility of these pre-trained representations in a variety of molecular property classification tasks, which we expect will benefit our proposed encoding approaches.

\subsubsection*{MoleculeNet Classification Results}
Our results are given for 6 binary classification tasks in Table \ref{tab:molnet_result}. We give results for HDC models with hypervector dimensionality $D=10,000$, as increasing the dimensionality to larger sizes (e.g. $1e^5, 1e^6$) tends to yield marginal improvement at best on most tasks considered. Our results suggest that the best overall HDC model is HDB-MoLFormer, which is based upon the representation extracted from MoLFormer\cite{Ross2022-ce} that is then randomly projected to the HDC representation. HDB-MoLFormer and is best in three of the 6 tasks among the HDC methods that we consider. The HDB-DECFP, which simply uses the representation generated directly from the ECFP algorithm\cite{Rogers2010-xp}, achieves competitive performance with HDB-MoLFormer on nearly each of the six tasks, while exceeding HDB-MoLFormer on three of six tasks though it is best only on the SIDER dataset. HDB-DECFP does not require the GPU for encoding the data into hypervectors, as opposed to our random projection-based approaches, allowing for significant energy savings (Table \ref{tab:encode_energy_costs}). Additionally, HDB-MoLFormer achieves SOA on two of the six tasks (BBBP, ClinTox) even when compared with the fine-tuned MoLFormer-XL\cite{Ross2022-ce}, demonstrating the ability of the approach to preserve learned substructure information provided by the more expensive pretraining. The HDB-Combo model, which combines the MoLFormer and DECFP representations (Fig. \ref{fig:hdb_summary_fig}) achieves generally high performance five of the six tasks (BBBP, Tox21, ClinTox, HIV, and SIDER) though it fails to achieve the best overall performance on any task. Our results further show that increasing the hypervector dimension fails to significantly increase the performance of the HDB-Combo model further on the MoleculeNet classification benchmarks (SI Table \ref{S-tab:molnet_result_combo}, SI Figures \ref{S-fig:roc_curve_bbbp}-\ref{S-fig:roc_curve_sider}).

\begin{table}[ht]
    \centering
    \begin{tabular}{c|c|c|c|c|c|c}
    \textbf{Method} & \textbf{BBBP} & \textbf{Tox21} & \textbf{ClinTox} & \textbf{HIV} & \textbf{BACE} & \textbf{SIDER}\\
    \hline
    \textbf{Molecules} & 2,039 & 7,831 & 1,478 & 41,127 & 1,513 & 1,427 \\
    \hline
    \textbf{Tasks} & 1 & 12 & 2 & 1 & 1 &27 \\
    \hline
      RF\cite{Ross2022-ce} & 71.4 & 76.9 & 71.3 & 78.1 & 86.7 & 68.4 \\
      SVM\cite{Ross2022-ce} & 72.9 & 81.8 & 66.9 & 79.2 & 86.2 & 68.2\\ 
      MLP & 79.0 & 67.2 & 82.2 & 73.1 & 70.3 & 58.6\\
      MGCN\cite{Lu2019-fr} & 85.0 & 70.7 & 63.4 & 73.8 & 73.4 & 55.2\\ 
      D-MPNN\cite{Yang2019-nv} & 71.2 & 68.9 & 90.5 & 75.0 & 85.3 & 63.2\\
      N-gram\cite{Liu2018-cn} & 91.2 & 76.9 & 85.5 & \textbf{83.0} & 87.6 & 63.2 \\
      GeomGCL\cite{Li2021-od} & - & \textbf{85.0} & 91.9 & - & - & 64.8 \\
       $\text{MolCLR}_{\text{GIN}}$\cite{Wang2022-hg} & 73.6 & 79.8 & 93.2 & 80.6 & \textbf{89.0} & 68.0 \\
      MoLFormer-XL\cite{Ross2022-ce} & \textbf{93.7} & 84.7 & \textbf{94.8} & 82.2 & 88.21 & \textbf{69.0} \\
      \hline
      MoleHD\cite{Ma2022-wu} & 84.4 & - & 98.7 & - & - & 56.6\\
      HDB-RPFP & 94.8 (0.3) & \textbf{70.8} (0.9) & 86.3 (4.0) & 71.8 (1.3) & 71.3 (0.7) & 55.2 (2.0)\\
      HDB-MolCLR & 66.8 (0.4) & 68.0 (0.8) & 71.2 (4.0) & 70.6 (0.7) & \textbf{82.4} (0.5) & 61.2 (1.9) \\
      HDB-MoLFormer & \textbf{99.2} (0.1) & 67.3 (1.0) & \textbf{98.8} (0.0)  & \textbf{79.2} (0.6) & 66.8 (0.4) & 55.4 (1.9)\\
      HDB-DECFP & 93.8 (0.2) & 69.6 (0.8)& 90.6 (4.0) & 77.8 (0.3) & 74.7 (1.1) & \textbf{61.4} (1.6)\\ 
    HDB-Combo &97.4 (0.3) &70.1 (1.2) &90.7 (3.4) &77.4 (0.8) & 67.0 (2.7) & 58.8 (2.8)\\ 
    \end{tabular}
    \caption{Comparison of supervised and self-supervised baselines on representative MoleculeNet benchmarks considered in previous work using the area under the curve of the receiver operating characteristic. All values are scaled by a factor of 100 for reader convenience. All methods are evaluated using scaffold splits to minimize the molecular similarity between the training and testing sets. All reported HDC models (HDBind and MoleHD\cite{Ma2022-wu}) use dimension $D=10$k. * denotes our implementation. `-` denotes no value reported in the original work. Values in parentheses denote standard deviation of the average of 10 trials per task in each dataset. Results above the horizontal line correspond to SOA supervised and self-supervised baselines, below correspond to HDC methods.}
    \label{tab:molnet_result}
\end{table}


\subsection*{LIT-PCBA}\label{subsec:lit_pcba_dataset}

\begin{figure}
    \centering
    \includegraphics[width=\textwidth]{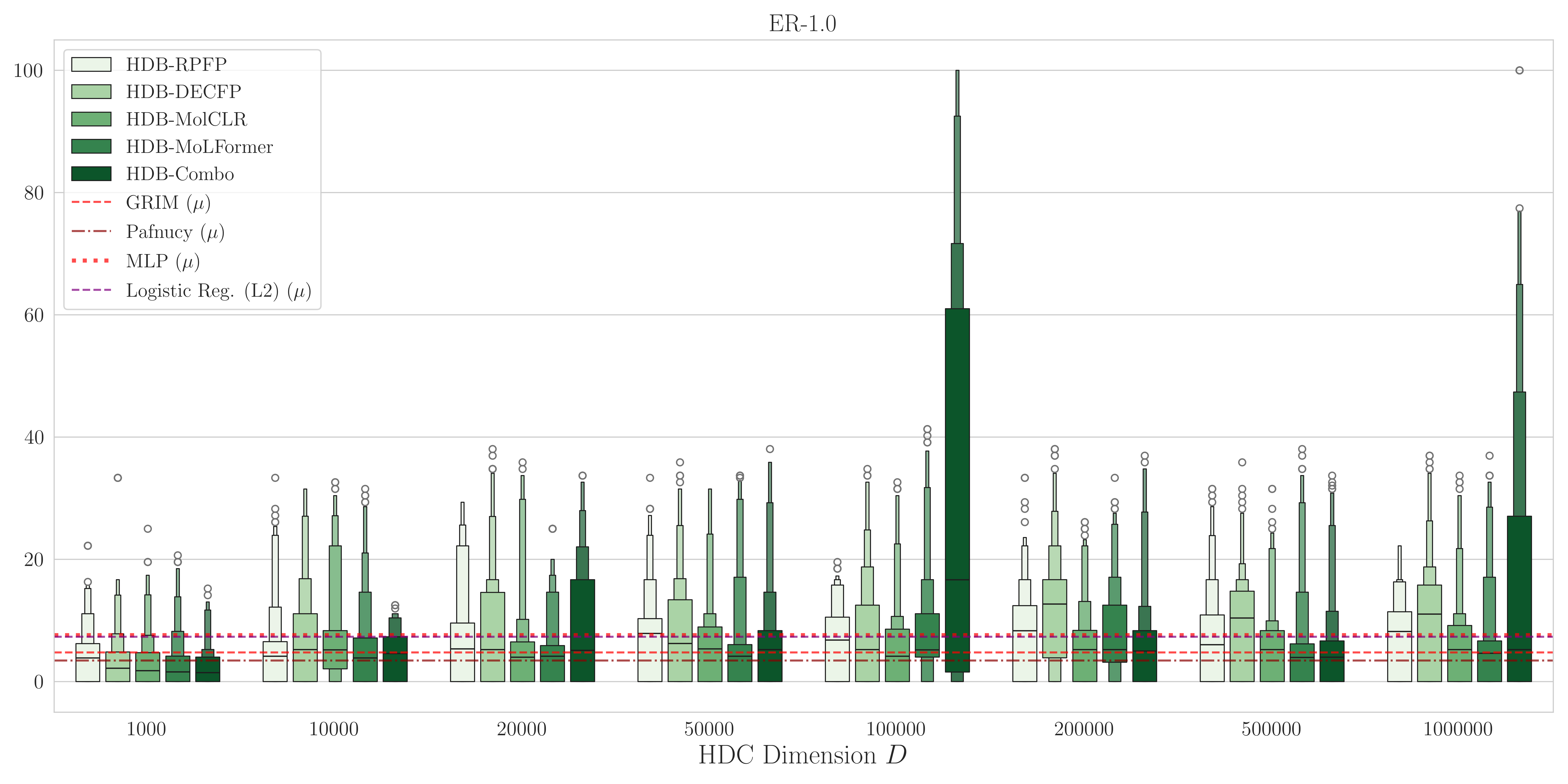}
    \caption{
    Boxenplots of the $\text{ER-}1\%$ roc-enrichment metric for HDBind models we present on the AVE split of the LIT-PCBA dataset. The red dashed line refers to the mean previously reported best overall (re-scoring) method on LIT-PCBA, GRIM\cite{Tran-Nguyen2021-jw, Desaphy2013-jt}. The dark red dotted dash line represents the previously reported Pafnucy 3D-CNN result on LIT-PCBA\cite{Stepniewska-Dziubinska2018-fo, Tran-Nguyen2021-jw}. The dotted red line denotes the mean $\text{ER-}1\%$ metric for our MLP baseline. The purple dashed line denotes our logistic regression baseline. For both Pafnucy and GRIM, we report the mean $\text{ER-}1\%$ over all 15 protein targets. For our MLP and logistic regression baselines, we report the mean over all 15 datasets and 10 random seeds. Additional sample sizes are included in SI Figures \ref{S-fig:er_01}-\ref{S-fig:er_5}.
    }
    \label{fig:lit-pcba_er1_boxenplot}
\end{figure}

\subsubsection*{Virtual molecular lead identification}
 The problem of virtual screening requires the specification of a scoring function that is applied to each of the candidate molecules, then these molecules are ranked accordingly then a filtered set above some threshold of the scoring function is selected for further processing with progressively more accurate but expensive algorithms. Scoring functions that approximate the experimental binding activity can be roughly divided into those that rely upon physics theory, machine learning, or some combination of the two\cite{Gentile2020-tj, Stafford2022-ov, Clyde2023-vk, Clyde2022-na, Lau2021-dr, Stevenson2021-nw}. A general workflow then first applies faster but less accurate docking methods, followed by more expensive and accurate calculations based on MM/GBSA or MD simulations\cite{Lau2021-dr}. 
 Physics-based methods such as \textit{molecular docking}~\cite{Trott2010-ij, Eberhardt2021-ql} are generally believed to be on the ``fast'' end of the spectrum of accuracy versus latency. More accurate methods including molecular mechanics/generalized Born surface area (MM/GBSA),\cite{Massova2000-qu, Greenidge2013-iz} which provides a more accurate binding energy calculation for a given docking pose, or binding free-energy calculations based upon intensive atomistic molecular dynamics (MD) simulations, are infeasible to run for even a relatively small number of candidate possibilities\cite{Wright2014-uq, Eberhardt2021-ql}. 
Benchmark datasets have long been used to validate a scoring function's ability to distinguish active versus inactive molecules for a given protein target\cite{Huang2006-gz, Mysinger2012-hn}. Recent research has identified limitations that have made these datasets trivial to learn thus overestimating the expected generalization performance when applied to real-world datasets\cite{Chaput2016-qr, Wallach2018-lf, Chen2019-gd, Sieg2019-gy, Tran-Nguyen2020-xq, Tran-Nguyen2021-jw}. 
The recently proposed LIT-PCBA\cite{Tran-Nguyen2020-xq} benchmark dataset is derived from high-confidence PubChem assay data  (7,761 actives and 382,674 unique inactives, 1:50 class ratio) and provides a rigorous test set constructed using the Atomwise-developed AVE (asymmetric validation embedding) bias-minimizing algorithm\cite{Jiang2021-wr, Tran-Nguyen2021-jw}. 
We additionally use a random stratified split of each protein-target specific dataset as a control with a 75\%/25\% train/test split ratio. To our knowledge, this represents the first demonstration of an HDC approach on a dataset of experimentally determined binding measurements of this scale of 100s of thousands\cite{Ma2022-wu}.
\subsubsection*{Enrichment results on LIT-PCBA}
In Figure \ref{fig:lit-pcba_er1_boxenplot}, we choose to report the roc-enrichment factor ($\text{ER-}1\%$) metric (eq. \ref{eq:roc-enrich_factor})\cite{Jain2007-di, Cleves2020-ye, Tran-Nguyen2021-jw}.
We consider two representative alternative approaches for molecular screening using either machine learning or physics-based knowledge, Pafnucy\cite{Stepniewska-Dziubinska2018-fo} and GRIM\cite{Desaphy2013-jt}. Pafnucy is a 3D Convolutional Neural Network (3D-CNN) trained on the PDBBind\cite{Su2019-ni} dataset to predict the binding affinity of a protein-ligand complex\cite{Stepniewska-Dziubinska2018-fo}. GRIM\cite{Desaphy2013-jt} is a fingerprint method that transforms the 3D atomic information, described using physics-based knowledge, in to a vector of 210 integers describing the molecular interaction which are then used as the basis of the GRscore. Each of these methods requires a molecular docking simulation to generate plausible 3D structures of the binding complex\cite{Jain2007-di}. Our proposed HDBind models considerably outperform our implementation of the MoleHD (using PyTorch) baseline with Smiles Pair Encoding (SPE)\cite{Li2021-sb} (SI Table \ref{S-tab:lit-pcba-er-mean}). In Figure \ref{fig:lit-pcba_er1_boxenplot} we give results compared to each of the representative methods we described. For dimension size $D<10$k, our HDBind methods generally perform competitively with the GRIM and Pafnucy approaches across each molecular encoding approach. For $D>10$k however, a noticeable improvement is observed for the HDB-Combo, which combines the graph structural information provided by the ECFP encoded into hypervectors (DECFP) with the pretrained representation extracted from the MoLFormer SMILES LLM\cite{Ross2022-ce}. Moreover, the performance of all HDBind molecular encoding methods tend to improve beyond the performance of GRIM and Pafnucy with increasing dimension size. To the best of our knowledge, our results, including HDB-Combo, represent the largest improvement in performance on this task that has been published to date\cite{Tran-Nguyen2021-jw}.

\subsubsection*{Classification results on LIT-PCBA}
\begin{figure}
    \centering
    \includegraphics[width=0.95\textwidth]{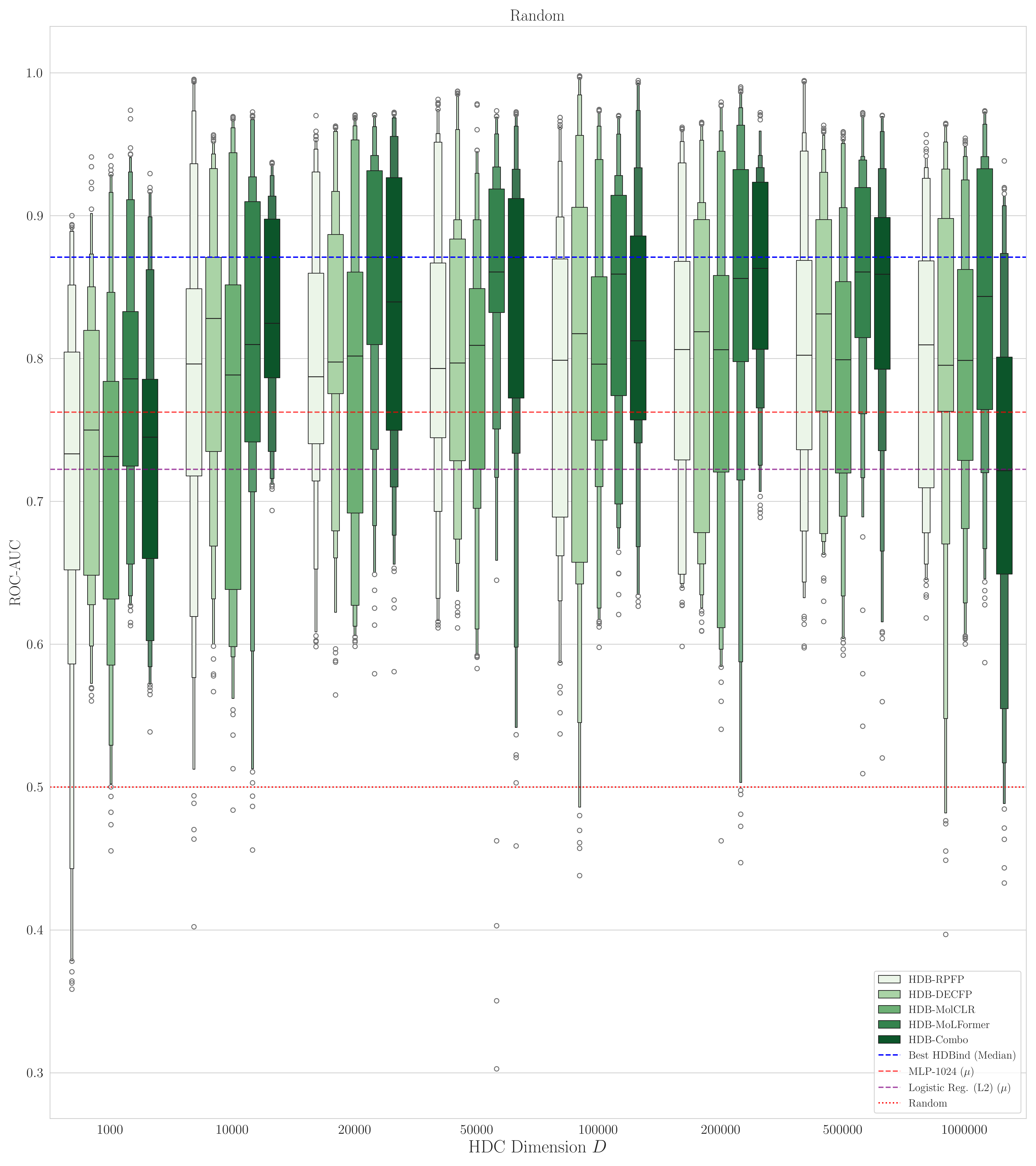}
    \caption{Boxenplot Comparison of ROC-AUC metric across different HDB model input representations and dimension size $D$ on our random split of the LIT-PCBA dataset. The red and purple dashed lines represents the mean roc-auc over all 15 datasets for our MLP and logistic regression baseline models respectively. The dotted red line denotes random performance. The blue dashed line corresponds to the best HDBind ROC-AUC distribution.
    }
    \label{fig:lit-pcba_rocauc_boxenplot_random}
\end{figure}

\begin{figure}
    \centering
    \includegraphics[width=0.95\textwidth]{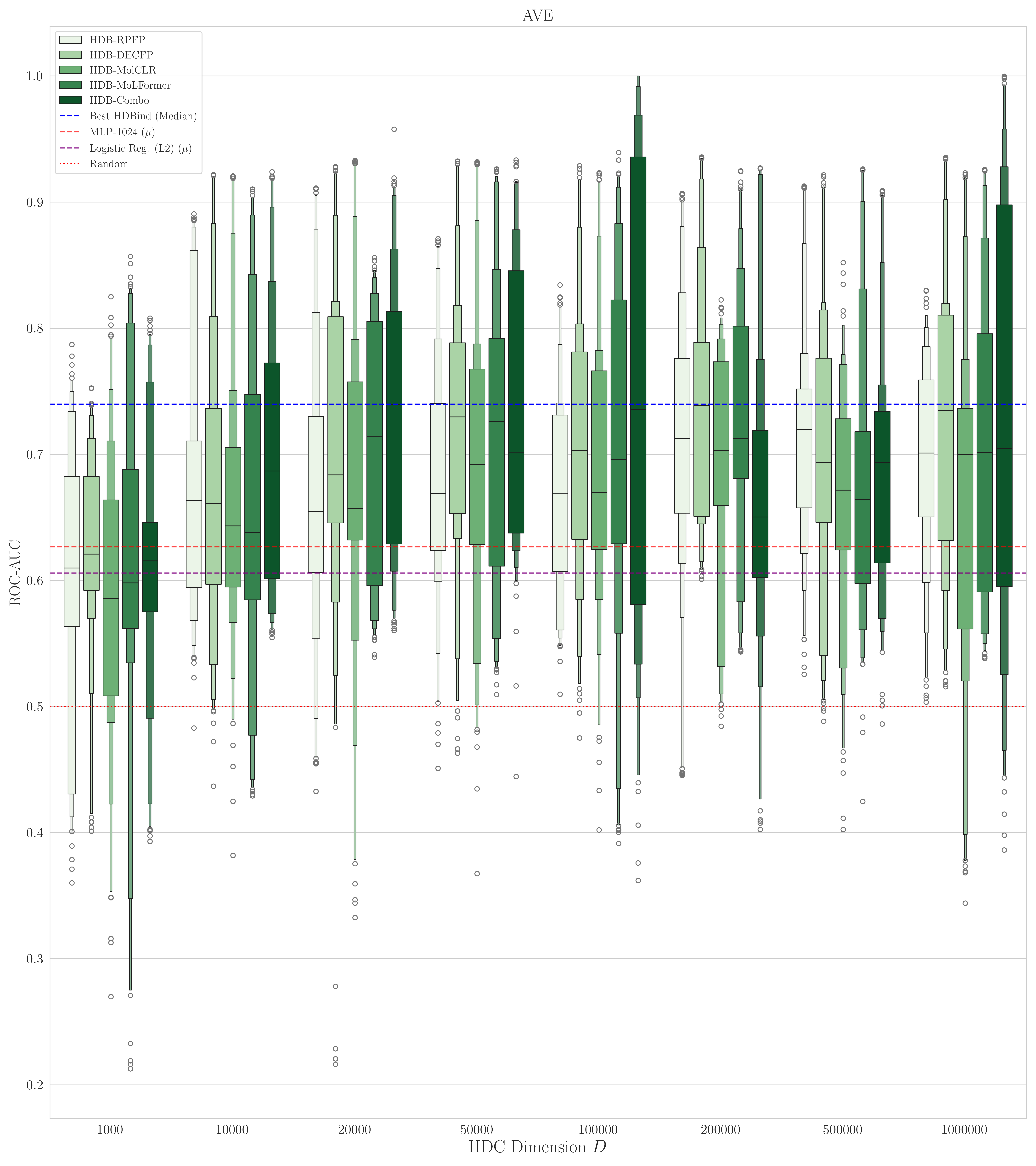}
    \caption{Boxenplot Comparison of ROC-AUC metric across different HDB model input representations and dimension size $D$ on the bias-minimizing AVE split of the LIT-PCBA dataset.
    The red and purple dashed lines represents the mean roc-auc over all 15 datasets for our MLP and logistic regression baseline models respectively. The dotted red line denotes random performance. The blue dashed line corresponds to the best HDBind ROC-AUC distribution.
    }
    \label{fig:lit-pcba_rocauc_boxenplot_ave}
\end{figure}

We choose the MLP as our baseline in order to compare against a standard approach for downstream prediction tasks that is relatively efficient compared to more complex approaches\cite{Gentile2020-tj, Jones2021-al, Ross2022-ce, Wang2022-hg} for which demonstrating energy efficiency would be trivial (Table \ref{tab:encode_energy_costs}). The MLP is trained directly on the ECFP representation to predict the binding activity of a drug molecule on each dataset, with no protein or 3D-structure information provided. The ECFP is generated using length 1024 and radius of 1. In this evaluation, we consider two splits of the dataset, a random stratified split, and the AVE split, to respectively assess model performance when making predictions on molecules \textit{similar} to the training set and when making predictions on molecules that are maximally \textit{dissimilar} to the training set.
In Figures \ref{fig:lit-pcba_rocauc_boxenplot_random} and \ref{fig:lit-pcba_rocauc_boxenplot_ave} we characterise the effect of hypervector dimension choice $D$ on classification performance of the active versus inactive molecules for both splits using the ROC-AUC metric. 
In the case of the random split, scaling $D$ beyond 10k however does not yield increasing returns as the models appear to saturate or even degrade performance with $D=1,000,000$. Despite this, for values of $D\geq10$k, nearly all models outperform our MLP baseline. However in the case of the AVE split, the benefit of increasing dimensionality is more pronounced as nearly every model considered benefits from larger vector representations, most noticeably the HDB-Combo method. Again, for values of $D\geq10$k, each our models outperform our MLP baseline model. We provide additional statistical significance of our results versus the baseline MLP model in SI Figures \ref{S-fig:stat_sig_er_1_random}-\ref{S-fig:stat_sig_roc_ave}.

\section*{Discussion}
\begin{figure}
    \centering
    \includegraphics[width=.8\textwidth]{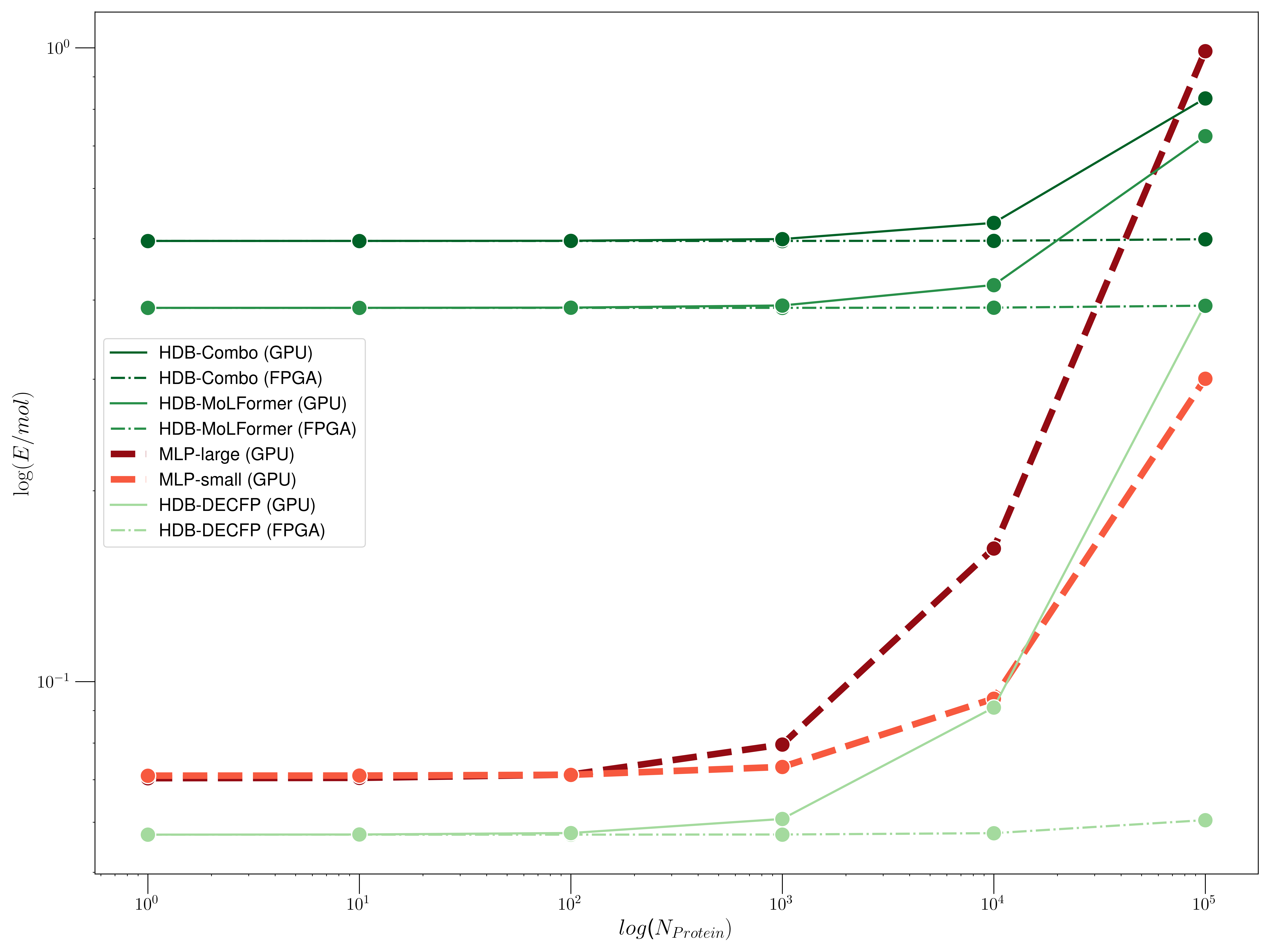}
    \caption{
    Energy usage of HDBind versus our largest and smallest MLP baselines, MLP-large and MLP-small respectively, versus number of protein targets screened with a fixed library of molecules. Energy is reported in terms of expenditure per molecule.
    We report the mean values per molecule using the HIV dataset from MoleculeNet\cite{Wu2018-or}. 
    For each method we include the feature extraction and encoding costs for each molecule.
    The HDB-DECFP model immediately outperforms all methods for screens involving a single protein on both the GPU and FPGA hardware. In particular, the HDB-DECFP with FPGA inference maintains the advantage over all screen sizes we consider (SI Table \ref{S-tab:energy_vs_target_num_table}). 
    The HDB-MoLFormer model pays a relatively high initial encoding cost that is amortized sufficiently to outperform the MLP-large baseline at the scale of 10s of thousands of protein targets on GPU and FPGA hardware. HDB-Combo pays the highest encoding cost overall, however at the scale of 10s of thousands of proteins, becomes more efficient than MLP-large when running inference on the FPGA. 
    }
    \label{fig:hdbind_energy}
\end{figure}

\subsection*{Energy efficiency as a performance metric}
Energy efficiency is often over-looked in machine learning based systems in general and specifically for virtual drug screening\cite{Schwartz2020-bl}. A current trend in development often results in the use of a select set of foundation models that are expensive to train, but provide highly informative feature representations that can be leveraged for specific tasks. Increasingly these models are learned using self-supervision and result in state of the art performance in molecular property prediction\cite{Ross2022-ce, Wang2022-hg}. Our setting assumes that for a given collection of drug-like molecules (purchasable libraries\cite{enamine_real}), a feature extraction step (compute ECFP, MoLFormer or MolCLR embedding, etc.) will be performed once, resulting in a fixed cost that can be amortized over time with multiple subsequent screens performed as novel proteins are encountered and potentially complex queries over the interaction space are performed (novel viruses, mutations of known proteins, comparison of activity for new chemistry, etc.). 

\subsection*{Feature extraction and encoding}
We provide data describing the efficiency of the various feature extraction techniques we consider (SI Tables \ref{S-tab:feature_latency} and \ref{S-tab:ecfp_energy_latency}). 
The ECFP was found to be the most efficient representation to compute.
Our results confirm that while the MoLFormer\cite{Ross2022-ce} embedding helps to achieve our best models on the LIT-PCBA dataset with respect to the ROC-AUC metric, the energy penalty per molecule paid to achieve these models limits their utility versus a baseline MLP model using an ECFP embedding (Table \ref{tab:encode_energy_costs}). We additionally consider the combination of the ECFP and MoLFormer methods by element-wise multiplication (i.e. binding) of the constituent hypervectors.
\subsection*{HDBind inference on FPGA hardware}
The popularity of GPUs lies within their superior performance in exploiting parallelism relative to CPUs, which is further enhanced by their relative ease of programming compared to other hardware platforms\cite{Jones2022-dr}. FPGAs however have found utility as target platforms for energy efficient algorithm implementations due to their relative high degree of flexibility available to developers to use on-chip resources. The price of the resource can be amortized over time by savings in energy costs versus GPU hardware implementations. In this context, we explored key components of HDBind individually, synthesizing and implementing them based on insights from HD2FPGA\cite{hd2fpga} using Vitis HLS 2021.2\cite{vitis}. 
This approach facilitated the evaluation of these components on the Xilinx Alveo U280 FPGA, enabling the assessment of potential energy efficiency improvements over traditional computing models.
The power measurements, obtained from Vitis Analyzer, provided data on the energy consumption of our implementations. While the algorithm remains consistent with that detailed in the Materials and Methods section, its implementation on FPGA, as guided by the architectural methodologies outlined in the HD2FPGA\cite{hd2fpga}, has afforded us a more granular approach to adjusting parallelism factors. This adaptability not only enhances the efficiency of our current implementations but also ensures that our approach can be scaled up for more capable future devices.

\subsection*{Energy analysis on GPU and FPGA}

\begin{figure}[H]
    \centering
    \includegraphics[width=\linewidth]{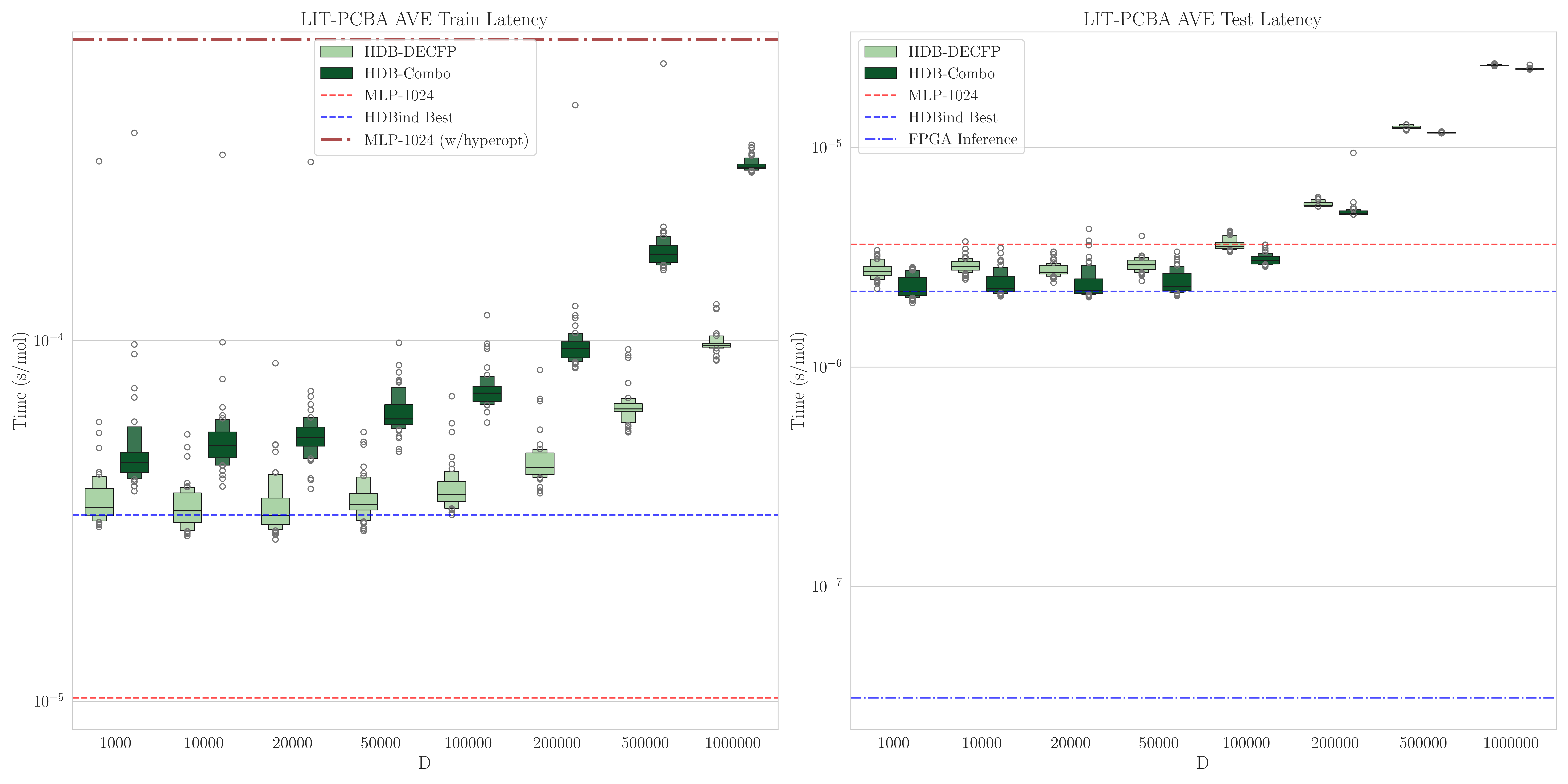}
    \caption{Processing latency measurements on the training (left) and testing (right) sets of LIT-PCBA. Times correspond to the GPU execution time per molecule measured using the PyTorch CUDA backend. (Left) Horizontal lines for MLP and HDBind correspond to the median time for training. The HDBind methods maintain similar training times for values of $D$ up to approximately 100k. When considering the overhead for hyperparameter optimization for the MLP, HDBind demonstrates improved latency for all values of $D$ considered in this study, including $D=1,000,000$. (Right) Horizontal lines for MLP and HDBind correspond to the median time for training. The horizontal line for FPGA inference represents the mean time per molecule. The HDBind methods maintain similar testing times for values of $D$ up to 100k. Inference on the FPGA for HDBind is over an order of magnitude faster than the MLP baseline.}
    \label{fig:train_and_test_times}
\end{figure}

Figure \ref{fig:hdbind_energy} illustrates the advantage of using HDBind when presented with increasingly large collections of protein targets. 
For all models, we consider the energy for encoding and testing steps. We measure the power required for encoding as the sum of the average power output of the CPU and GPU. For testing, HDBind uses either a custom kernel for similarity search on the GPU\cite{Kang2022-rv} or FPGA\cite{Zhang2023-yy}. We measure the power required for testing using the average power output of the GPU or FPGA in isolation.
HDB-DECFP is the most efficient encoding method overall and is expectantly similar to the energy required for the MLP encodings as each are simply computing the ECFP representation (Table \ref{tab:encode_energy_costs}). HDB-MoLFormer and HDB-Combo each require extraction of the LLM embedding on the GPU which imposes a relatively large overall encoding penalty ($\sim 0.39$ J/mol) that requires larger numbers of proteins (10s of thousands) to be screened before the energy efficiency improvement of the testing step is realized versus the MLP baseline models (Table \ref{tab:encode_energy_costs}).
Compared to inference on the GPU, the inference energy efficiency improves by a factor of $4.5\times$ when considering the FPGA for inference. The improvement on FPGA is also considerable compared to both the largest and smallest MLP architectures (approx. $12.2\times$, $3.1 \times$ respectively) (Table \ref{tab:test_energy_costs}). 
As a virtual screen campaign scales beyond the consideration of a single protein, for a fixed library of molecules, the improvement in energy usage that is attained by HDBind on the FPGA grows by several orders of magnitude compared to the GPU-based HDBind implementation and the MLP baselines considered in our work. In Figure \ref{fig:train_and_test_times} we give the latency of each method for the training and testing steps, normalized per molecule. When considering the overhead incurred from hyperparameter optimization of the MLP baseline, training an HDBind model becomes approximately an order of magnitude more efficient for most values of $D$ that we considered. For testing, the HDBind models demonstrate a slight improvement over the MLP for values of $D$ approaching 100k. When considering the FPGA for testing, the advantage grows approximately by an order of magnitude over the GPU-accelerated MLP.
Given the inherent hardware-friendly characteristics of our implementation and the encouraging outcomes, advancing towards Application Specific Integrated Circuit (ASIC) development is promising, offering substantial benefits in efficiency and performance.

\begin{table}[ht]
    \centering
    \begin{tabular}{c|c}
    \textbf{Method} & \textbf{Encode (J/mol)}\\ \hline
        HDB-DECFP & 0.06 \\
        HDB-MoLFormer & 0.39\\
        HDB-Combo & 0.50\\
        MLP-small & 0.07\\
        MLP-large &  0.07\\
    \end{tabular}
    \caption{
    Encode energy estimates for each processor choice for the HDBind. We calculate power using the sum of the average CPU and GPU power output for all models.
    We include the energy for all feature extraction steps for all models (including MLP baselines) in addition to the encoding costs for HDBind models. Models that only require ECFP computation possess the lowest energy penalties (MLP baslines and HDB-DECFP).
    }
    \label{tab:encode_energy_costs}
\end{table}

\begin{table}[ht]
    \centering
    \begin{tabular}{c|c|c|c}
    \textbf{Method} & \textbf{Device} & \textbf{Test (J/mol)}& \textbf{Improvement Factor}\\ \hline

        HDBind & FPGA\cite{Zhang2023-yy} & .75 & 12.2\\
        HDBind & GPU\cite{Kang2022-rv} & 3.37 &  2.7\\
        MLP-small & GPU & 2.30 & 4.0 \\
        MLP-large & GPU &9.18 & 1.0 \\
        
    \end{tabular}
    \caption{
    Test energy estimated for each processor choice for HDBind inference\cite{Kang2022-rv, Zhang2023-yy} versus the MLP baseline models. Values are scaled by $10^{-6}$ for reader convenience.
    }
    \label{tab:test_energy_costs}
\end{table}

\section*{Conclusion}
We have demonstrated the first comprehensive study of structure-based molecular encoding methods for HDC that additionally demonstrate consistent improvement over the competing SOA SMILES-based approaches. We additionally are the first to consider the use of molecular foundation models with HDC that leverage self-supervised learning on large unlabelled collections of SMILES strings and molecular graphs as input, demonstrating an improvement over SOA and ECFP-based approaches. Additionally we compare the performance of all methods on a broad collection of quantitative and qualitative molecular property prediction tasks in the well-studied MoleculeNet. We are the first to demonstrate the viability of HDC on a challenging benchmark protein-drug activity dataset, LIT-PCBA, which outperform physics-based molecular docking while being competitive with our MLP baseline method. The analysis of the computational burden and energy usage show clear advantages to using HDC approaches. 
We leave it to future work to improve the decision boundary learned by the HDC model, one approach employs metric learning to pretrain the projection layer, as considered in previous HDC works\cite{Xu2023-zg}. Improving the latency of the feature extraction and encoding steps can help improve the efficiency of the overall system\cite{Gu2023-iv}. In the case of the self-supervised representations we considered, MolCLR and MoLFormer, improving the energy efficiency of these steps will lower the threshold at which these methods will become more attractive compared to our MLP baseline.
Our code is available for use with the publicly available datasets, to enable reproduction of our study at \url{https://github.com/LLNL/hdbind}.

\bibliography{bibliography}

@article{liaw2018tune,
    title={Tune: A Research Platform for Distributed Model Selection and Training},
    author={Liaw, Richard and Liang, Eric and Nishihara, Robert
            and Moritz, Philipp and Gonzalez, Joseph E and Stoica, Ion},
    journal={arXiv preprint arXiv:1807.05118},
    year={2018}
}

@ARTICLE{Rahimi2017-ox,
  title    = "High-dimensional computing as a nanoscalable paradigm",
  author   = "Rahimi, Abbas and Datta, Sohum and Kleyko, D and Frady, E P and
              Olshausen, B and Kanerva, P and Rabaey, J",
  journal  = "IEEE Trans. Circuits Syst. I Regul. Pap.",
  volume   =  64,
  pages    = "2508--2521",
  month    =  jun,
  year     =  2017
}

@INPROCEEDINGS{Abhijith2021-fi,
  title     = "Neuromorphic High Dimensional Computing Architecture for
               Classification Applications",
  booktitle = "2021 {IEEE} International Conference on Nanoelectronics,
               Nanophotonics, Nanomaterials, Nanobioscience \& Nanotechnology
               ({5NANO})",
  author    = "Abhijith, M and Nair, Devika R",
  publisher = "IEEE",
  pages     = "1--10",
  month     =  apr,
  year      =  2021
}

@ARTICLE{Xu2023-iq,
  title    = "{HyperSpec}: Ultrafast Mass Spectra Clustering in
              Hyperdimensional Space",
  author   = "Xu, Weihong and Kang, Jaeyoung and Bittremieux, Wout and Moshiri,
              Niema and Rosing, Tajana",
  journal  = "J. Proteome Res.",
  volume   =  22,
  number   =  6,
  pages    = "1639--1648",
  month    =  jun,
  year     =  2023,
  language = "en"
}

@ARTICLE{Weininger1988-mh,
  title     = "{SMILES}, a chemical language and information system. 1.
               Introduction to methodology and encoding rules",
  author    = "Weininger, David",
  journal   = "J. Chem. Inf. Comput. Sci.",
  publisher = "American Chemical Society",
  volume    =  28,
  number    =  1,
  pages     = "31--36",
  month     =  feb,
  year      =  1988
}

@ARTICLE{Rumelhart1986-ul,
  title     = "Learning representations by back-propagating errors",
  author    = "Rumelhart, David E and Hinton, Geoffrey E and Williams, Ronald J",
  journal   = "Nature",
  publisher = "Nature Publishing Group",
  volume    =  323,
  number    =  6088,
  pages     = "533--536",
  month     =  oct,
  year      =  1986,
  language  = "en"
}

@INPROCEEDINGS{Zhang2023-yy,
  title     = "{HD2FPGA}: Automated Framework for Accelerating Hyperdimensional
               Computing on {FPGAs}",
  booktitle = "2023 24th International Symposium on Quality Electronic Design
               ({ISQED})",
  author    = "Zhang, Tinaqi and Salamat, Sahand and Khaleghi, Behnam and
               Morris, Justin and Aksanli, Baris and Rosing, Tajana Simunic",
  publisher = "IEEE",
  pages     = "1--9",
  month     =  apr,
  year      =  2023
}

@ARTICLE{Kang2022-rv,
  title     = "{OpenHD}: A {GPU-Powered} Framework for Hyperdimensional
               Computing",
  author    = "Kang, Jaeyoung and Khaleghi, Behnam and Rosing, Tajana and Kim,
               Yeseong",
  journal   = "IEEE Trans. Comput.",
  publisher = "Institute of Electrical and Electronics Engineers (IEEE)",
  volume    =  71,
  number    =  11,
  pages     = "2753--2765",
  month     =  nov,
  year      =  2022,
  copyright = "https://ieeexplore.ieee.org/Xplorehelp/downloads/license-information/IEEE.html"
}

@INPROCEEDINGS{Dutta2022-kl,
  title     = "{HDnn-PIM}: Efficient in Memory Design of Hyperdimensional
               Computing with Feature Extraction",
  booktitle = "Proceedings of the Great Lakes Symposium on {VLSI} 2022",
  author    = "Dutta, Arpan and Gupta, Saransh and Khaleghi, Behnam and
               Chandrasekaran, Rishikanth and Xu, Weihong and Rosing, Tajana",
  publisher = "Association for Computing Machinery",
  pages     = "281--286",
  series    = "GLSVLSI '22",
  month     =  jun,
  year      =  2022,
  address   = "New York, NY, USA",
  location  = "Irvine, CA, USA"
}

@ARTICLE{Jain2007-di,
  title    = "{Surflex-Dock} 2.1: robust performance from ligand energetic
              modeling, ring flexibility, and knowledge-based search",
  author   = "Jain, Ajay N",
  journal  = "J. Comput. Aided Mol. Des.",
  volume   =  21,
  number   =  5,
  pages    = "281--306",
  month    =  may,
  year     =  2007,
  language = "en"
}

@ARTICLE{Desaphy2013-jt,
  title    = "Encoding protein-ligand interaction patterns in fingerprints and
              graphs",
  author   = "Desaphy, J{\'e}r{\'e}my and Raimbaud, Eric and Ducrot, Pierre and
              Rognan, Didier",
  journal  = "J. Chem. Inf. Model.",
  volume   =  53,
  number   =  3,
  pages    = "623--637",
  month    =  mar,
  year     =  2013,
  language = "en"
}

@ARTICLE{Jain2008-nr,
  title    = "Recommendations for evaluation of computational methods",
  author   = "Jain, Ajay N and Nicholls, Anthony",
  journal  = "J. Comput. Aided Mol. Des.",
  volume   =  22,
  number   = "3-4",
  pages    = "133--139",
  month    =  mar,
  year     =  2008,
  language = "en"
}

@ARTICLE{Cleves2020-ye,
  title    = "Structure- and {Ligand-Based} Virtual Screening on {DUD-E+}:
              Performance Dependence on Approximations to the Binding Pocket",
  author   = "Cleves, Ann E and Jain, Ajay N",
  journal  = "J. Chem. Inf. Model.",
  volume   =  60,
  number   =  9,
  pages    = "4296--4310",
  month    =  sep,
  year     =  2020,
  language = "en"
}

@MISC{vitis,
  title        = "{AMD} Technical Information Portal",
  howpublished = "\url{https://docs.amd.com/r/en-US/ug1399-vitis-hls}",
  note         = "Accessed: 2024-5-30",
  language     = "en"
}

@MISC{variorum,
authors = {Brink, Stephanie and Patki, Tapasya and Rountree, Barry and Marathe, Aniruddha and Shoga, Kathleen and Green, Elena},
  title       = " Variorum: Vendor-Agnostic Computing Power Management",
  institution = "Lawrence Livermore National Laboratory",
  language    = "en",
url = {https://ipo.llnl.gov/sites/default/files/2023-08/Final_variorum-rnd-100-award.pdf}
}

@ARTICLE{Baek2021-wq,
  title    = "Accurate prediction of protein structures and interactions using
              a three-track neural network",
  author   = "Baek, Minkyung and DiMaio, Frank and Anishchenko, Ivan and
              Dauparas, Justas and Ovchinnikov, Sergey and Lee, Gyu Rie and
              Wang, Jue and Cong, Qian and Kinch, Lisa N and Schaeffer, R
              Dustin and Mill{\'a}n, Claudia and Park, Hahnbeom and Adams,
              Carson and Glassman, Caleb R and DeGiovanni, Andy and Pereira,
              Jose H and Rodrigues, Andria V and van Dijk, Alberdina A and
              Ebrecht, Ana C and Opperman, Diederik J and Sagmeister, Theo and
              Buhlheller, Christoph and Pavkov-Keller, Tea and Rathinaswamy,
              Manoj K and Dalwadi, Udit and Yip, Calvin K and Burke, John E and
              Garcia, K Christopher and Grishin, Nick V and Adams, Paul D and
              Read, Randy J and Baker, David",
  journal  = "Science",
  volume   =  373,
  number   =  6557,
  pages    = "871--876",
  month    =  aug,
  year     =  2021,
  language = "en"
}

@ARTICLE{Grygorenko2020-nw,
  title    = "Generating Multibillion Chemical Space of Readily Accessible
              Screening Compounds",
  author   = "Grygorenko, Oleksandr O and Radchenko, Dmytro S and Dziuba, Igor
              and Chuprina, Alexander and Gubina, Kateryna E and Moroz, Yurii S",
  journal  = "iScience",
  volume   =  23,
  number   =  11,
  pages    = "101681",
  month    =  nov,
  year     =  2020,
  language = "en"
}

@ARTICLE{You2019-an,
  title         = "Large Batch Optimization for Deep Learning: Training {BERT}
                   in 76 minutes",
  author        = "You, Yang and Li, Jing and Reddi, Sashank and Hseu, Jonathan
                   and Kumar, Sanjiv and Bhojanapalli, Srinadh and Song,
                   Xiaodan and Demmel, James and Keutzer, Kurt and Hsieh,
                   Cho-Jui",
  month         =  apr,
  year          =  2019,
  archivePrefix = "arXiv",
  primaryClass  = "cs.LG",
  journal       = "arXiv",
  eprint        = "1904.00962"
}

@ARTICLE{Kingma2014-ti,
  title         = "Adam: A Method for Stochastic Optimization",
  author        = "Kingma, Diederik P and Ba, Jimmy",
  month         =  dec,
  year          =  2014,
  archivePrefix = "arXiv",
  primaryClass  = "cs.LG",
  journal       = "arXiv",
  eprint        = "1412.6980"
}

@ARTICLE{Gu2023-iv,
  title         = "{MiniLLM}: Knowledge Distillation of large language models",
  author        = "Gu, Yuxian and Dong, Li and Wei, Furu and Huang, Minlie",
  month         =  jun,
  year          =  2023,
  copyright     = "http://creativecommons.org/licenses/by/4.0/",
  archivePrefix = "arXiv",
  journal       = "arXiv",
  primaryClass  = "cs.CL",
  eprint        = "2306.08543"
}

@ARTICLE{Li2021-sb,
  title    = "{SMILES} Pair Encoding: A {Data-Driven} Substructure Tokenization
              Algorithm for Deep Learning",
  author   = "Li, Xinhao and Fourches, Denis",
  journal  = "J. Chem. Inf. Model.",
  volume   =  61,
  number   =  4,
  pages    = "1560--1569",
  month    =  apr,
  year     =  2021,
  language = "en"
}

@ARTICLE{Eberhardt2021-ql,
  title    = "{AutoDock} Vina 1.2.0: New Docking Methods, Expanded Force Field,
              and Python Bindings",
  author   = "Eberhardt, Jerome and Santos-Martins, Diogo and Tillack, Andreas
              F and Forli, Stefano",
  journal  = "J. Chem. Inf. Model.",
  volume   =  61,
  number   =  8,
  pages    = "3891--3898",
  month    =  aug,
  year     =  2021,
  language = "en"
}

@ARTICLE{Huang2006-gz,
  title    = "Benchmarking sets for molecular docking",
  author   = "Huang, Niu and Shoichet, Brian K and Irwin, John J",
  journal  = "J. Med. Chem.",
  volume   =  49,
  number   =  23,
  pages    = "6789--6801",
  month    =  nov,
  year     =  2006,
  language = "en"
}

@ARTICLE{Clyde2022-na,
  title    = "{High-Throughput} Virtual Screening and Validation of a
              {SARS-CoV-2} Main Protease Noncovalent Inhibitor",
  author   = "Clyde, Austin and Galanie, Stephanie and Kneller, Daniel W and
              Ma, Heng and Babuji, Yadu and Blaiszik, Ben and Brace, Alexander
              and Brettin, Thomas and Chard, Kyle and Chard, Ryan and Coates,
              Leighton and Foster, Ian and Hauner, Darin and Kertesz, Vilmos
              and Kumar, Neeraj and Lee, Hyungro and Li, Zhuozhao and Merzky,
              Andre and Schmidt, Jurgen G and Tan, Li and Titov, Mikhail and
              Trifan, Anda and Turilli, Matteo and Van Dam, Hubertus and
              Chennubhotla, Srinivas C and Jha, Shantenu and Kovalevsky, Andrey
              and Ramanathan, Arvind and Head, Martha S and Stevens, Rick",
  journal  = "J. Chem. Inf. Model.",
  volume   =  62,
  number   =  1,
  pages    = "116--128",
  month    =  jan,
  year     =  2022,
  language = "en"
}

@ARTICLE{Stafford2022-ov,
  title    = "{AtomNet} {PoseRanker}: Enriching Ligand Pose Quality for Dynamic
              Proteins in Virtual {High-Throughput} Screens",
  author   = "Stafford, Kate A and Anderson, Brandon M and Sorenson, Jon and
              van den Bedem, Henry",
  journal  = "J. Chem. Inf. Model.",
  volume   =  62,
  number   =  5,
  pages    = "1178--1189",
  month    =  mar,
  year     =  2022,
  language = "en"
}

@INPROCEEDINGS{Xu2023-zg,
  title     = "{HyperMetric}: Robust Hyperdimensional Computing on Error-prone
               Memories using Metric Learning",
  booktitle = "2023 {IEEE} 41st International Conference on Computer Design
               ({ICCD})",
  author    = "{Xu} and {Swaminathan} and {Pinge} and {Fuhrman} and {Rosing}",
  volume    =  0,
  pages     = "243--246",
  month     =  nov,
  year      =  2023
}

@ARTICLE{Zdrazil2024-xc,
  title    = "The {ChEMBL} Database in 2023: a drug discovery platform spanning
              multiple bioactivity data types and time periods",
  author   = "Zdrazil, Barbara and Felix, Eloy and Hunter, Fiona and Manners,
              Emma J and Blackshaw, James and Corbett, Sybilla and de Veij,
              Marleen and Ioannidis, Harris and Lopez, David Mendez and
              Mosquera, Juan F and Magarinos, Maria Paula and Bosc, Nicolas and
              Arcila, Ricardo and Kizil{\"o}ren, Tevfik and Gaulton, Anna and
              Bento, A Patr{\'\i}cia and Adasme, Melissa F and Monecke, Peter
              and Landrum, Gregory A and Leach, Andrew R",
  journal  = "Nucleic Acids Res.",
  volume   =  52,
  number   = "D1",
  pages    = "D1180--D1192",
  month    =  jan,
  year     =  2024,
  language = "en"
}

@INPROCEEDINGS{Devlin2019-vh,
  title     = "{{BERT}}: Pre-training of Deep Bidirectional Transformers for
               Language Understanding",
  booktitle = "Proceedings of the 2019 Conference of the North {A}merican
               Chapter of the Association for Computational Linguistics: Human
               Language Technologies, Volume 1 (Long and Short Papers)",
  author    = "Devlin, Jacob and Chang, Ming-Wei and Lee, Kenton and Toutanova,
               Kristina",
  editor    = "Burstein, Jill and Doran, Christy and Solorio, Thamar",
  publisher = "Association for Computational Linguistics",
  pages     = "4171--4186",
  month     =  jun,
  year      =  2019,
  address   = "Minneapolis, Minnesota"
}

@ARTICLE{Liu2019-lz,
  title         = "{RoBERTa}: A Robustly Optimized {BERT} Pretraining Approach",
  author        = "Liu, Yinhan and Ott, Myle and Goyal, Naman and Du, Jingfei
                   and Joshi, Mandar and Chen, Danqi and Levy, Omer and Lewis,
                   Mike and Zettlemoyer, Luke and Stoyanov, Veselin",
  month         =  jul,
  year          =  2019,
  archivePrefix = "arXiv",
  journal = "arXiv",
  primaryClass  = "cs.CL",
  eprint        = "1907.11692"
}

@ARTICLE{Kim2023-vo,
  title    = "{PubChem} 2023 update",
  author   = "Kim, Sunghwan and Chen, Jie and Cheng, Tiejun and Gindulyte, Asta
              and He, Jia and He, Siqian and Li, Qingliang and Shoemaker,
              Benjamin A and Thiessen, Paul A and Yu, Bo and Zaslavsky, Leonid
              and Zhang, Jian and Bolton, Evan E",
  journal  = "Nucleic Acids Res.",
  volume   =  51,
  number   = "D1",
  pages    = "D1373--D1380",
  month    =  jan,
  year     =  2023,
  language = "en"
}

@ARTICLE{Chen2020-ub,
  title         = "A Simple Framework for Contrastive Learning of Visual
                   Representations",
  author        = "Chen, Ting and Kornblith, Simon and Norouzi, Mohammad and
                   Hinton, Geoffrey",
  month         =  feb,
  year          =  2020,
  archivePrefix = "arXiv",
  journal = "arXiv",
  primaryClass  = "cs.LG",
  eprint        = "2002.05709"
}

@ARTICLE{Chithrananda2020-el,
  title         = "{ChemBERTa}: {Large-Scale} {Self-Supervised} Pretraining for
                   Molecular Property Prediction",
  author        = "Chithrananda, Seyone and Grand, Gabriel and Ramsundar,
                   Bharath",
  month         =  oct,
  year          =  2020,
  archivePrefix = "arXiv",
  journal = "arXiv",
  primaryClass  = "cs.LG",
  eprint        = "2010.09885"
}

@ARTICLE{Li2021-od,
  title         = "{GeomGCL}: Geometric Graph Contrastive Learning for
                   Molecular Property Prediction",
  author        = "Li, Shuangli and Zhou, Jingbo and Xu, Tong and Dou, Dejing
                   and Xiong, Hui",
  month         =  sep,
  year          =  2021,
  archivePrefix = "arXiv",
  journal = "arXiv",
  primaryClass  = "cs.LG",
  eprint        = "2109.11730"
}

@ARTICLE{Liu2018-cn,
  title    = "N-gram graph: Simple unsupervised representation for graphs, with
              applications to molecules",
  author   = "Liu, Shengchao and Demirel, M F and Liang, Yingyu",
  journal  = "Adv. Neural Inf. Process. Syst.",
  pages    = "8464--8476",
  month    =  jun,
  year     =  2018
}

@ARTICLE{Yang2019-nv,
  title    = "Analyzing Learned Molecular Representations for Property
              Prediction",
  author   = "Yang, Kevin and Swanson, Kyle and Jin, Wengong and Coley, Connor
              and Eiden, Philipp and Gao, Hua and Guzman-Perez, Angel and
              Hopper, Timothy and Kelley, Brian and Mathea, Miriam and Palmer,
              Andrew and Settels, Volker and Jaakkola, Tommi and Jensen, Klavs
              and Barzilay, Regina",
  journal  = "J. Chem. Inf. Model.",
  volume   =  59,
  number   =  8,
  pages    = "3370--3388",
  month    =  aug,
  year     =  2019,
  language = "en"
}

@ARTICLE{Lu2019-fr,
  title    = "Molecular Property Prediction: A Multilevel Quantum Interactions
              Modeling Perspective",
  author   = "Lu, Chengqiang and Liu, Qi and Wang, Chao and Huang, Zhenya and
              Lin, Peize and He, Lixin",
  journal  = "AAAI",
  volume   =  33,
  number   =  01,
  pages    = "1052--1060",
  month    =  jul,
  year     =  2019,
  language = "en"
}

@INPROCEEDINGS{Bingham2001-rk,
  title     = "Random projection in dimensionality reduction: applications to
               image and text data",
  booktitle = "Proceedings of the seventh {ACM} {SIGKDD} international
               conference on Knowledge discovery and data mining",
  author    = "Bingham, Ella and Mannila, Heikki",
  publisher = "Association for Computing Machinery",
  pages     = "245--250",
  series    = "KDD '01",
  month     =  aug,
  year      =  2001,
  address   = "New York, NY, USA",
  location  = "San Francisco, California"
}

@INCOLLECTION{Kainen1997-cf,
  title     = "Utilizing Geometric Anomalies of High Dimension: When Complexity
               Makes Computation Easier",
  booktitle = "Computer Intensive Methods in Control and Signal Processing: The
               Curse of Dimensionality",
  author    = "Kainen, Paul C",
  editor    = "K{\'a}rn{\'y}, Miroslav and Warwick, Kevin",
  publisher = "Birkh{\"a}user Boston",
  pages     = "283--294",
  year      =  1997,
  address   = "Boston, MA"
}

@MISC{alphafold_database,
  title        = "{AlphaFold} Database",
  author       = "Database, Alphafold Protein Structure",
  howpublished = "\url{https://alphafold.ebi.ac.uk/}",
  note         = "Accessed: 2023-11-13"
}

@ARTICLE{Yu2023-nt,
  title    = "{Uni-Dock}: {GPU-Accelerated} Docking Enables Ultralarge Virtual
              Screening",
  author   = "Yu, Yuejiang and Cai, Chun and Wang, Jiayue and Bo, Zonghua and
              Zhu, Zhengdan and Zheng, Hang",
  journal  = "J. Chem. Theory Comput.",
  volume   =  19,
  number   =  11,
  pages    = "3336--3345",
  month    =  jun,
  year     =  2023,
  language = "en"
}

@ARTICLE{Pinge2023-wn,
  title         = "{SpecHD}: Hyperdimensional Computing Framework for
                   {FPGA-based} Mass Spectrometry Clustering",
  author        = "Pinge, Sumukh and Xu, Weihong and Kang, Jaeyoung and Zhang,
                   Tianqi and Moshiri, Neima and Bittremieux, Wout and Rosing,
                   Tajana",
  month         =  nov,
  year          =  2023,
  archivePrefix = "arXiv",
  primaryClass  = "q-bio.QM",
  eprint        = "2311.12874",
  journal       = "arXiv"
}

@ARTICLE{Schwartz2020-bl,
  title     = "Green {AI}",
  author    = "Schwartz, Roy and Dodge, Jesse and Smith, Noah A and Etzioni,
               Oren",
  journal   = "Commun. ACM",
  publisher = "Association for Computing Machinery",
  volume    =  63,
  number    =  12,
  pages     = "54--63",
  month     =  nov,
  year      =  2020,
  address   = "New York, NY, USA"
}

@ARTICLE{Volkov2022-vc,
  title     = "On the Frustration to Predict Binding Affinities from
               {Protein--Ligand} Structures with Deep Neural Networks",
  author    = "Volkov, Mikhail and Turk, Joseph-Andr{\'e} and Drizard, Nicolas
               and Martin, Nicolas and Hoffmann, Brice and Gaston-Math{\'e},
               Yann and Rognan, Didier",
  journal   = "J. Med. Chem.",
  publisher = "American Chemical Society",
  month     =  may,
  year      =  2022
}

@ARTICLE{Wallach2018-lf,
  title    = "Most {Ligand-Based} Classification Benchmarks Reward Memorization
              Rather than Generalization",
  author   = "Wallach, Izhar and Heifets, Abraham",
  journal  = "J. Chem. Inf. Model.",
  volume   =  58,
  number   =  5,
  pages    = "916--932",
  month    =  may,
  year     =  2018,
  language = "en"
}

@ARTICLE{Minnich2020-iz,
  title    = "{AMPL}: A {Data-Driven} Modeling Pipeline for Drug Discovery",
  author   = "Minnich, Amanda J and McLoughlin, Kevin and Tse, Margaret and
              Deng, Jason and Weber, Andrew and Murad, Neha and Madej, Benjamin
              D and Ramsundar, Bharath and Rush, Tom and Calad-Thomson, Stacie
              and Brase, Jim and Allen, Jonathan E",
  journal  = "J. Chem. Inf. Model.",
  volume   =  60,
  number   =  4,
  pages    = "1955--1968",
  month    =  apr,
  year     =  2020,
  language = "en"
}

@ARTICLE{Jones2021-al,
  title    = "Improved {Protein-Ligand} Binding Affinity Prediction with
              {Structure-Based} Deep Fusion Inference",
  author   = "Jones, Derek and Kim, Hyojin and Zhang, Xiaohua and Zemla, Adam
              and Stevenson, Garrett and Bennett, W F Drew and Kirshner, Daniel
              and Wong, Sergio E and Lightstone, Felice C and Allen, Jonathan E",
  journal  = "J. Chem. Inf. Model.",
  volume   =  61,
  number   =  4,
  pages    = "1583--1592",
  month    =  apr,
  year     =  2021,
  language = "en"
}

@ARTICLE{Su2019-ni,
  title    = "Comparative Assessment of Scoring Functions: The {CASF-2016}
              Update",
  author   = "Su, Minyi and Yang, Qifan and Du, Yu and Feng, Guoqin and Liu,
              Zhihai and Li, Yan and Wang, Renxiao",
  journal  = "J. Chem. Inf. Model.",
  volume   =  59,
  number   =  2,
  pages    = "895--913",
  month    =  feb,
  year     =  2019,
  language = "en"
}

@ARTICLE{Rogers2010-xp,
  title    = "Extended-connectivity fingerprints",
  author   = "Rogers, David and Hahn, Mathew",
  journal  = "J. Chem. Inf. Model.",
  volume   =  50,
  number   =  5,
  pages    = "742--754",
  month    =  may,
  year     =  2010,
  language = "en"
}

@INPROCEEDINGS{Ma2021-xu,
  title     = "{MoleHD}: Efficient Drug Discovery using Brain Inspired
               Hyperdimensional Computing",
  booktitle = "2022 {IEEE} International Conference on Bioinformatics and
               Biomedicine ({BIBM})",
  author    = "Ma, Dongning and Thapa, Rahul and Jiao, Xun",
  pages     = "390--393",
  month     =  dec,
  year      =  2022
}

@ARTICLE{Kang2022-dac,
title="XCelHD: An Efficient GPU-Powered Hyperdimensional Computing
with Parallelized Training",
author="Kang, Jaeyoung and  Khaleghi, Behnam and Kim, Yeseong and Rosing, Tajana",
year=2022,
journal="The 27th Asia and South Pacific Design Automation Conference",}

@ARTICLE{Thomas2021-uo,
  title     = "A Theoretical Perspective on Hyperdimensional Computing",
  author    = "Thomas, Anthony and Dasgupta, Sanjoy and Rosing, Tajana",
  journal   = "J. Artif. Intell. Res.",
  publisher = "AI Access Foundation",
  volume    =  72,
  pages     = "215--249",
  month     =  dec,
  year      =  2021,
  address   = "El Segundo, CA, USA"
}

@ARTICLE{Lau2021-dr,
  title    = "Discovery of {Small-Molecule} Inhibitors of {SARS-CoV-2} Proteins
              Using a Computational and Experimental Pipeline",
  author   = "Lau, Edmond Y and Negrete, Oscar A and Bennett, W F Drew and
              Bennion, Brian J and Borucki, Monica and Bourguet, Feliza and
              Epstein, Aidan and Franco, Magdalena and Harmon, Brooke and He,
              Stewart and Jones, Derek and Kim, Hyojin and Kirshner, Daniel and
              Lao, Victoria and Lo, Jacky and McLoughlin, Kevin and Mosesso,
              Richard and Murugesh, Deepa K and Saada, Edwin A and Segelke,
              Brent and Stefan, Maxwell A and Stevenson, Garrett A and Torres,
              Marisa W and Weilhammer, Dina R and Wong, Sergio and Yang, Yue
              and Zemla, Adam and Zhang, Xiaohua and Zhu, Fangqiang and Allen,
              Jonathan E and Lightstone, Felice C",
  journal  = "Front Mol Biosci",
  volume   =  8,
  pages    = "678701",
  month    =  jul,
  year     =  2021,
  language = "en"
}

@INPROCEEDINGS{Stevenson2021-nw,
  title     = "High-throughput virtual screening of small molecule inhibitors
               for {SARS-CoV-2} protein targets with deep fusion models",
  booktitle = "Proceedings of the International Conference for High Performance
               Computing, Networking, Storage and Analysis",
  author    = "Stevenson, Garrett A and Jones, Derek and Kim, Hyojin and
               Bennett, W F Drew and Bennion, Brian J and Borucki, Monica and
               Bourguet, Feliza and Epstein, Aidan and Franco, Magdalena and
               Harmon, Brooke and He, Stewart and Katz, Max P and Kirshner,
               Daniel and Lao, Victoria and Lau, Edmond Y and Lo, Jacky and
               McLoughlin, Kevin and Mosesso, Richard and Murugesh, Deepa K and
               Negrete, Oscar A and Saada, Edwin A and Segelke, Brent and
               Stefan, Maxwell and Torres, Marisa W and Weilhammer, Dina and
               Wong, Sergio and Yang, Yue and Zemla, Adam and Zhang, Xiaohua
               and Zhu, Fangqiang and Lightstone, Felice C and Allen, Jonathan
               E",
  publisher = "Association for Computing Machinery",
  number    = "Article 74",
  pages     = "1--13",
  series    = "SC '21",
  month     =  nov,
  year      =  2021,
  address   = "New York, NY, USA",
  location  = "St. Louis, Missouri"
}

@ARTICLE{Wu2018-or,
  title    = "{MoleculeNet}: a benchmark for molecular machine learning",
  author   = "Wu, Zhenqin and Ramsundar, Bharath and Feinberg, Evan N and
              Gomes, Joseph and Geniesse, Caleb and Pappu, Aneesh S and
              Leswing, Karl and Pande, Vijay",
  journal  = "Chem. Sci.",
  volume   =  9,
  number   =  2,
  pages    = "513--530",
  month    =  jan,
  year     =  2018,
  language = "en"
}

@MISC{Landrum2021-zy,
  title  = "rdkit/rdkit: 2021\_09\_2 (Q3 2021) Release",
  author = "Landrum, Greg and Tosco, Paolo and Kelley, Brian and {Ric} and
            {sriniker} and {gedeck} and Vianello, Riccardo and
            {NadineSchneider} and Kawashima, Eisuke and Dalke, Andrew and Dan,
            N and Cole, Brian and Swain, Matt and Turk, Samo and Cosgrove,
            David and {AlexanderSavelyev} and Vaucher, Alain and
            W{\'o}jcikowski, Maciej and Jones, Gareth and Probst, Daniel and
            Scalfani, Vincent F and Godin, Guillaume and Pahl, Axel and
            Berenger, Francois and {JLVarjo} and {strets} and {JP} and
            {DoliathGavid} and Sforna, Gianluca and Jensen, Jan Holst",
  month  =  oct,
  year   =  2021
}

@ARTICLE{Karunaratne2019-zu,
  title         = "In-memory hyperdimensional computing",
  author        = "Karunaratne, Geethan and Le Gallo, Manuel and Cherubini,
                   Giovanni and Benini, Luca and Rahimi, Abbas and Sebastian,
                   Abu",
  month         =  jun,
  year          =  2019,
  archivePrefix = "arXiv",
  primaryClass  = "cs.ET",
  eprint        = "1906.01548",
  journal       = "arXiv",
}

@INPROCEEDINGS{Kanerva2000-uw,
  title     = "Random indexing of text samples for latent semantic analysis",
  booktitle = "Proceedings of the Annual Meeting of the Cognitive Science
               Society",
  author    = "Kanerva, Pentii and Kristoferson, Jan and Holst, Anders",
  volume    =  22,
  year      =  2000
}

@ARTICLE{Rahimi2019-vb,
  title    = "Efficient Biosignal Processing Using Hyperdimensional Computing:
              Network Templates for Combined Learning and Classification of
              {ExG} Signals",
  author   = "Rahimi, Abbas and Kanerva, Pentti and Benini, Luca and Rabaey,
              Jan M",
  journal  = "Proc. IEEE",
  volume   =  107,
  number   =  1,
  pages    = "123--143",
  month    =  jan,
  year     =  2019,
}

@INPROCEEDINGS{Burrello2019-qf,
  title     = "Laelaps: An {Energy-Efficient} Seizure Detection Algorithm from
               Long-term Human {iEEG} Recordings without False Alarms",
  booktitle = "2019 Design, Automation Test in Europe Conference Exhibition
               ({DATE})",
  author    = "Burrello, Alessio and Cavigelli, Lukas and Schindler, Kaspar and
               Benini, Luca and Rahimi, Abbas",
  pages     = "752--757",
  month     =  mar,
  year      =  2019,
}

@ARTICLE{Rasanen2016-ac,
  title    = "Sequence Prediction With Sparse Distributed Hyperdimensional
              Coding Applied to the Analysis of Mobile Phone Use Patterns",
  author   = "Rasanen, Okko J and Saarinen, Jukka P",
  journal  = "IEEE Trans Neural Netw Learn Syst",
  volume   =  27,
  number   =  9,
  pages    = "1878--1889",
  month    =  sep,
  year     =  2016,
  language = "en"
}

@ARTICLE{Mitrokhin2019-op,
  title    = "Learning sensorimotor control with neuromorphic sensors: Toward
              hyperdimensional active perception",
  author   = "Mitrokhin, A and Sutor, P and Ferm{\"u}ller, C and Aloimonos, Y",
  journal  = "Sci Robot",
  volume   =  4,
  number   =  30,
  month    =  may,
  year     =  2019,
  language = "en"
}

@ARTICLE{Jones2022-dr,
  title     = "Accelerators for Classical Molecular Dynamics Simulations of
               Biomolecules",
  author    = "Jones, Derek and Allen, Jonathan E and Yang, Yue and Drew
               Bennett, William F and Gokhale, Maya and Moshiri, Niema and
               Rosing, Tajana S",
  journal   = "J. Chem. Theory Comput.",
  publisher = "American Chemical Society",
  volume    =  18,
  number    =  7,
  pages     = "4047--4069",
  month     =  jul,
  year      =  2022,
  language  = "en"
}

@ARTICLE{Schneider2020-wz,
  title    = "Rethinking drug design in the artificial intelligence era",
  author   = "Schneider, Petra and Walters, W Patrick and Plowright, Alleyn T
              and Sieroka, Norman and Listgarten, Jennifer and Goodnow, Jr,
              Robert A and Fisher, Jasmin and Jansen, Johanna M and Duca,
              Jos{\'e} S and Rush, Thomas S and Zentgraf, Matthias and Hill,
              John Edward and Krutoholow, Elizabeth and Kohler, Matthias and
              Blaney, Jeff and Funatsu, Kimito and Luebkemann, Chris and
              Schneider, Gisbert",
  journal  = "Nat. Rev. Drug Discov.",
  volume   =  19,
  number   =  5,
  pages    = "353--364",
  month    =  may,
  year     =  2020,
  language = "en"
}

@ARTICLE{Wright2014-uq,
  title    = "Computing Clinically Relevant Binding Free Energies of {HIV-1}
              Protease Inhibitors",
  author   = "Wright, David W and Hall, Benjamin A and Kenway, Owain A and Jha,
              Shantenu and Coveney, Peter V",
  journal  = "J. Chem. Theory Comput.",
  volume   =  10,
  number   =  3,
  pages    = "1228--1241",
  month    =  mar,
  year     =  2014,
  language = "en"
}

@ARTICLE{Kanerva2009-sz,
  title     = "Hyperdimensional computing: An introduction to computing in
               distributed representation with high-dimensional random vectors",
  author    = "Kanerva, Pentti",
  journal   = "Cognit. Comput.",
  publisher = "Springer Science and Business Media LLC",
  volume    =  1,
  number    =  2,
  pages     = "139--159",
  month     =  jun,
  year      =  2009,
  language  = "en"
}

@ARTICLE{Plate1995-gz,
  title    = "Holographic reduced representations",
  author   = "Plate, T A",
  journal  = "IEEE Trans. Neural Netw.",
  volume   =  6,
  number   =  3,
  pages    = "623--641",
  year     =  1995,
  language = "en"
}

@ARTICLE{Mysinger2012-hn,
  title    = "Directory of useful decoys, enhanced ({DUD-E)}: better ligands
              and decoys for better benchmarking",
  author   = "Mysinger, Michael M and Carchia, Michael and Irwin, John J and
              Shoichet, Brian K",
  journal  = "J. Med. Chem.",
  volume   =  55,
  number   =  14,
  pages    = "6582--6594",
  month    =  jul,
  year     =  2012,
  language = "en"
}

@ARTICLE{Ge2020-dp,
  title    = "Classification Using Hyperdimensional Computing: A Review",
  author   = "Ge, Lulu and Parhi, Keshab K",
  journal  = "IEEE Circuits and Systems Magazine",
  volume   =  20,
  number   =  2,
  pages    = "30--47",
  year     =  2020
}

@ARTICLE{Yu2022-no,
  title         = "Understanding Hyperdimensional Computing for Parallel
                   {Single-Pass} Learning",
  author        = "Yu, Tao and Zhang, Yichi and Zhang, Zhiru and De Sa,
                   Christopher",
  month         =  feb,
  year          =  2022,
  archivePrefix = "arXiv",
  primaryClass  = "cs.LG",
  eprint        = "2202.04805",
  journal       = "arXiv"
}

@INCOLLECTION{Duvenaud2015-kl,
  title     = "Convolutional Networks on Graphs for Learning Molecular
               Fingerprints",
  booktitle = "Advances in Neural Information Processing Systems 28",
  author    = "Duvenaud, David K and Maclaurin, Dougal and Iparraguirre, Jorge
               and Bombarell, Rafael and Hirzel, Timothy and Aspuru-Guzik, Alan
               and Adams, Ryan P",
  editor    = "Cortes, C and Lawrence, N D and Lee, D D and Sugiyama, M and
               Garnett, R",
  publisher = "Curran Associates, Inc.",
  pages     = "2224--2232",
  year      =  2015
}

@ARTICLE{Sun2019-fa,
  title         = "{InfoGraph}: Unsupervised and Semi-supervised {Graph-Level}
                   Representation Learning via Mutual Information Maximization",
  author        = "Sun, Fan-Yun and Hoffmann, Jordan and Verma, Vikas and Tang,
                   Jian",
  month         =  jul,
  year          =  2019,
  archivePrefix = "arXiv",
  journal       = "arXiv",
  primaryClass  = "cs.LG",
  eprint        = "1908.01000"
}

@ARTICLE{Li2021-pf,
  title    = "{SMILES} Pair Encoding: A {Data-Driven} Substructure Tokenization
              Algorithm for Deep Learning",
  author   = "Li, Xinhao and Fourches, Denis",
  journal  = "J. Chem. Inf. Model.",
  volume   =  61,
  number   =  4,
  pages    = "1560--1569",
  month    =  apr,
  year     =  2021,
  language = "en"
}

@ARTICLE{Morgan1965-zp,
  title     = "The Generation of a Unique Machine Description for Chemical
               {Structures-A} Technique Developed at Chemical Abstracts Service",
  author    = "Morgan, H L",
  journal   = "J. Chem. Doc.",
  publisher = "American Chemical Society",
  volume    =  5,
  number    =  2,
  pages     = "107--113",
  month     =  may,
  year      =  1965
}

@ARTICLE{Tran-Nguyen2020-xq,
  title    = "{LIT-PCBA}: An Unbiased Data Set for Machine Learning and Virtual
              Screening",
  author   = "Tran-Nguyen, Viet-Khoa and Jacquemard, C{\'e}lien and Rognan,
              Didier",
  journal  = "J. Chem. Inf. Model.",
  month    =  apr,
  year     =  2020,
  language = "en"
}

@ARTICLE{Chaput2016-qr,
  title    = "Benchmark of four popular virtual screening programs:
              construction of the active/decoy dataset remains a major
              determinant of measured performance",
  author   = "Chaput, Ludovic and Martinez-Sanz, Juan and Saettel, Nicolas and
              Mouawad, Liliane",
  journal  = "J. Cheminform.",
  volume   =  8,
  pages    = "56",
  month    =  oct,
  year     =  2016,
  language = "en"
}

@ARTICLE{Chen2019-gd,
  title    = "Hidden bias in the {DUD-E} dataset leads to misleading
              performance of deep learning in structure-based virtual screening",
  author   = "Chen, Lieyang and Cruz, Anthony and Ramsey, Steven and Dickson,
              Callum J and Duca, Jose S and Hornak, Viktor and Koes, David R
              and Kurtzman, Tom",
  journal  = "PLoS One",
  volume   =  14,
  number   =  8,
  pages    = "e0220113",
  month    =  aug,
  year     =  2019,
  language = "en"
}

@ARTICLE{Sieg2019-gy,
  title    = "In Need of Bias Control: Evaluating Chemical Data for Machine
              Learning in {Structure-Based} Virtual Screening",
  author   = "Sieg, Jochen and Flachsenberg, Florian and Rarey, Matthias",
  journal  = "J. Chem. Inf. Model.",
  volume   =  59,
  number   =  3,
  pages    = "947--961",
  month    =  mar,
  year     =  2019,
  language = "en"
}

@ARTICLE{Jiang2021-wr,
  title    = "{InteractionGraphNet}: A Novel and Efficient Deep Graph
              Representation Learning Framework for Accurate {Protein-Ligand}
              Interaction Predictions",
  author   = "Jiang, Dejun and Hsieh, Chang-Yu and Wu, Zhenxing and Kang, Yu
              and Wang, Jike and Wang, Ercheng and Liao, Ben and Shen, Chao and
              Xu, Lei and Wu, Jian and Cao, Dongsheng and Hou, Tingjun",
  journal  = "J. Med. Chem.",
  volume   =  64,
  number   =  24,
  pages    = "18209--18232",
  month    =  dec,
  year     =  2021,
  language = "en"
}

@ARTICLE{Tran-Nguyen2021-jw,
  title    = "True Accuracy of Fast Scoring Functions to Predict
              {High-Throughput} Screening Data from Docking Poses: The Simpler
              the Better",
  author   = "Tran-Nguyen, Viet-Khoa and Bret, Guillaume and Rognan, Didier",
  journal  = "J. Chem. Inf. Model.",
  volume   =  61,
  number   =  6,
  pages    = "2788--2797",
  month    =  jun,
  year     =  2021,
  language = "en"
}

@ARTICLE{Gentile2020-tj,
  title    = "Deep Docking: A Deep Learning Platform for Augmentation of
              Structure Based Drug Discovery",
  author   = "Gentile, Francesco and Agrawal, Vibudh and Hsing, Michael and
              Ton, Anh-Tien and Ban, Fuqiang and Norinder, Ulf and Gleave,
              Martin E and Cherkasov, Artem",
  journal  = "ACS Cent Sci",
  volume   =  6,
  number   =  6,
  pages    = "939--949",
  month    =  jun,
  year     =  2020,
  language = "en"
}

@MISC{enamine_real,
  title        = "{REAL} Compounds - Enamine",
  howpublished = "\url{https://enamine.net/compound-collections/real-compounds}",
  note         = "Accessed: 2021-10-5"
}

@ARTICLE{Bender2005-vc,
  title    = "A discussion of measures of enrichment in virtual screening:
              comparing the information content of descriptors with increasing
              levels of sophistication",
  author   = "Bender, Andreas and Glen, Robert C",
  journal  = "J. Chem. Inf. Model.",
  volume   =  45,
  number   =  5,
  pages    = "1369--1375",
  month    =  sep,
  year     =  2005,
  language = "en"
}

@ARTICLE{Clyde2023-vk,
  title    = "{AI-accelerated} protein-ligand docking for {SARS-CoV-2} is
              100-fold faster with no significant change in detection",
  author   = "Clyde, Austin and Liu, Xuefeng and Brettin, Thomas and Yoo,
              Hyunseung and Partin, Alexander and Babuji, Yadu and Blaiszik,
              Ben and Mohd-Yusof, Jamaludin and Merzky, Andre and Turilli,
              Matteo and Jha, Shantenu and Ramanathan, Arvind and Stevens, Rick",
  journal  = "Sci. Rep.",
  volume   =  13,
  number   =  1,
  pages    = "2105",
  month    =  feb,
  year     =  2023,
  language = "en"
}

@ARTICLE{Trott2010-ij,
  title    = "{AutoDock} Vina: improving the speed and accuracy of docking with
              a new scoring function, efficient optimization, and
              multithreading",
  author   = "Trott, Oleg and Olson, Arthur J",
  journal  = "J. Comput. Chem.",
  volume   =  31,
  number   =  2,
  pages    = "455--461",
  month    =  jan,
  year     =  2010,
  language = "en"
}

@ARTICLE{Greenidge2013-iz,
  title    = "{MM/GBSA} binding energy prediction on the {PDBbind} data set:
              successes, failures, and directions for further improvement",
  author   = "Greenidge, Paulette A and Kramer, Christian and Mozziconacci,
              Jean-Christophe and Wolf, Romain M",
  
  journal  = "J. Chem. Inf. Model.",
  volume   =  53,
  number   =  1,
  pages    = "201--209",
  month    =  jan,
  year     =  2013,
  language = "en"
}

@ARTICLE{Massova2000-qu,
  title    = "Combined molecular mechanical and continuum solvent approach
              ({MM-PBSA/GBSA}) to predict ligand binding",
  author   = "Massova, Irina and Kollman, Peter A",
  
  journal  = "Perspect. Drug Discov. Des.",
  volume   =  18,
  number   =  1,
  pages    = "113--135",
  month    =  jun,
  year     =  2000
}

@ARTICLE{Stepniewska-Dziubinska2018-fo,
  title    = "Development and evaluation of a deep learning model for
              protein-ligand binding affinity prediction",
  author   = "Stepniewska-Dziubinska, Marta M and Zielenkiewicz, Piotr and
              Siedlecki, Pawel",
  
  journal  = "Bioinformatics",
  volume   =  34,
  number   =  21,
  pages    = "3666--3674",
  month    =  nov,
  year     =  2018,
  language = "en"
}

@INPROCEEDINGS{Kang2022-pm,
  title     = "{RelHD}: A Graph-based Learning on {FeFET} with Hyperdimensional
               Computing",
  booktitle = "2022 {IEEE} 40th International Conference on Computer Design
               ({ICCD})",
  author    = "Kang, Jaeyoung and Zhou, Minxuan and Bhansali, Abhinav and Xu,
               Weihong and Thomas, Anthony and Rosing, Tajana",
  pages     = "553--560",
  month     =  oct,
  year      =  2022
}

@INPROCEEDINGS{Dasgupta2000-hb,
  title     = "Experiments with random projection",
  booktitle = "Proceedings of the Sixteenth conference on Uncertainty in
               artificial intelligence",
  author    = "Dasgupta, Sanjoy",
  publisher = "Morgan Kaufmann Publishers Inc.",
  pages     = "143--151",
  series    = "UAI'00",
  month     =  jun,
  year      =  2000,
  address   = "San Francisco, CA, USA",
  location  = "Stanford, California"
}

@ARTICLE{Kazemi2022-zw,
  title    = "Achieving software-equivalent accuracy for hyperdimensional
              computing with ferroelectric-based in-memory computing",
  author   = "Kazemi, Arman and M{\"u}ller, Franz and Sharifi, Mohammad Mehdi
              and Errahmouni, Hamza and Gerlach, Gerald and K{\"a}mpfe, Thomas
              and Imani, Mohsen and Hu, Xiaobo Sharon and Niemier, Michael",
  journal  = "Sci. Rep.",
  volume   =  12,
  number   =  1,
  pages    = "19201",
  month    =  nov,
  year     =  2022,
  language = "en"
}

@ARTICLE{Ma2022-wu,
  title         = "Hyperdimensional Computing vs. Neural Networks: Comparing
                   Architecture and Learning Process",
  author        = "Ma, Dongning and Jiao, Xun",
  month         =  jul,
  year          =  2022,
  archivePrefix = "arXiv",
  primaryClass  = "cs.NE",
  eprint        = "2207.12932",
  journal       = "arXiv"
}

@ARTICLE{Ross2022-ce,
  title     = "Large-scale chemical language representations capture molecular
               structure and properties",
  author    = "Ross, Jerret and Belgodere, Brian and Chenthamarakshan, Vijil
               and Padhi, Inkit and Mroueh, Youssef and Das, Payel",
  journal   = "Nature Machine Intelligence",
  publisher = "Nature Publishing Group",
  volume    =  4,
  number    =  12,
  pages     = "1256--1264",
  month     =  dec,
  year      =  2022,
  language  = "en"
}

@ARTICLE{Wang2022-hg,
  title     = "Molecular contrastive learning of representations via graph
               neural networks",
  author    = "Wang, Yuyang and Wang, Jianren and Cao, Zhonglin and Barati
               Farimani, Amir",
  journal   = "Nature Machine Intelligence",
  publisher = "Nature Publishing Group",
  volume    =  4,
  number    =  3,
  pages     = "279--287",
  month     =  mar,
  year      =  2022,
  language  = "en"
}

@ARTICLE{Salamat2020-ld,
  title     = "Accelerating Hyperdimensional Computing on {FPGAs} by Exploiting
               Computational Reuse",
  author    = "Salamat, Sahand and Imani, Mohsen and Rosing, Tajana",
  journal   = "IEEE Trans. Comput.",
  publisher = "IEEE",
  volume    =  69,
  number    =  8,
  pages     = "1159--1171",
  month     =  aug,
  year      =  2020
}

@INPROCEEDINGS{Gupta2020-cp,
  title     = "{THRIFTY}: training with hyperdimensional computing across flash
               hierarchy",
  booktitle = "Proceedings of the 39th International Conference on
               {Computer-Aided} Design",
  author    = "Gupta, Saransh and Morris, Justin and Imani, Mohsen and
               Ramkumar, Ranganathan and Yu, Jeffrey and Tiwari, Aniket and
               Aksanli, Baris and Rosing, Tajana {\v S}imuni{\'c}",
  publisher = "Association for Computing Machinery",
  number    = "Article 27",
  pages     = "1--9",
  series    = "ICCAD '20",
  month     =  nov,
  year      =  2020,
  address   = "New York, NY, USA",
  location  = "Virtual Event, USA"
}

@INPROCEEDINGS{hd2fpga,
  author={Zhang, Tinaqi and Salamat, Sahand and Khaleghi, Behnam and Morris, Justin and Aksanli, Baris and Rosing, Tajana Simunic},
  booktitle={2023 24th International Symposium on Quality Electronic Design (ISQED)}, 
  title={HD2FPGA: Automated Framework for Accelerating Hyperdimensional Computing on FPGAs}, 
  year={2023},
  volume={},
  number={},
  pages={1-9},
  doi={10.1109/ISQED57927.2023.10129332}}



\section*{Acknowledgements}
This research was supported by funds from the UC National Laboratory Fees Research Program of the University of California, Grant Number L23GF6259. This work was supported in part by CRISP and PRISM, centers in JUMP 1.0 and 2.0, SRC programs sponsored by DARPA, SRC \#236160. Computing support for this work came from the Lawrence Livermore National Laboratory (LLNL) Institutional Computing Grand Challenge program. Part of this research was also supported by the American Heart Association under CRADA TC02274-4. Funding in part by DTRA project HDTRA1036045. All work performed at Lawrence Livermore National Laboratory is performed under the auspices of the U.S. Department of Energy under Contract DE-AC52-07NA27344, LLNL-JRNL-847376-DRAFT. We thank Michael K. Gilson (UCSD), Rose Yu (UCSD), Stewart He (LLNL), Dan Kirshner (LLNL), Kevin McLoughlin (LLNL), and Amanda Paulson (UCSF) for their helpful feedback in the development of this work.

\section*{Author contributions statement}

D.J., J.A, N.M, and T.R. conceived the experiments. D.J. conducted the experiments and wrote the draft manuscript. D.J., J.A., N.M., and T.R. edited the manuscript. X.Z and B.B. contributed molecular docking simulations and their data. J.K., B.K., and W.X. contributed code. S.P. contributed hardware energy efficiency measurements. All authors read and approved the manuscript. 

\section*{Data availability}
All code is publicly available on github \url{https://github.com/LLNL/hdbind}. All MoleculeNet data is available at \url{https://moleculenet.org/} and all LIT-PCBA data is available at \url{https://drugdesign.unistra.fr/LIT-PCBA/}. Additional reasonable requests can be made by contacting the corresponding authors.

\section*{Additional information}


The authors declare no competing interests.
\end{document}


\section*{Supplementary Information}

\begin{figure}[H]
    \centering
    \includegraphics[width=\linewidth]{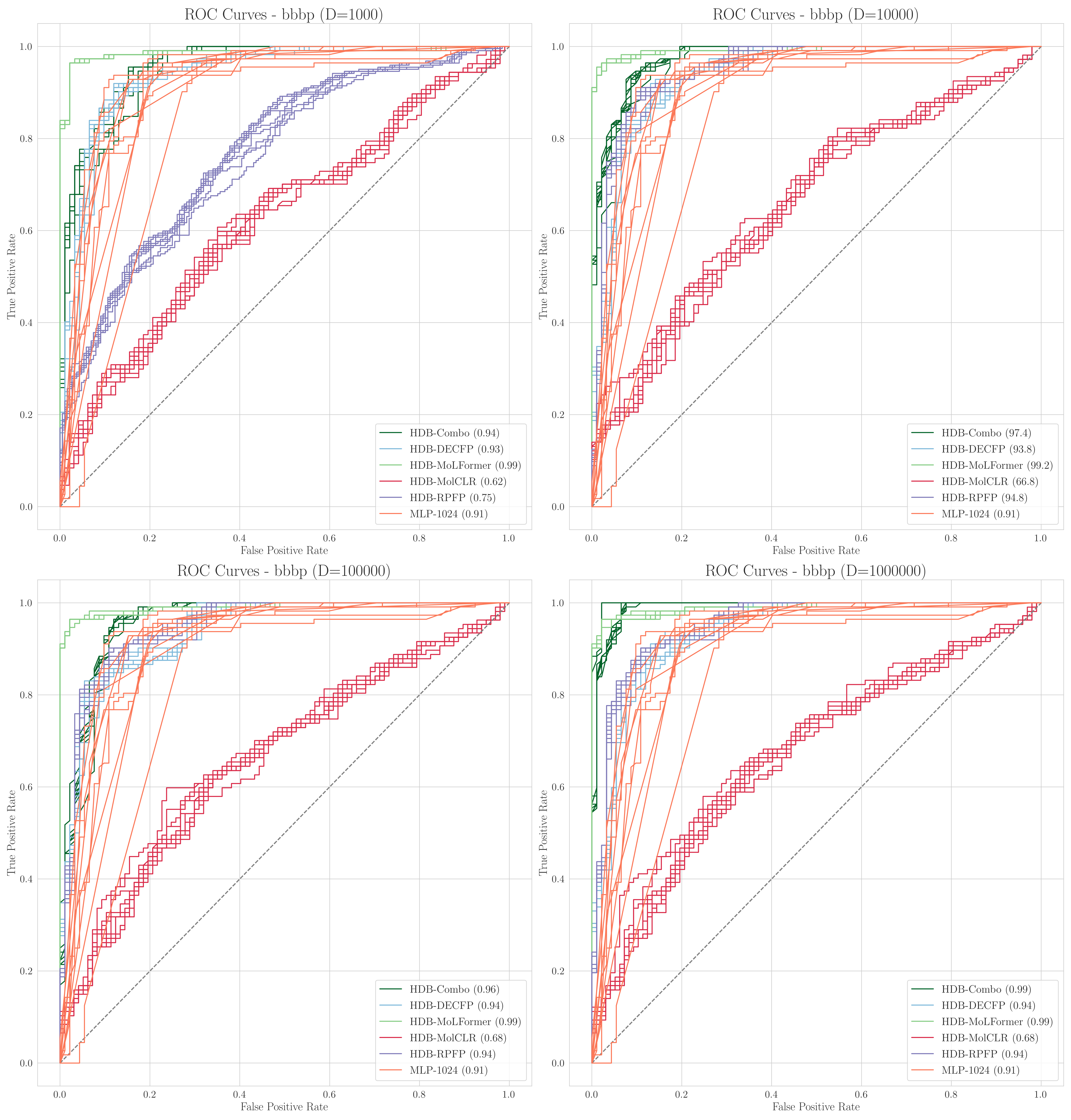}
    \caption{ROC Curves for the BBBP dataset. Values inside of `()' denote the mean ROC-AUC score, measured over 10 trials.}
    \label{fig:roc_curve_bbbp}
\end{figure}

\begin{figure}[H]
    \centering
    \includegraphics[width=\linewidth]{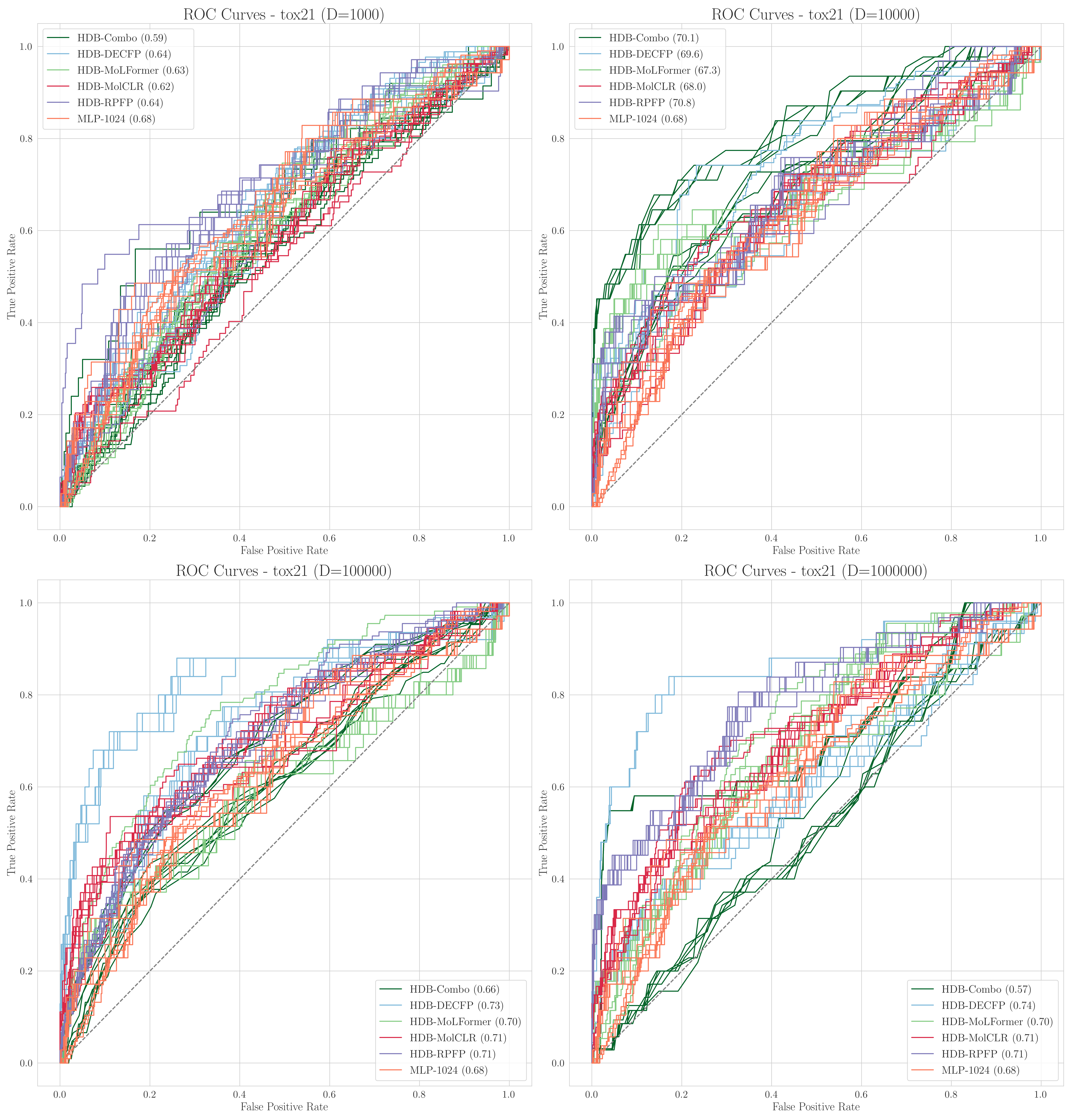}
    \caption{ROC Curves for the Tox21 dataset. Values inside of `()' denote the mean ROC-AUC score, measured over 10 trials.}
    \label{fig:roc_curve_tox21}
\end{figure}

\begin{figure}[H]
    \centering
    \includegraphics[width=\linewidth]{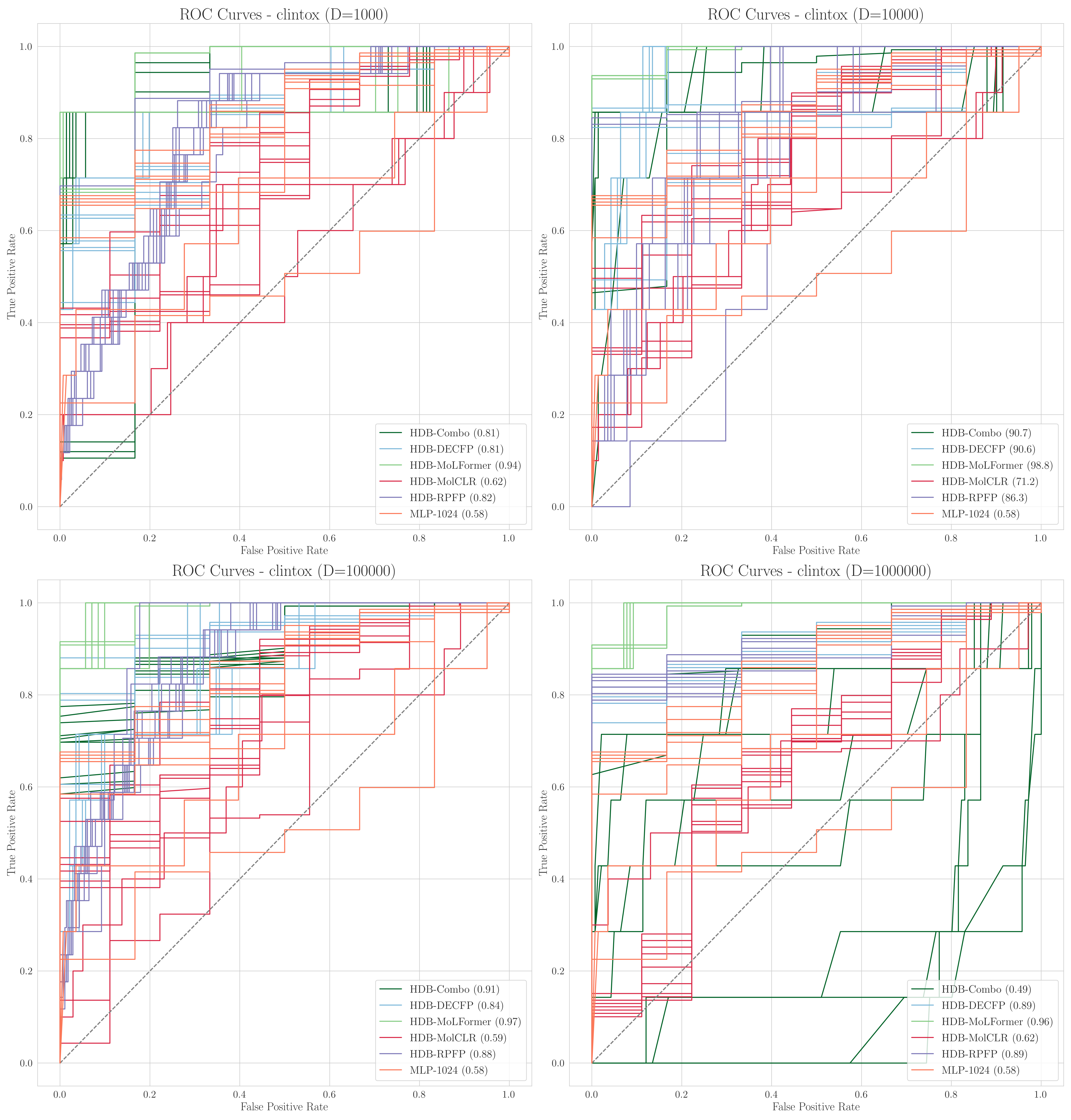}
    \caption{ROC Curves for the Clintox dataset. Values inside of `()' denote the mean ROC-AUC score, measured over 10 trials.}
    \label{fig:roc_curve_clintox}
\end{figure}

\begin{figure}[H]
    \centering
    \includegraphics[width=\linewidth]{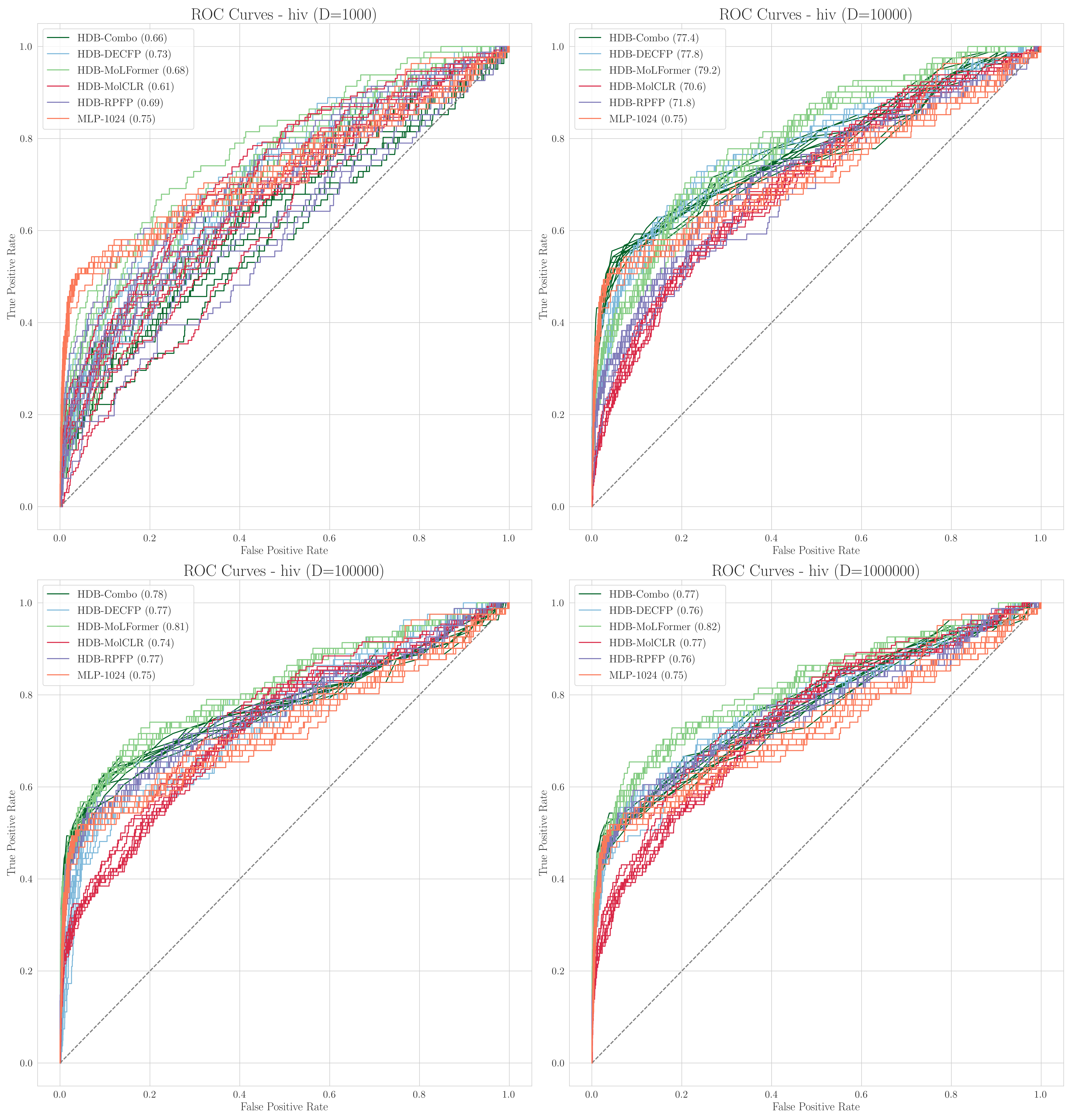}
    \caption{ROC Curves for the HIV dataset. Values inside of `()' denote the mean ROC-AUC score, measured over 10 trials.}
    \label{fig:roc_curve_hiv}
\end{figure}

\begin{figure}[H]
    \centering
    \includegraphics[width=\linewidth]{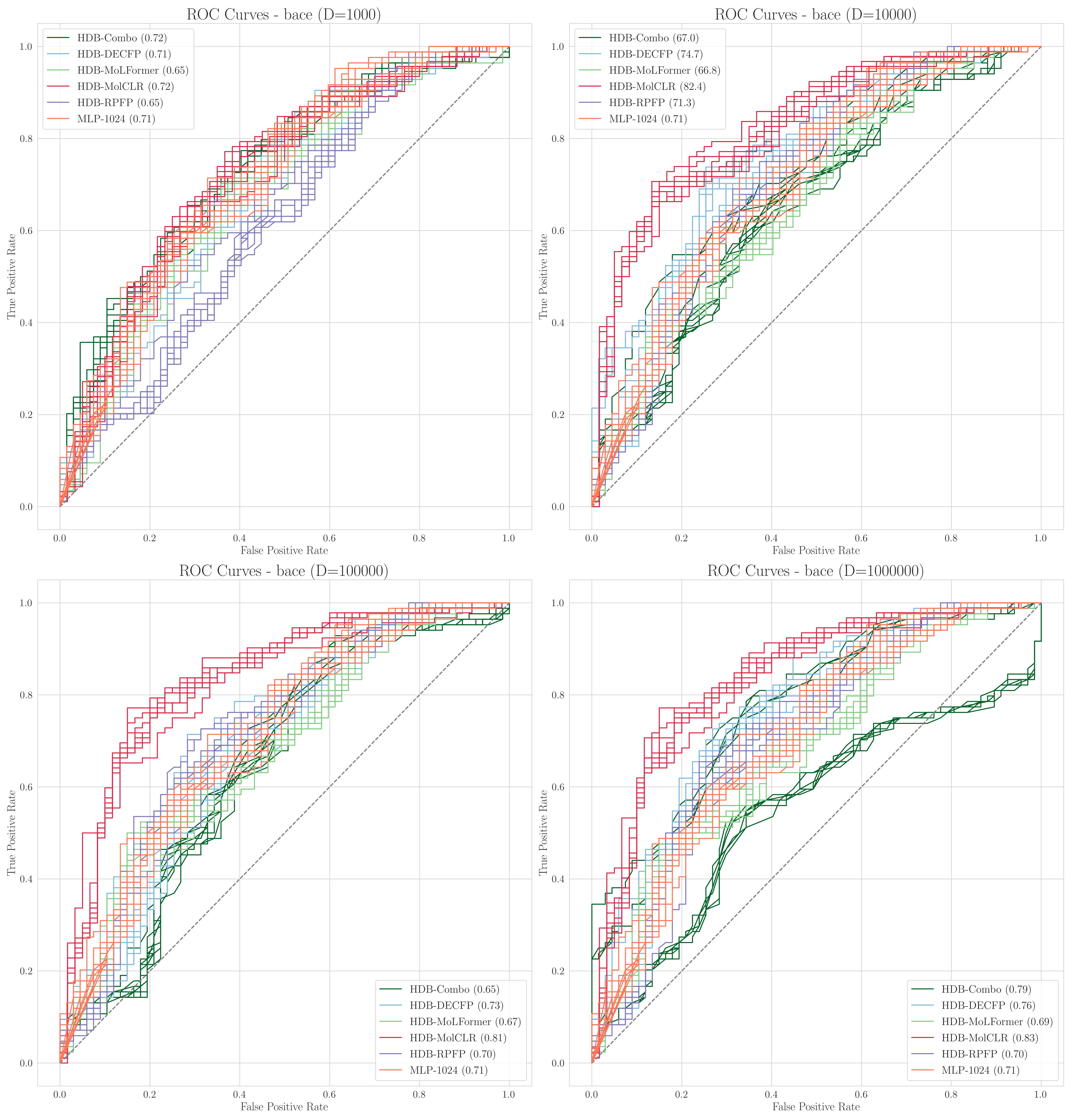}
    \caption{ROC Curves for the Bace dataset. Values inside of `()' denote the mean ROC-AUC score, measured over 10 trials.}
    \label{fig:roc_curve_bace}
\end{figure}

\begin{figure}[H]
    \centering
    \includegraphics[width=\linewidth]{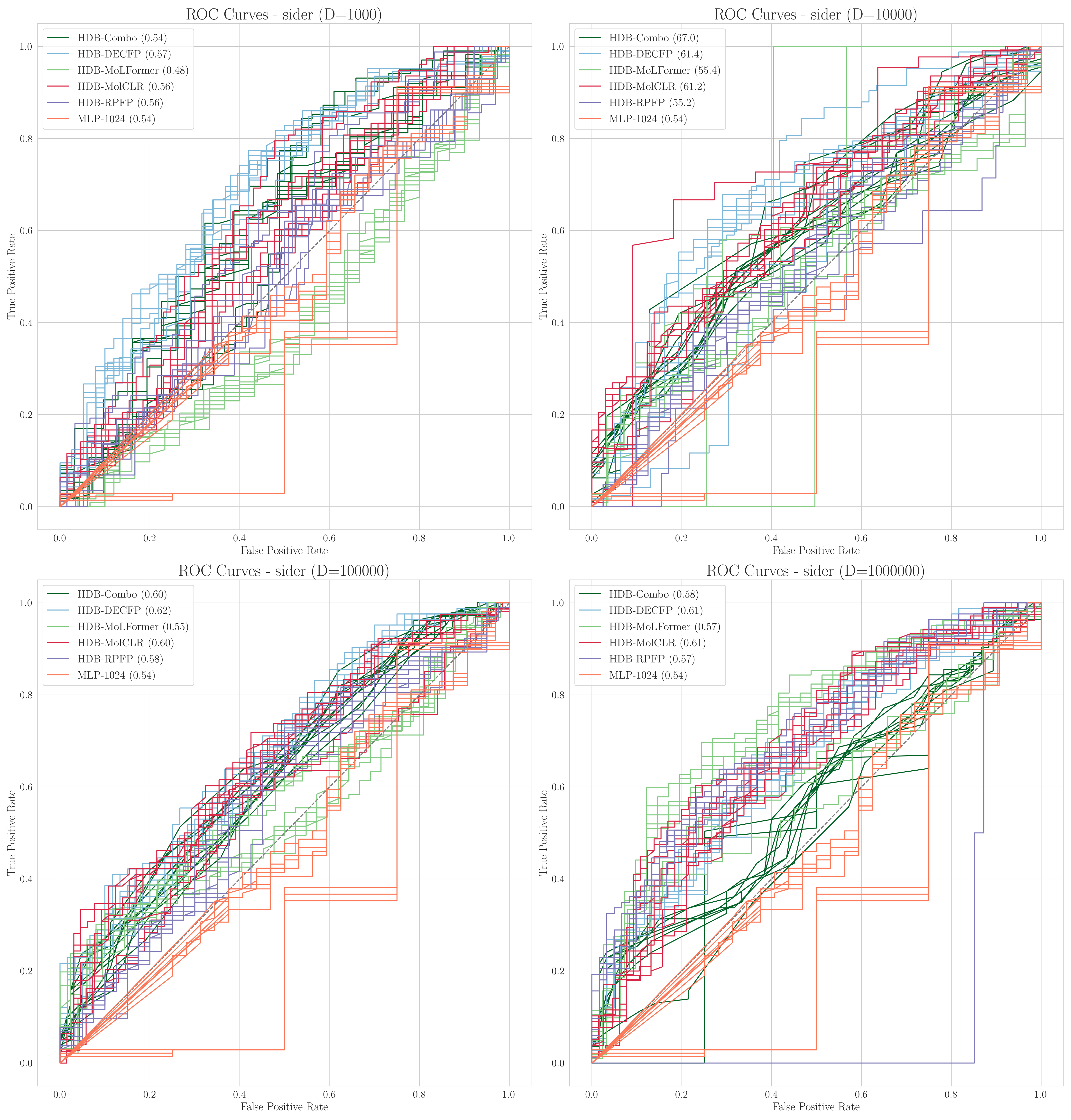}
    \caption{ROC Curves for the SIDER dataset. Values inside of `()' denote the mean ROC-AUC score, measured over 10 trials.}
    \label{fig:roc_curve_sider}
\end{figure}

\begin{table}[H]
    \centering
    \begin{tabular}{c|c|c|c|c|c|c}
    \textbf{Method} & \textbf{BBBP} & \textbf{Tox21} & \textbf{ClinTox} & \textbf{HIV} & \textbf{BACE} & \textbf{SIDER}\\
    \hline
    \textbf{Molecules} & 2,039 & 7,831 & 1,478 & 41,127 & 1,513 & 1,427 \\
    \hline
    \textbf{Tasks} & 1 & 12 & 2 & 1 & 1 &27 \\
    \hline

      HDB-DECFP (1k-1) & 93.6 (0.1) & 65.8 (1.6) & 87.4 (3.9) & 73.4 (2.3) & 68.0 (1.8) & 57.4 (2.0)\\
        HDB-DECFP (1k-2) & 93.7 (0.2) & 67.4 (1.1) & 87.6 (2.6) & 72.1 (2.1) & 70.9 (1.2) & 55.9 (2.0)\\
        HDB-DECFP (1k-4) & 94.2 (0.5) & 63.4 (1.4) & 81.9 (1.8) & 68.9 (2.3) & 68.6 (0.8) &60.4 (2.1)\\
      HDB-DECFP (1024-1) & 92.5 (0.5) & 66.3 (1.0) & 85.0 (4.7) & 69.2 (2.2) & 68.6 (2.4) & 57.3 (1.9)\\
      HDB-DECFP (1024-2) & 93.7 (0.4)& 72.4 (1.0) & 79.2 (3.4) & 68.9 (2.2) & 74.1 (0.4) & 55.3 (1.5)\\
      HDB-DECFP (1024-4) & 92.1 (0.5) & 65.3 (1.3) & 66.4 (2.4) & 69.4 (1.8) & 71.6 (0.8) & 56.4 (1.4) \\
      HDB-DECFP (2048-1) & 93.1 (0.3) & 72.1 (0.8) & 88.3 (4.6) & 70.7 (1.3) & 70.6 (2.3) & 60.6 (1.5) \\
      HDB-DECFP (2048-2) & \textbf{94.3} (0.2) & 70.5 (0.7)& 86.2 (4.4) & 71.3 (1.3) & 73.8 (1.3) & 61.1 (2.0) \\
      HDB-DECFP (2048-4) & 94.0 (0.1) & 73.3 (0.7) & 74.6 (2.6) & 65.8 (0.8) & 72.6 (1.1) & 58.5 (1.8)\\
      HDB-DECFP (10k-1) & 93.9 (0.2) & 72.5 (0.9) & 88.4 (4.8) & 75.7 (1.0) & 68.4 (1.4) & 61.9 (1.6)\\
      HDB-DECFP (10k-2) & 93.8 (0.2) & 69.6 (0.8)& 90.6 (4.0) & 77.8 (0.3) & 74.7 (1.1) & 61.4 (1.6)\\
        HDB-DECFP (10k-4) & 94.1 (0.2) & 68.7 (1.1) & 83.0 (3.6) & 77.9 (0.2)& 73.2 (1.1)& 60.3 (1.3)\\
      HDB-DECFP (100k-1) & 93.8 (0.2) & 76.1 (0.5) & 89.3 (3.1) & 75.3 (0.4) & 68.9 (1.7) & 61.8 (1.5)
      \\
      HDB-DECFP (100k-2) & 93.8 (0.2)& 73.0 (1.0) & 90.1 (1.4) &77.6 (0.4) & 73.8 (1.1) & 61.8 (1.2)
      \\
      HDB-DECFP (100k-4) & 94.2 (0.2) & 73.8 (0.8) & 85.6 (4.6) & 77.9 (0.3) & 73.8 (1.4)& 59.5 (1.4)
      \\
        HDB-DECFP (1m-1) & 93.8 (0.2) & 73.0 (0.8) & \textbf{93.8} (1.8) & 75.8 (1.2) & 69.0 (1.9) & 59.0 (1.4)\\
        HDB-DECFP (1m-2) & 93.8 (0.1)&75.0 (0.9) & 90.0 (0.8) & 78.0 (0.3) & \textbf{74.8} (1.1) & \textbf{63.1} (1.0)\\
        HDB-DECFP (1m-4) & \textbf{94.3} (0.2) & \textbf{79.5} (0.6)&86.7 (0.9) & \textbf{78.8} (0.3) & 74.2 (1.6) & 61.2 (1.3)\\

    \end{tabular}
    \caption{Comparison of HDB-DECFP models with varied ECFP length and radius parameters on MoleculeNet classification benchmarks. The standard deviation generally decreases with smaller radius parameter values while performance tends to increase with larger length parameter values. For $D=1,000,000$, the best model is achieved for each task, with an exception for BBBP where a value of $D=2048$ and ecfp radius of 2 tie the respective model using $D=1,000,000$ and an ecfp radius of 4.}
    \label{tab:molnet_result_decfp}
\end{table}

\begin{table}[H]
    \centering
    \begin{tabular}{c|c|c|c|c|c|c}
    \textbf{Method} & \textbf{BBBP} & \textbf{Tox21} & \textbf{ClinTox} & \textbf{HIV} & \textbf{BACE} & \textbf{SIDER}\\
    \hline
    \textbf{Molecules} & 2,039 & 7,831 & 1,478 & 41,127 & 1,513 & 1,427 \\
    \hline
    \textbf{Tasks} & 1 & 12 & 2 & 1 & 1 &27 \\
    \hline
    HDB-MoLFormer (1k) & 98.6 (0.0) & 64.1 (1.3) & 92.1 (0.9) & 72.7 (3.0) & \textbf{68.6} (1.3) & 48.1 (1.8)\\
      HDB-MoLFormer (10k) & \textbf{99.2} (0.0) & 67.3 (1.0)& \textbf{98.8} (0.0) &79.2 (0.6) & 66.8 (0.4) & 55.4 (1.9)\\
      HDB-MoLFormer (100k) &99.1 (0.1) & 71.3 (0.9)& 98.4 (0.7)& 81.2 (0.3)&68.2 (0.5) &56.5 (2.1) \\
      HDB-MoLFormer (1m) & 99.0 (0.1) & \textbf{71.5} (1.0)& 98.4 (0.8) & \textbf{81.6} (0.3) & 68.3 (0.4) & \textbf{58.1} (2.1)\\

    \end{tabular}
    \caption{Comparison of HDB-MoLFormer models on MoleculeNet classification benchmarks with varied hypervector dimension size $D$. Generally, increased dimension bring about improved roc-auc scores.}
    \label{tab:molnet_result_molformer}
\end{table}

\begin{table}[H]
    \centering
    \begin{tabular}{c|c|c|c|c|c|c}
    \textbf{Method} & \textbf{BBBP} & \textbf{Tox21} & \textbf{ClinTox} & \textbf{HIV} & \textbf{BACE} & \textbf{SIDER}\\
    \hline
    \textbf{Molecules} & 2,039 & 7,831 & 1,478 & 41,127 & 1,513 & 1,427 \\
    \hline
    \textbf{Tasks} & 1 & 12 & 2 & 1 & 1 &27 \\
    \hline
    HDB-Combo (1k) &  95.7 (0.5) & 60.4 (1.3) & 86.2 (1.7) & 64.0 (3.2) & \textbf{72.4} (0.3) & 54.9 (1.9)\\
      HDB-Combo (10k) &97.4 (0.3) & \textbf{70.1} (1.2) &90.7 (3.4) &77.4 (0.8) & \textbf{67.0} (2.7) & 58.8 (2.8)\\
      HDB-Combo (100k) & 95.9 (0.4) &69.7 (1.5)& \textbf{91.3} (1.9) & \textbf{77.4} (0.3) & 65.3 (0.6)& \textbf{59.5} (2.7) \\
      HDB-Combo (1m) & \textbf{99.1} (0.3) & 58.1 (0.9) & 69.6 (16.5) & 77.2 (0.8) & 64.1 (10.5) & 59.0 (2.1) \\

    \end{tabular}
    \caption{Comparison of HDB-Combo models on MoleculeNet classification benchmarks with varied ECFP length and radius parameters. The standard deviation generally decreases with smaller radius parameter values while performance tends to increase with larger length parameter values.}
    \label{tab:molnet_result_combo}
\end{table}

\begin{table}[H]
    \centering
    \begin{tabular}{c|c|c|c}

                \textbf{Model} & \textbf{$D$} & \textbf{ER-1\% (r)} & \textbf{ER-1\% (a)}\\ \hline
        MoleHD\textsuperscript{*} & 10k & 15.12 (1.65) & \cellcolor{red!25} 2.63 (0.70)\\
         HDB-RPFP & 10k & 25.76 (2.59) & \cellcolor{yellow!25}5.46 (1.87)\\
        HDB-RPFP & 100k & 25.31 (2.90) & \cellcolor{yellow!25}6.84 (1.43)\\ 
        HDB-RPFP & 1m & 25.50 (2.51) & \cellcolor{yellow!25}7.73 (1.81)\\

         HDB-MolCLR & 10k & 24.55 (4.27) & \cellcolor{yellow!25}7.99 (1.06)\\
         HDB-MolCLR & 100k & 28.40 (3.93) & \cellcolor{yellow!25}6.02 (1.36)\\
         HDB-MolCLR & 1m &  28.44 (4.30) & \cellcolor{yellow!25}6.91 (0.80)\\
         HDB-MoLFormer & 10k & 28.24 (2.18) & \cellcolor{yellow!25}5.72 (0.95)\\
         HDB-MoLFormer & 100k & 35.81 (3.86) & \cellcolor{yellow!25}8.15 (1.80)\\
         HDB-MoLFormer & 1m & \textbf{36.69} (3.43) & \cellcolor{yellow!25}6.87 (0.70)\\
         HDB-DECFP (r=1) & 10k & 28.19 (2.74) & \cellcolor{yellow!25}6.74 (1.80)\\
         HDB-DECFP (r=2) & 10k & 30.70 (3.07) & \cellcolor{yellow!25}8.02 (1.06)\\
         HDB-DECFP (r=4) & 10k & 32.15 (2.87)  & \cellcolor{yellow!25}7.85 (0.95)\\
         HDB-DECFP (r=1) & 100k & 28.35 (3.67) & \cellcolor{yellow!25}8.33 (1.59)\\
         HDB-DECFP (r=2) & 100k & 32.28 (2.51) & \cellcolor{yellow!25} 6.75 (0.88)\\
         HDB-DECFP (r=4) & 100k & 31.80 (2.68) & \cellcolor{yellow!25}8.46 (1.22)\\         
         HDB-DECFP (r=1) & 1m & 29.32 (3.69) & \cellcolor{yellow!25}8.81 (1.68)\\
         HDB-DECFP (r=2) & 1m & 33.55 (2.34) & \cellcolor{green!25}9.80 (1.24)\\
         HDB-DECFP (r=4) & 1m & 34.83 (2.01) & \cellcolor{green!25}10.46 (1.78) \\
         HDB-Combo & 10k & 35.82 (5.24) & \cellcolor{red!25}4.72 (0.97)\\
         HDB-Combo & 100k & 32.03 (2.12) & \cellcolor{green!25}\textbf{30.27} (2.10)\\
         HDB-Combo & 1m & 16.49 (9.44) & \cellcolor{green!25} 17.84 (8.49)\\

         \hline
         MLP\textsuperscript{*} & - & 30.79 (5.04) & 7.61 (1.84)\\
         Pafnucy\cite{Stepniewska-Dziubinska2018-fo, Tran-Nguyen2021-jw} & - & - &3.46 (1.97)\\
         GRIM\cite{Desaphy2013-jt, Tran-Nguyen2021-jw} & - &- &4.78 (3.11) \\

    \end{tabular}
    \caption{LIT-PCBA roc-enrichment ($\text{ER}$-$1\%$) factor metrics, averaged over 15 target datasets where each value is reported as the mean over 10 random seeds. `r' denotes random split and `a' denotes ave bias minimizing split provided by the authors\cite{Tran-Nguyen2020-xq}. Values correspond to the $\text{ER}$-$1\%$ metrics, averaged over all 15 targets using 10 random seeds for each. Values inside of the parentheses represent the standard deviation over the 10 random seeds for each model, averaged over the 15 protein targets. * denotes our implementation of the MoleHD and MLP baselines.
    Bold indicates best overall model and best HDC model. Values highlighted in green represent those that have statistically significant improvement in $\text{ER}$-$1\%$ upon the best previously reported method, GRIM\cite{Desaphy2013-jt, Tran-Nguyen2021-jw}. Values highlighted in yellow represent those models for which the mean $\text{ER}$-$1\%$ is higher than the mean GRIM score. Values in red represent those that have mean $\text{ER}$-$1\%$ lower than GRIM\cite{Desaphy2013-jt,Tran-Nguyen2021-jw}.
    }
    \label{tab:lit-pcba-er-mean}
\end{table}

\begin{table}[H]
    \centering
    \begin{tabular}{c|c|c|c|c|c}
    \textbf{Representation} & \textbf{\#Parameters} & \textbf{Device} & \textbf{Type} & \textbf{Time (mol/s)} & \textbf{Molecules/day} \\
\hline
    ECFP\cite{Rogers2010-xp} & - & IBM Power 9 & CPU & 10,000 & 864,000,000 \\
    MoLFormer\cite{Ross2022-ce} & 46,781,184 & Nvidia V100 (16GB) & GPU & 692 & 59,828,514\\

    MolCLR\cite{Wang2022-hg} & 2,404,196 & Nvidia V100 (16GB) & GPU & 6,250 & 540,000,000\\
    \end{tabular}
    \caption{Latency measurements for the different molecular feature extractors considered in this work. All measurement taken on a single Lassen compute node. A single GPU is used for MoLFormer and MolCLR extraction.}
    \label{tab:feature_latency}
\end{table}

\begin{table}[H]
    \centering
    \begin{tabular}{c|c|c|c}
    \textbf{length} & \textbf{radius} & \textbf{Time (mol/s)} & \textbf{$E$ (J/mol.)} \\
    \hline
                1000 & 1 & 2628 & 0.062  \\
        1000 & 2 & 2301 & 0.070  \\
        1000 & 4 & 1844 & 0.088  \\
        10000 & 1 & 2333 & 0.069 \\
        10000 & 2 & 2072 & 0.078 \\
        10000 & 4 & 1704 & 0.096  \\
        100000 & 1 & 1024 & 0.159 \\
        100000 & 2 & 975 & 0.169\\
        100000 & 4 & 888 & 0.185\\
        1000000 & 1 & 155 & 1.059\\
        1000000 & 2 & 155 & 1.060\\
        1000000 & 4 & 153 & 1.079\\

    \end{tabular}
    \caption{ECFP latency and energy analysis with respect to length and radius.}
    \label{tab:ecfp_energy_latency}
\end{table}

\begin{table}[H]
    \centering
    \begin{tabular}{c|c|c}
    \textbf{Model} & \textbf{Layers} & \textbf{Trainable parameters} \\ \hline
        MLP-small & (1024, 128), (128,2)  & 131,458\\
         MLP-large & (1024, 512), (512, 256), (256,128),(128,2) & 688,384
    \end{tabular}
    \caption{Description of the representative large and small MLP baseline models. For each set of parentheses (i.e. layer), the first number denotes the input size and the second denotes the output size.}
    \label{tab:mlp_training_hyperparams}
\end{table}

\begin{table}[H]
    \centering

\begin{tabular}{llrrrrrrrr}
\toprule
                                 Model & Device &  Encode (J/mol) &  Test (J/mol) $\times 10^{-6}$ &     1 &    10 &   100 &  1000 &  10000 &  100000 \\
\midrule

         HDB-DECFP &    GPU &           0.057 &  3.37 & \cellcolor{green!25}0.057 & \cellcolor{green!25}0.057 & \cellcolor{green!25}0.058 & \cellcolor{green!25}0.061 &  \cellcolor{green!25}0.091 &   \cellcolor{yellow!25}0.394\\
         HDB-MoLFormer &    GPU &           0.389 &  3.37 & \cellcolor{red!25}0.389 & \cellcolor{red!25}0.389 & \cellcolor{red!25}0.389 & \cellcolor{red!25}0.392 &  \cellcolor{red!25}0.422 &   \cellcolor{yellow!25}0.725\\
        HDB-Combo &    GPU &           0.496 &  3.37 & \cellcolor{red!25}0.496 & \cellcolor{red!25}0.496 & \cellcolor{red!25}0.496 & \cellcolor{red!25}0.499 &  \cellcolor{red!25}0.529 &   \cellcolor{yellow!25}0.832\\
                      MLP-small &    GPU &           0.071 &  2.30 & 0.071 & 0.071 & 0.071 & 0.073 &  0.094 &   0.300\\
                      MLP-large &    GPU &           0.071 &  9.18 & 0.071 & 0.071 & 0.072 & 0.080 &  0.163 &   0.989\\
                HDB-DECFP &   FPGA &           0.057 &  .75 & \cellcolor{green!25}0.057 & \cellcolor{green!25}0.057 & \cellcolor{green!25}0.057 & \cellcolor{green!25}0.058 &  \cellcolor{green!25}0.065 &   \cellcolor{green!25}0.132 \\
                HDB-MoLFormer &   FPGA &           0.389 &  .75 & \cellcolor{red!25}0.389 & \cellcolor{red!25}0.389 & \cellcolor{red!25}0.389 & \cellcolor{red!25}0.390 &  \cellcolor{red!25}0.396 &   \cellcolor{yellow!25}0.464\\
       HDB-Combo &   FPGA &           0.496 &  .75 & \cellcolor{red!25}0.496 & \cellcolor{red!25}0.496 & \cellcolor{red!25}0.496 & \cellcolor{red!25}0.497 &  \cellcolor{red!25}0.503 &   \cellcolor{yellow!25}0.571\\
\bottomrule
\end{tabular}

    \caption{Energy profiling results for HDBind models and our smallest and largest MLP baselines, respectively MLP-small and MLP-large. All values correspond to CPU and GPU power terms provided by variorum\cite{variorum}. All steps of feature extraction are included for encoding the input, such as extraction of the MoLFormer embedding itself as well as calculation of ECFPs. All HDC models use $D=1$k to facilitate comparison to FPGA baseline, which is relatively constrained in memory resources to the CPU and GPU. Green denotes models for which screening a particular number of proteins can be done with less energy than our MLP-small baseline. Values in yellow correspond to those models for which at a particular point can be done with less energy than MLP-large but still require more energy than MLP-small. Values in red correspond to those models for which screening require more energy than either of the MLP-small or MLP-large baselines.}
    \label{tab:energy_vs_target_num_table}
\end{table}

\begin{table}[H]
    \centering
\begin{tabular}{lllrrrrrrrrr}
\toprule
 &  &  & \multicolumn{4}{r}{roc-auc} & \multicolumn{4}{r}{er-1.0} & Actives \\
 &  &  & mean & std & min & max & mean & std & min & max & mean \\
model & D & target &  &  &  &  &  &  &  &  &  \\
\midrule

\multirow[t]{26}{*}{HDB-DECFP} & \multirow[t]{13}{*}{100000} & ADRB2 & 0.67 & 0.14 & 0.41 & 0.80 & 0.00 & 0.00 & 0.00 & 0.00 & 17.00 \\
 &  & ALDH1 & 0.71 & 0.04 & 0.65 & 0.76 & 10.99 & 3.90 & 4.99 & 15.18 & 5363.00 \\
 &  & FEN1 & 0.78 & 0.10 & 0.69 & 0.93 & 23.91 & 6.76 & 15.22 & 34.78 & 360.00 \\
 &  & GBA & 0.73 & 0.06 & 0.67 & 0.82 & 14.47 & 4.34 & 4.88 & 21.95 & 163.00 \\
 &  & IDH1 & 0.74 & 0.03 & 0.67 & 0.80 & 9.63 & 3.84 & 0.00 & 11.11 & 39.00 \\
 &  & KAT2A & 0.64 & 0.02 & 0.59 & 0.67 & 11.88 & 3.68 & 4.17 & 18.75 & 194.00 \\
 &  & MAPK1 & 0.66 & 0.03 & 0.60 & 0.70 & 3.64 & 1.94 & 0.00 & 9.09 & 308.00 \\
 &  & MTORC1 & 0.61 & 0.07 & 0.48 & 0.68 & 3.19 & 1.79 & 0.00 & 4.17 & 97.00 \\
 &  & OPRK1 & 0.46 & 0.10 & 0.35 & 0.61 & 0.00 & 0.00 & 0.00 & 0.00 & 24.00 \\
 &  & PKM2 & 0.67 & 0.07 & 0.60 & 0.77 & 7.89 & 3.30 & 1.47 & 13.24 & 546.00 \\
 &  & PPARG & 0.72 & 0.03 & 0.68 & 0.79 & 11.11 & 15.98 & 0.00 & 33.33 & 24.00 \\
 &  & TP53 & 0.62 & 0.03 & 0.55 & 0.68 & 10.18 & 5.16 & 5.26 & 15.79 & 64.00 \\
 &  & VDR & 0.73 & 0.05 & 0.67 & 0.80 & 10.44 & 1.16 & 7.88 & 12.73 & 655.00 \\
\cline{2-12}
 & \multirow[t]{13}{*}{1000000} & ADRB2 & 0.77 & 0.04 & 0.66 & 0.82 & 0.00 & 0.00 & 0.00 & 0.00 & 17.00 \\
 &  & ALDH1 & 0.71 & 0.04 & 0.64 & 0.76 & 11.10 & 4.16 & 4.61 & 15.55 & 5363.00 \\
 &  & FEN1 & 0.89 & 0.05 & 0.81 & 0.94 & 31.96 & 4.80 & 21.74 & 39.13 & 360.00 \\
 &  & GBA & 0.73 & 0.08 & 0.57 & 0.82 & 16.34 & 5.21 & 7.32 & 24.39 & 163.00 \\
 &  & IDH1 & 0.77 & 0.04 & 0.70 & 0.83 & 15.19 & 8.99 & 0.00 & 22.22 & 39.00 \\
 &  & KAT2A & 0.64 & 0.03 & 0.59 & 0.68 & 12.64 & 3.83 & 6.25 & 18.75 & 194.00 \\
 &  & MAPK1 & 0.67 & 0.02 & 0.64 & 0.70 & 4.68 & 1.69 & 2.60 & 10.39 & 308.00 \\
 &  & MTORC1 & 0.64 & 0.04 & 0.56 & 0.69 & 3.33 & 1.70 & 0.00 & 4.17 & 97.00 \\
 &  & OPRK1 & 0.52 & 0.01 & 0.49 & 0.54 & 0.00 & 0.00 & 0.00 & 0.00 & 24.00 \\
 &  & PKM2 & 0.71 & 0.07 & 0.60 & 0.78 & 8.55 & 3.05 & 1.47 & 13.24 & 546.00 \\
 &  & PPARG & 0.70 & 0.01 & 0.68 & 0.73 & 9.44 & 8.40 & 0.00 & 16.67 & 24.00 \\
 &  & TP53 & 0.66 & 0.05 & 0.60 & 0.74 & 17.89 & 5.63 & 5.26 & 26.32 & 64.00 \\
 &  & VDR & 0.78 & 0.04 & 0.71 & 0.81 & 12.10 & 1.78 & 8.48 & 14.55 & 655.00 \\

\bottomrule
\end{tabular}

    \caption{ROC-AUC and enrichment factor in true positives at a 1\% false positive rate (er-1) metrics for HDB-DECFP across all 15 LIT-PCBA target sets.}
    \label{tab:large_model_metrics-decfp}
\end{table}

\begin{table}[H]
    \centering

\begin{tabular}{lllrrrrrrrrr}
\toprule
 &  &  & \multicolumn{4}{r}{roc-auc} & \multicolumn{4}{r}{er-1.0} & Actives \\
 &  &  & mean & std & min & max & mean & std & min & max & mean \\
model & D & target &  &  &  &  &  &  &  &  &  \\
\midrule
\multirow[t]{26}{*}{HDB-RPFP} & \multirow[t]{13}{*}{100000} & ADRB2 & 0.56 & 0.01 & 0.55 & 0.58 & 0.00 & 0.00 & 0.00 & 0.00 & 17.00 \\
 &  & ALDH1 & 0.74 & 0.00 & 0.73 & 0.74 & 9.08 & 0.69 & 7.59 & 9.82 & 5363.00 \\
 &  & FEN1 & 0.82 & 0.01 & 0.80 & 0.83 & 17.17 & 1.68 & 15.22 & 19.57 & 360.00 \\
 &  & GBA & 0.74 & 0.00 & 0.73 & 0.74 & 13.66 & 1.26 & 12.20 & 14.63 & 163.00 \\
 &  & IDH1 & 0.65 & 0.02 & 0.60 & 0.67 & 0.00 & 0.00 & 0.00 & 0.00 & 39.00 \\
 &  & KAT2A & 0.65 & 0.01 & 0.64 & 0.66 & 8.75 & 2.15 & 6.25 & 12.50 & 194.00 \\
 &  & MAPK1 & 0.69 & 0.00 & 0.68 & 0.69 & 3.97 & 0.66 & 2.60 & 5.19 & 308.00 \\
 &  & MTORC1 & 0.55 & 0.02 & 0.51 & 0.58 & 2.08 & 2.20 & 0.00 & 4.17 & 97.00 \\
 &  & OPRK1 & 0.56 & 0.01 & 0.55 & 0.58 & 0.00 & 0.00 & 0.00 & 0.00 & 24.00 \\
 &  & PKM2 & 0.74 & 0.01 & 0.73 & 0.75 & 9.12 & 1.05 & 7.35 & 11.03 & 546.00 \\
 &  & PPARG & 0.70 & 0.02 & 0.69 & 0.75 & 15.00 & 5.27 & 0.00 & 16.67 & 24.00 \\
 &  & TP53 & 0.67 & 0.03 & 0.62 & 0.70 & 12.63 & 3.68 & 5.26 & 15.79 & 64.00 \\
 &  & VDR & 0.69 & 0.02 & 0.66 & 0.71 & 9.58 & 0.80 & 8.48 & 10.91 & 655.00 \\
\cline{2-12}
 & \multirow[t]{13}{*}{1000000} & ADRB2 & 0.59 & 0.01 & 0.57 & 0.60 & 0.00 & 0.00 & 0.00 & 0.00 & 17.00 \\
 &  & ALDH1 & 0.76 & 0.00 & 0.76 & 0.76 & 10.56 & 0.57 & 9.75 & 11.46 & 5363.00 \\
 &  & FEN1 & 0.80 & 0.01 & 0.78 & 0.81 & 15.33 & 1.08 & 13.04 & 16.30 & 360.00 \\
 &  & GBA & 0.65 & 0.01 & 0.62 & 0.66 & 4.88 & 0.00 & 4.88 & 4.88 & 163.00 \\
 &  & IDH1 & 0.78 & 0.02 & 0.73 & 0.79 & 20.00 & 4.68 & 11.11 & 22.22 & 39.00 \\
 &  & KAT2A & 0.67 & 0.00 & 0.66 & 0.67 & 10.21 & 0.66 & 8.33 & 10.42 & 194.00 \\
 &  & MAPK1 & 0.70 & 0.01 & 0.69 & 0.70 & 5.06 & 0.41 & 3.90 & 5.19 & 308.00 \\
 &  & MTORC1 & 0.64 & 0.01 & 0.62 & 0.66 & 0.00 & 0.00 & 0.00 & 0.00 & 97.00 \\
 &  & OPRK1 & 0.70 & 0.01 & 0.69 & 0.71 & 1.67 & 5.27 & 0.00 & 16.67 & 24.00 \\
 &  & PKM2 & 0.74 & 0.01 & 0.71 & 0.75 & 9.12 & 1.05 & 7.35 & 10.29 & 546.00 \\
 &  & PPARG & 0.80 & 0.02 & 0.76 & 0.83 & 11.67 & 8.05 & 0.00 & 16.67 & 24.00 \\
 &  & TP53 & 0.72 & 0.02 & 0.68 & 0.74 & 14.74 & 2.22 & 10.53 & 15.79 & 64.00 \\
 &  & VDR & 0.76 & 0.01 & 0.75 & 0.77 & 10.36 & 1.05 & 9.09 & 12.12 & 655.00 \\

\bottomrule
\end{tabular}

    \caption{ROC-AUC and enrichment factor in true positives at a 1\% false positive rate (er-1) metrics for HDB-RPFP across all 15 LIT-PCBA target sets.}
    \label{tab:large_model_metrics-rpfp}
\end{table}

\begin{table}[H]
    \centering

\begin{tabular}{lllrrrrrrrrr}
\toprule
 &  &  & \multicolumn{4}{r}{roc-auc} & \multicolumn{4}{r}{er-1.0} & Actives \\
 &  &  & mean & std & min & max & mean & std & min & max & mean \\
model & D & target &  &  &  &  &  &  &  &  &  \\
\midrule
\multirow[t]{26}{*}{HDB-MolCLR} & \multirow[t]{13}{*}{100000} & ADRB2 & 0.48 & 0.04 & 0.40 & 0.53 & 0.00 & 0.00 & 0.00 & 0.00 & 17.00 \\
 &  & ALDH1 & 0.72 & 0.01 & 0.70 & 0.72 & 8.47 & 0.59 & 7.07 & 9.15 & 5363.00 \\
 &  & FEN1 & 0.92 & 0.01 & 0.90 & 0.92 & 30.22 & 1.90 & 26.09 & 32.61 & 360.00 \\
 &  & GBA & 0.78 & 0.01 & 0.77 & 0.80 & 10.24 & 2.24 & 7.32 & 14.63 & 163.00 \\
 &  & IDH1 & 0.77 & 0.01 & 0.76 & 0.78 & 1.11 & 3.51 & 0.00 & 11.11 & 39.00 \\
 &  & KAT2A & 0.66 & 0.00 & 0.65 & 0.67 & 4.58 & 1.64 & 2.08 & 6.25 & 194.00 \\
 &  & MAPK1 & 0.71 & 0.01 & 0.71 & 0.72 & 6.75 & 1.02 & 5.19 & 9.09 & 308.00 \\
 &  & MTORC1 & 0.64 & 0.02 & 0.61 & 0.67 & 2.92 & 3.43 & 0.00 & 8.33 & 97.00 \\
 &  & OPRK1 & 0.63 & 0.02 & 0.61 & 0.66 & 0.00 & 0.00 & 0.00 & 0.00 & 24.00 \\
 &  & PKM2 & 0.70 & 0.00 & 0.69 & 0.70 & 2.35 & 0.90 & 1.47 & 3.68 & 546.00 \\
 &  & PPARG & 0.58 & 0.01 & 0.57 & 0.58 & 0.00 & 0.00 & 0.00 & 0.00 & 24.00 \\
 &  & TP53 & 0.66 & 0.02 & 0.61 & 0.68 & 6.84 & 2.54 & 5.26 & 10.53 & 64.00 \\
 &  & VDR & 0.78 & 0.01 & 0.77 & 0.79 & 10.36 & 0.60 & 9.70 & 10.91 & 655.00 \\
\cline{2-12}
 & \multirow[t]{13}{*}{1000000} & ADRB2 & 0.48 & 0.05 & 0.38 & 0.56 & 0.00 & 0.00 & 0.00 & 0.00 & 17.00 \\
 &  & ALDH1 & 0.73 & 0.00 & 0.72 & 0.74 & 8.49 & 0.56 & 7.43 & 9.22 & 5363.00 \\
 &  & FEN1 & 0.92 & 0.00 & 0.91 & 0.92 & 30.33 & 2.64 & 25.00 & 33.70 & 360.00 \\
 &  & GBA & 0.78 & 0.01 & 0.77 & 0.79 & 11.71 & 1.92 & 9.76 & 14.63 & 163.00 \\
 &  & IDH1 & 0.74 & 0.01 & 0.71 & 0.75 & 11.11 & 0.00 & 11.11 & 11.11 & 39.00 \\
 &  & KAT2A & 0.71 & 0.00 & 0.71 & 0.72 & 4.58 & 0.88 & 4.17 & 6.25 & 194.00 \\
 &  & MAPK1 & 0.69 & 0.00 & 0.68 & 0.70 & 6.75 & 0.82 & 5.19 & 7.79 & 308.00 \\
 &  & MTORC1 & 0.54 & 0.01 & 0.51 & 0.56 & 3.75 & 3.65 & 0.00 & 8.33 & 97.00 \\
 &  & OPRK1 & 0.67 & 0.01 & 0.66 & 0.69 & 0.00 & 0.00 & 0.00 & 0.00 & 24.00 \\
 &  & PKM2 & 0.70 & 0.00 & 0.69 & 0.71 & 3.68 & 0.60 & 2.94 & 4.41 & 546.00 \\
 &  & PPARG & 0.56 & 0.00 & 0.55 & 0.56 & 0.00 & 0.00 & 0.00 & 0.00 & 24.00 \\
 &  & TP53 & 0.71 & 0.01 & 0.70 & 0.73 & 5.26 & 0.00 & 5.26 & 5.26 & 64.00 \\
 &  & VDR & 0.77 & 0.01 & 0.76 & 0.79 & 10.06 & 0.96 & 9.09 & 12.12 & 655.00 \\

\bottomrule
\end{tabular}

    \caption{ROC-AUC and enrichment factor in true positives at a 1\% false positive rate (er-1) metrics for HDB-MolCLR across all 15 LIT-PCBA target sets.}
    \label{tab:large_model_metrics-molclr}
\end{table}

\begin{table}[H]
    \centering

\begin{tabular}{lllrrrrrrrrr}
\toprule
 &  &  & \multicolumn{4}{r}{roc-auc} & \multicolumn{4}{r}{er-1.0} & Actives \\
 &  &  & mean & std & min & max & mean & std & min & max & mean \\
model & D & target &  &  &  &  &  &  &  &  &  \\
\midrule
\multirow[t]{26}{*}{HDB-MoLFormer} & \multirow[t]{13}{*}{100000} & ADRB2 & 0.91 & 0.02 & 0.88 & 0.94 & 5.00 & 10.54 & 0.00 & 25.00 & 17.00 \\
 &  & ALDH1 & 0.69 & 0.00 & 0.68 & 0.69 & 4.38 & 0.17 & 4.17 & 4.61 & 5363.00 \\
 &  & FEN1 & 0.92 & 0.01 & 0.91 & 0.92 & 38.04 & 2.51 & 34.78 & 41.30 & 360.00 \\
 &  & GBA & 0.82 & 0.01 & 0.80 & 0.83 & 14.63 & 2.82 & 7.32 & 17.07 & 163.00 \\
 &  & IDH1 & 0.86 & 0.01 & 0.83 & 0.87 & 10.00 & 3.51 & 0.00 & 11.11 & 39.00 \\
 &  & KAT2A & 0.64 & 0.01 & 0.63 & 0.65 & 3.75 & 1.32 & 0.00 & 4.17 & 194.00 \\
 &  & MAPK1 & 0.70 & 0.00 & 0.69 & 0.71 & 6.23 & 1.19 & 3.90 & 7.79 & 308.00 \\
 &  & MTORC1 & 0.54 & 0.03 & 0.48 & 0.57 & 2.50 & 2.15 & 0.00 & 4.17 & 97.00 \\
 &  & OPRK1 & 0.76 & 0.03 & 0.70 & 0.80 & 16.67 & 0.00 & 16.67 & 16.67 & 24.00 \\
 &  & PKM2 & 0.76 & 0.00 & 0.75 & 0.76 & 6.76 & 0.76 & 5.15 & 7.35 & 546.00 \\
 &  & PPARG & 0.58 & 0.03 & 0.53 & 0.61 & 0.00 & 0.00 & 0.00 & 0.00 & 24.00 \\
 &  & TP53 & 0.63 & 0.00 & 0.63 & 0.64 & 5.26 & 0.00 & 5.26 & 5.26 & 64.00 \\
 &  & VDR & 0.73 & 0.01 & 0.72 & 0.75 & 5.45 & 0.76 & 3.64 & 6.06 & 655.00 \\
\cline{2-12}
 & \multirow[t]{13}{*}{1000000} & ADRB2 & 0.61 & 0.04 & 0.56 & 0.66 & 0.00 & 0.00 & 0.00 & 0.00 & 17.00 \\
 &  & ALDH1 & 0.72 & 0.00 & 0.72 & 0.72 & 6.29 & 0.22 & 5.95 & 6.62 & 5363.00 \\
 &  & FEN1 & 0.92 & 0.00 & 0.92 & 0.93 & 33.26 & 1.47 & 31.52 & 36.96 & 360.00 \\
 &  & GBA & 0.81 & 0.01 & 0.79 & 0.82 & 18.54 & 1.71 & 17.07 & 21.95 & 163.00 \\
 &  & IDH1 & 0.89 & 0.01 & 0.87 & 0.90 & 11.11 & 0.00 & 11.11 & 11.11 & 39.00 \\
 &  & KAT2A & 0.72 & 0.00 & 0.71 & 0.73 & 5.62 & 2.21 & 2.08 & 8.33 & 194.00 \\
 &  & MAPK1 & 0.69 & 0.00 & 0.69 & 0.69 & 5.45 & 0.55 & 5.19 & 6.49 & 308.00 \\
 &  & MTORC1 & 0.58 & 0.01 & 0.57 & 0.59 & 6.25 & 2.20 & 4.17 & 8.33 & 97.00 \\
 &  & OPRK1 & 0.84 & 0.01 & 0.83 & 0.85 & 0.00 & 0.00 & 0.00 & 0.00 & 24.00 \\
 &  & PKM2 & 0.71 & 0.01 & 0.70 & 0.72 & 3.16 & 1.04 & 0.74 & 4.41 & 546.00 \\
 &  & PPARG & 0.55 & 0.00 & 0.55 & 0.56 & 0.00 & 0.00 & 0.00 & 0.00 & 24.00 \\
 &  & TP53 & 0.62 & 0.00 & 0.61 & 0.63 & 5.26 & 0.00 & 5.26 & 5.26 & 64.00 \\
 &  & VDR & 0.71 & 0.02 & 0.68 & 0.74 & 4.06 & 1.14 & 3.03 & 6.67 & 655.00 \\
\bottomrule
\end{tabular}

    \caption{ROC-AUC and enrichment factor in true positives at a 1\% false positive rate (er-1) metrics for HDB-MoLFormer across all 15 LIT-PCBA target sets.}
    \label{tab:large_model_metrics-molformer}
\end{table}

\begin{table}[H]
    \centering
\begin{tabular}{lllrrrrrrrrr}
\toprule
 &  &  & \multicolumn{4}{r}{roc-auc} & \multicolumn{4}{r}{er-1.0} & Actives \\
 &  &  & mean & std & min & max & mean & std & min & max & mean \\
model & D & target &  &  &  &  &  &  &  &  &  \\
\midrule

\multirow[t]{26}{*}{HDB-Combo} & \multirow[t]{13}{*}{100000} & ADRB2 & 1.00 & 0.00 & 1.00 & 1.00 & 100.00 & 0.00 & 100.00 & 100.00 & 17.00 \\
 &  & ALDH1 & 0.53 & 0.01 & 0.51 & 0.55 & 1.53 & 0.30 & 1.14 & 2.11 & 5363.00 \\
 &  & FEN1 & 0.97 & 0.00 & 0.96 & 0.97 & 73.55 & 3.12 & 65.88 & 76.09 & 360.00 \\
 &  & GBA & 0.73 & 0.02 & 0.68 & 0.74 & 35.09 & 5.37 & 24.39 & 39.02 & 163.00 \\
 &  & IDH1 & 0.44 & 0.05 & 0.36 & 0.52 & 0.75 & 2.37 & 0.00 & 7.50 & 39.00 \\
 &  & KAT2A & 0.86 & 0.01 & 0.84 & 0.87 & 19.36 & 2.11 & 15.57 & 22.89 & 194.00 \\
 &  & MAPK1 & 0.97 & 0.00 & 0.97 & 0.97 & 60.71 & 2.76 & 57.14 & 66.23 & 308.00 \\
 &  & MTORC1 & 0.94 & 0.00 & 0.93 & 0.94 & 66.12 & 2.70 & 60.86 & 70.83 & 97.00 \\
 &  & OPRK1 & 0.92 & 0.01 & 0.90 & 0.93 & 16.67 & 0.00 & 16.67 & 16.67 & 24.00 \\
 &  & PKM2 & 0.93 & 0.02 & 0.90 & 0.95 & 60.96 & 9.60 & 42.70 & 72.63 & 546.00 \\
 &  & PPARG & 0.69 & 0.04 & 0.62 & 0.74 & 0.00 & 0.00 & 0.00 & 0.00 & 24.00 \\
 &  & TP53 & 0.64 & 0.01 & 0.63 & 0.67 & 4.85 & 1.31 & 1.11 & 5.26 & 64.00 \\
 &  & VDR & 0.57 & 0.01 & 0.56 & 0.58 & 2.12 & 0.63 & 1.21 & 3.27 & 655.00 \\
\cline{2-12}
 & \multirow[t]{13}{*}{1000000} & ADRB2 & 0.97 & 0.04 & 0.89 & 1.00 & 67.20 & 37.38 & 0.00 & 100.00 & 17.00 \\
 &  & ALDH1 & 0.54 & 0.02 & 0.53 & 0.56 & 1.16 & 0.18 & 0.82 & 1.34 & 5363.00 \\
 &  & FEN1 & 0.81 & 0.11 & 0.62 & 0.90 & 14.91 & 12.96 & 0.30 & 27.17 & 360.00 \\
 &  & GBA & 0.69 & 0.10 & 0.55 & 0.79 & 19.44 & 16.02 & 0.00 & 36.59 & 163.00 \\
 &  & IDH1 & 0.48 & 0.05 & 0.41 & 0.56 & 0.11 & 0.36 & 0.00 & 1.13 & 39.00 \\
 &  & KAT2A & 0.79 & 0.11 & 0.53 & 0.87 & 12.31 & 9.89 & 0.00 & 22.09 & 194.00 \\
 &  & MAPK1 & 0.86 & 0.09 & 0.69 & 0.94 & 20.64 & 11.46 & 2.49 & 40.86 & 308.00 \\
 &  & MTORC1 & 0.90 & 0.01 & 0.89 & 0.91 & 37.59 & 12.01 & 12.50 & 50.00 & 97.00 \\
 &  & OPRK1 & 0.68 & 0.20 & 0.39 & 0.94 & 13.64 & 7.25 & 0.00 & 19.73 & 24.00 \\
 &  & PKM2 & 0.94 & 0.02 & 0.92 & 0.96 & 58.95 & 6.47 & 50.74 & 65.62 & 546.00 \\
 &  & PPARG & 0.64 & 0.11 & 0.45 & 0.81 & 0.00 & 0.00 & 0.00 & 0.00 & 24.00 \\
 &  & TP53 & 0.65 & 0.07 & 0.54 & 0.74 & 2.63 & 2.77 & 0.00 & 5.26 & 64.00 \\
 &  & VDR & 0.56 & 0.08 & 0.48 & 0.64 & 2.03 & 0.45 & 1.08 & 2.42 & 655.00 \\
\bottomrule
\end{tabular}

    \caption{ROC-AUC and enrichment factor in true positives at a 1\% false positive rate (er-1) metrics for HDB-Combo across all 15 LIT-PCBA target sets.}
    \label{tab:large_model_metrics-combo}
\end{table}

\begin{table}[H]
    \centering
    \begin{tabular}{c|c|c|c|c}
  \textbf{D} & \textbf{Single core (train)} & \textbf{Single core (test)} &\textbf{Max core (train)} & \textbf{Max core (test)} \\ \hline
     1000	&2.3	&2.0&	2.2& 4.5 \\
10000	&16.2	&14.0 &	3.9&	4.5 \\
100000&	118.2&	97.7&	19.5&	23.0 \\

\end{tabular}
    \caption{HDBind processing latency speedup on GPU versus CPU (single core, max cores) for training and testing. The advantage of the GPU grows with increasing dimension sizes $D$. We set the PyTorch thread count to 43, i.e. the number of physical cores - 1.}
    \label{tab:hdbind_hardware_speedup}
\end{table}

\begin{figure}
    \centering
    \includegraphics[width=\linewidth]{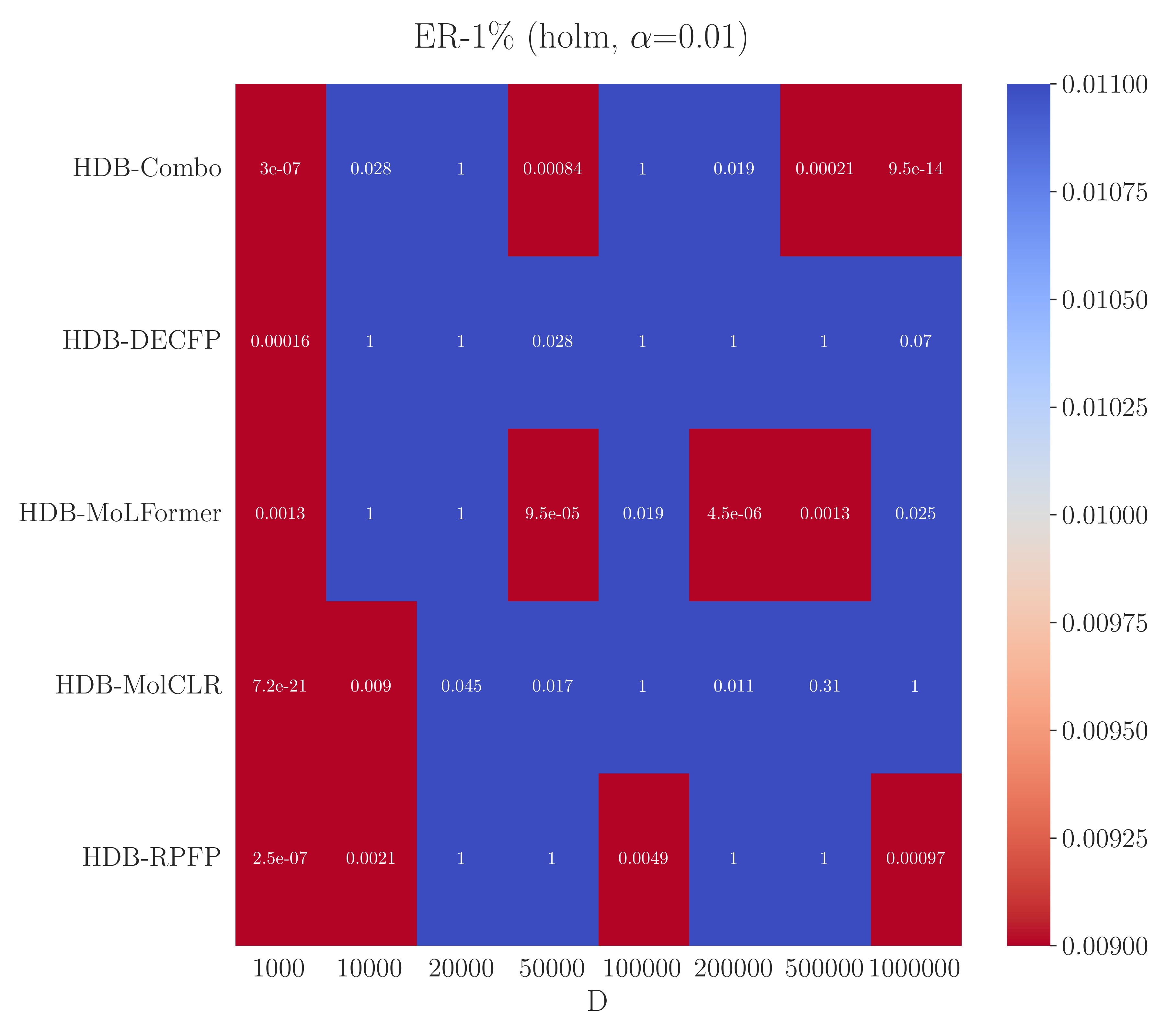}
    \caption{Paired $t$-test of HDBind models compared to MLP baseline models on the random split using the roc-auc metric distributions over all datasets (15 protein targets) and random seeds (10). Red cells indicate a statistically significant difference in the means of the respective roc-auc metric distributions (model, $D$) and blue cells indicate a non-significant difference. in The holm step-down correction is applied for multiple hypothesis testing and we use $\alpha=0.01$. }
    \label{fig:stat_sig_er_1_random}
\end{figure}

\begin{figure}
    \centering
    \includegraphics[width=\linewidth]{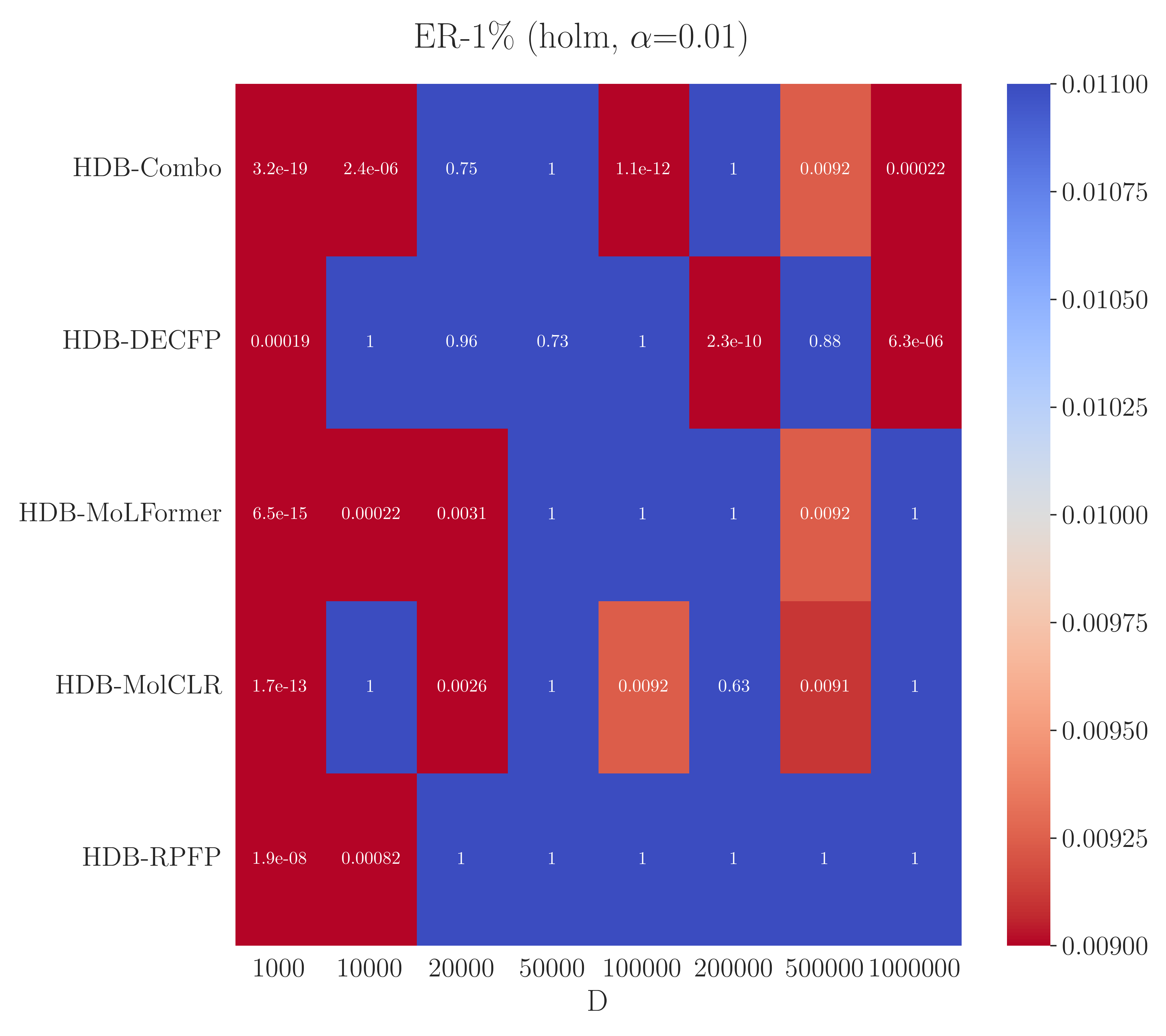}
    \caption{Paired $t$-test of HDBind models compared to MLP baseline models on the AVE bias-minimizing split using the roc-auc metric distributions over all datasets (15 protein targets) and random seeds (10). Red cells indicate a statistically significant difference in the means of the respective roc-auc metric distributions (model, $D$) and blue cells indicate a non-significant difference. in The holm step-down correction is applied for multiple hypothesis testing and we use $\alpha=0.01$.}
    \label{fig:stat_sig_er_1_ave}
\end{figure}

\begin{figure}
    \centering
    \includegraphics[width=\linewidth]{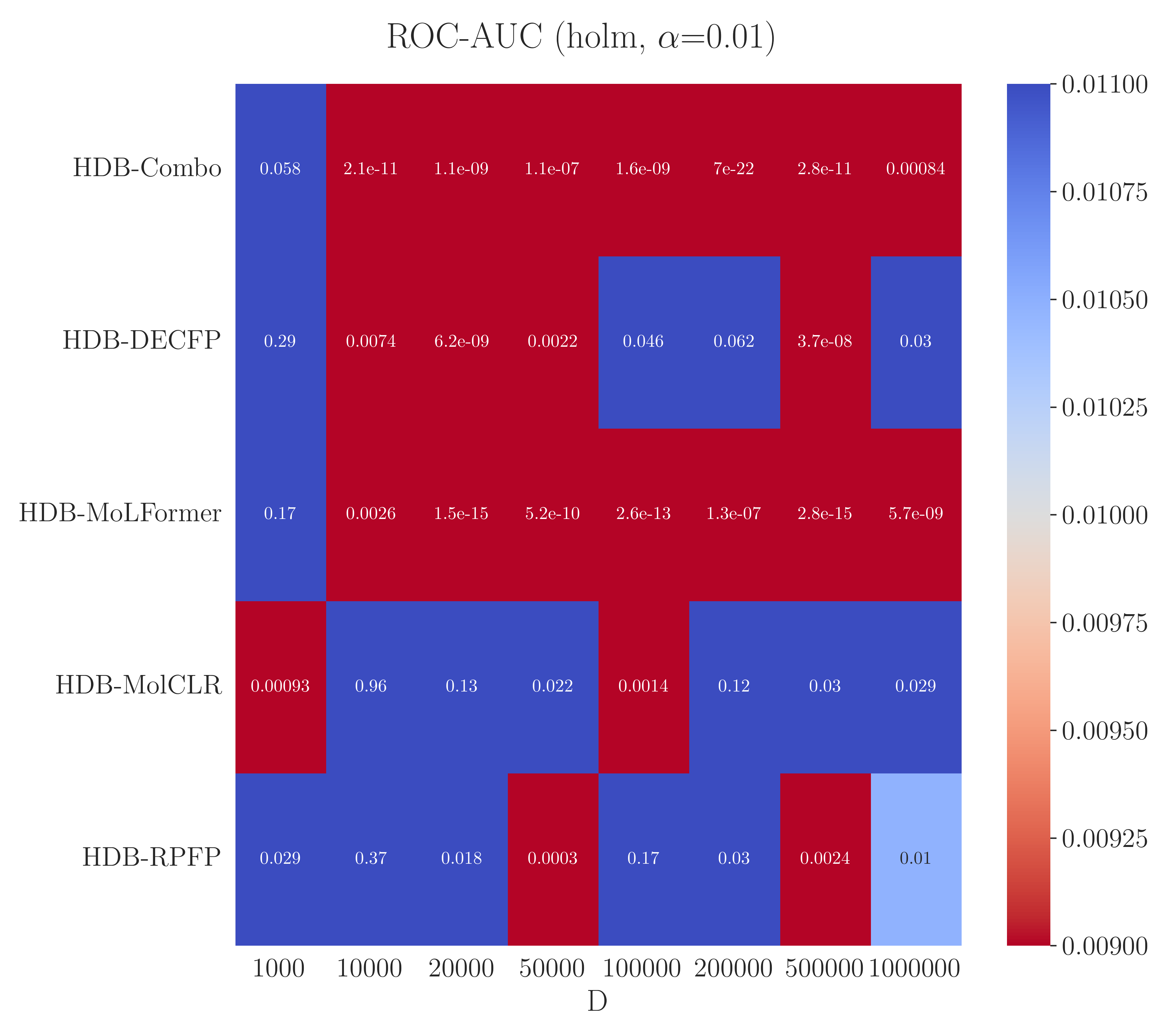}
    \caption{Paired $t$-test of HDBind models compared to MLP baseline models on the random split using the roc-auc metric distributions over all datasets (15 protein targets) and random seeds (10). Red cells indicate a statistically significant difference in the means of the respective roc-auc metric distributions (model, $D$) and blue cells indicate a non-significant difference. in The holm step-down correction is applied for multiple hypothesis testing and we use $\alpha=0.01$.}
    \label{fig:stat_sig_roc_random}
\end{figure}

\begin{figure}
    \centering
    \includegraphics[width=\linewidth]{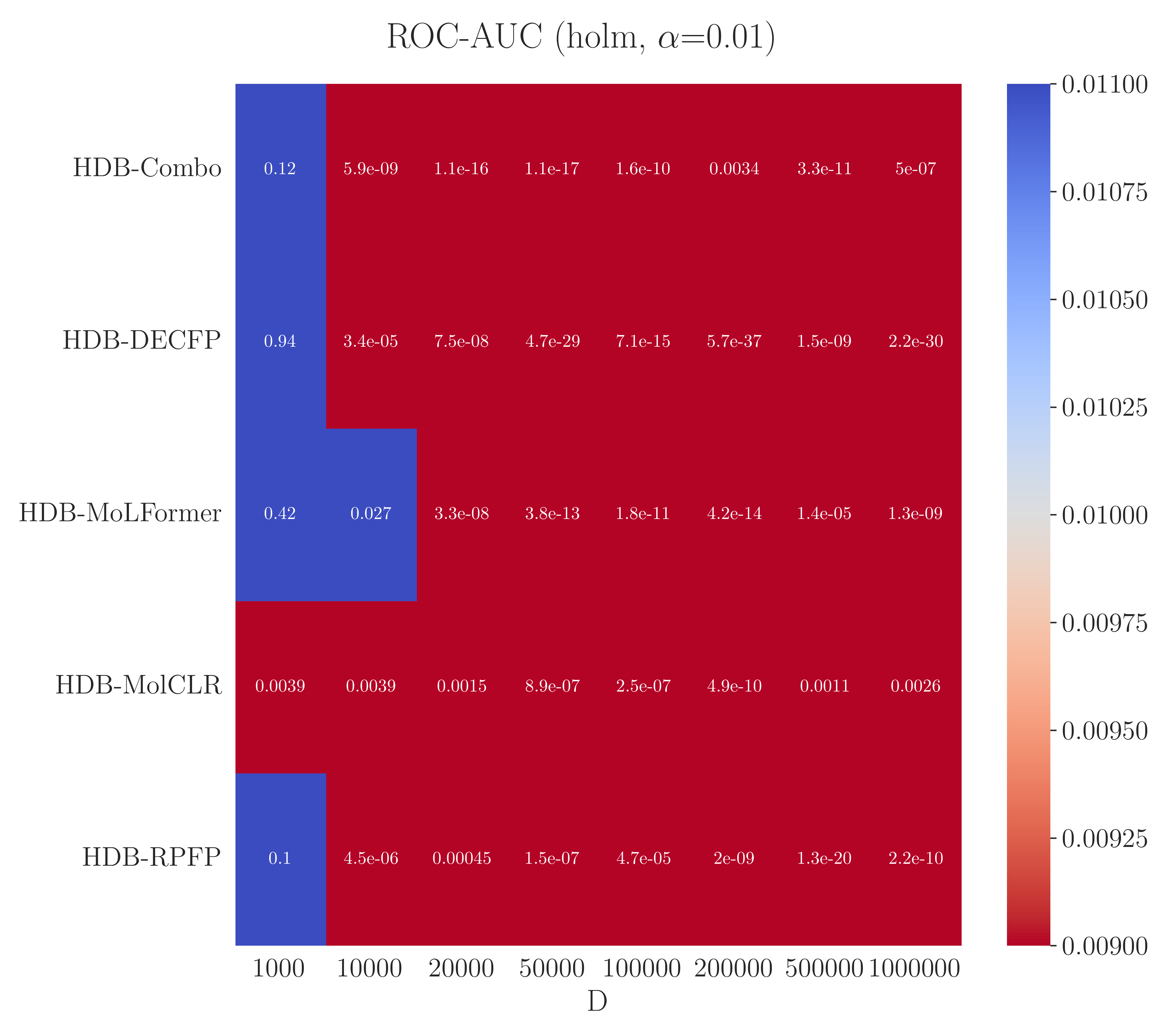}
    \caption{Paired $t$-test of HDBind models compared to MLP baseline models on the AVE bias-minimizing split using the roc-auc metric distributions over all datasets (15 protein targets) and random seeds (10). Red cells indicate a statistically significant difference in the means of the respective roc-auc metric distributions (model, $D$) and blue cells indicate a non-significant difference. in The holm step-down correction is applied for multiple hypothesis testing and we use $\alpha=0.01$.}
    \label{fig:stat_sig_roc_ave}
\end{figure}

\begin{figure}
    \centering
    \includegraphics[width=\linewidth]{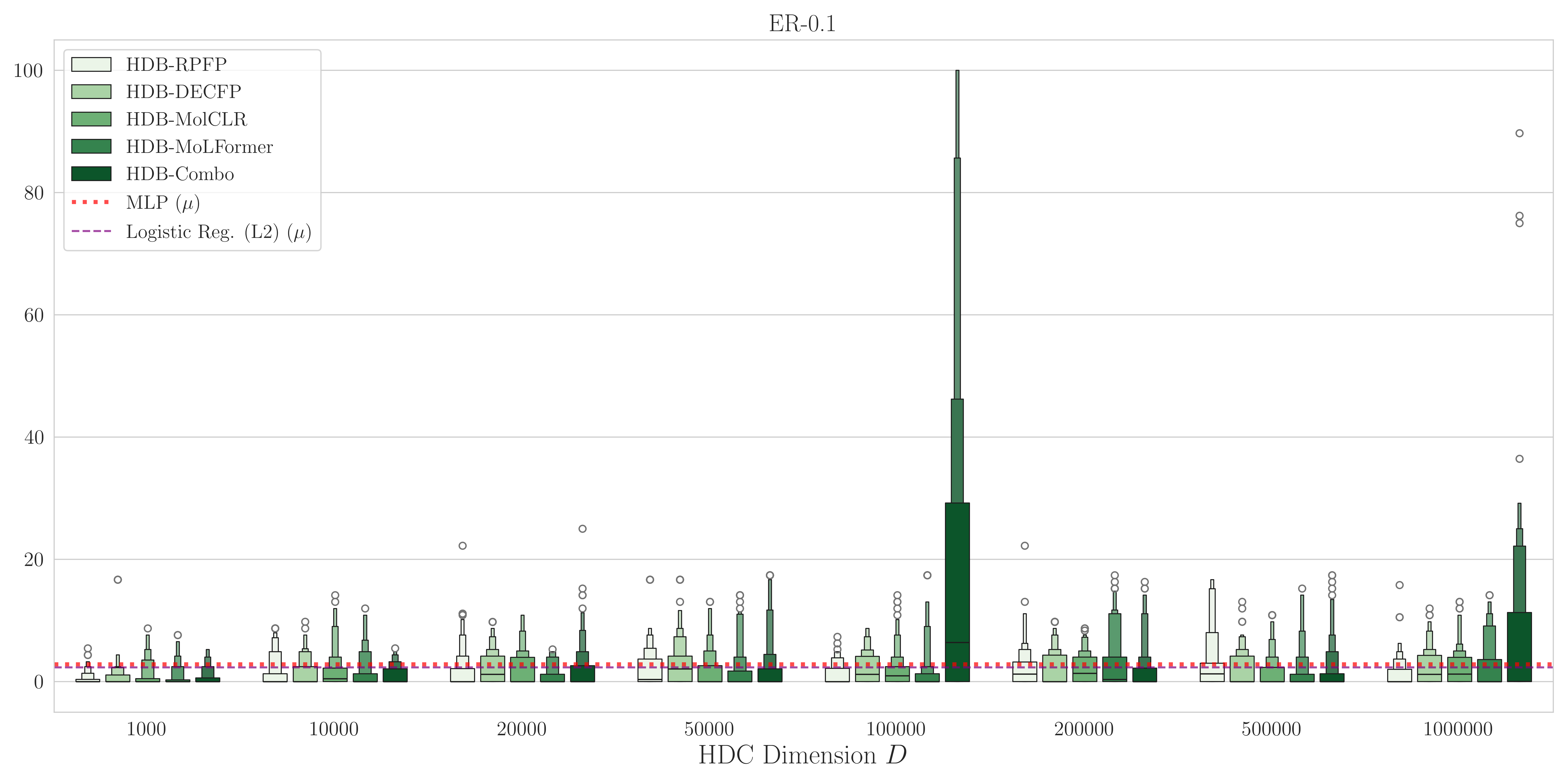}
    \caption{Distribution of the roc-enrichment with FPR=1/1000, compared to MLP and logistic regression baseline models.}
    \label{fig:er_01}
\end{figure}

\begin{figure}
    \centering
    \includegraphics[width=\linewidth]{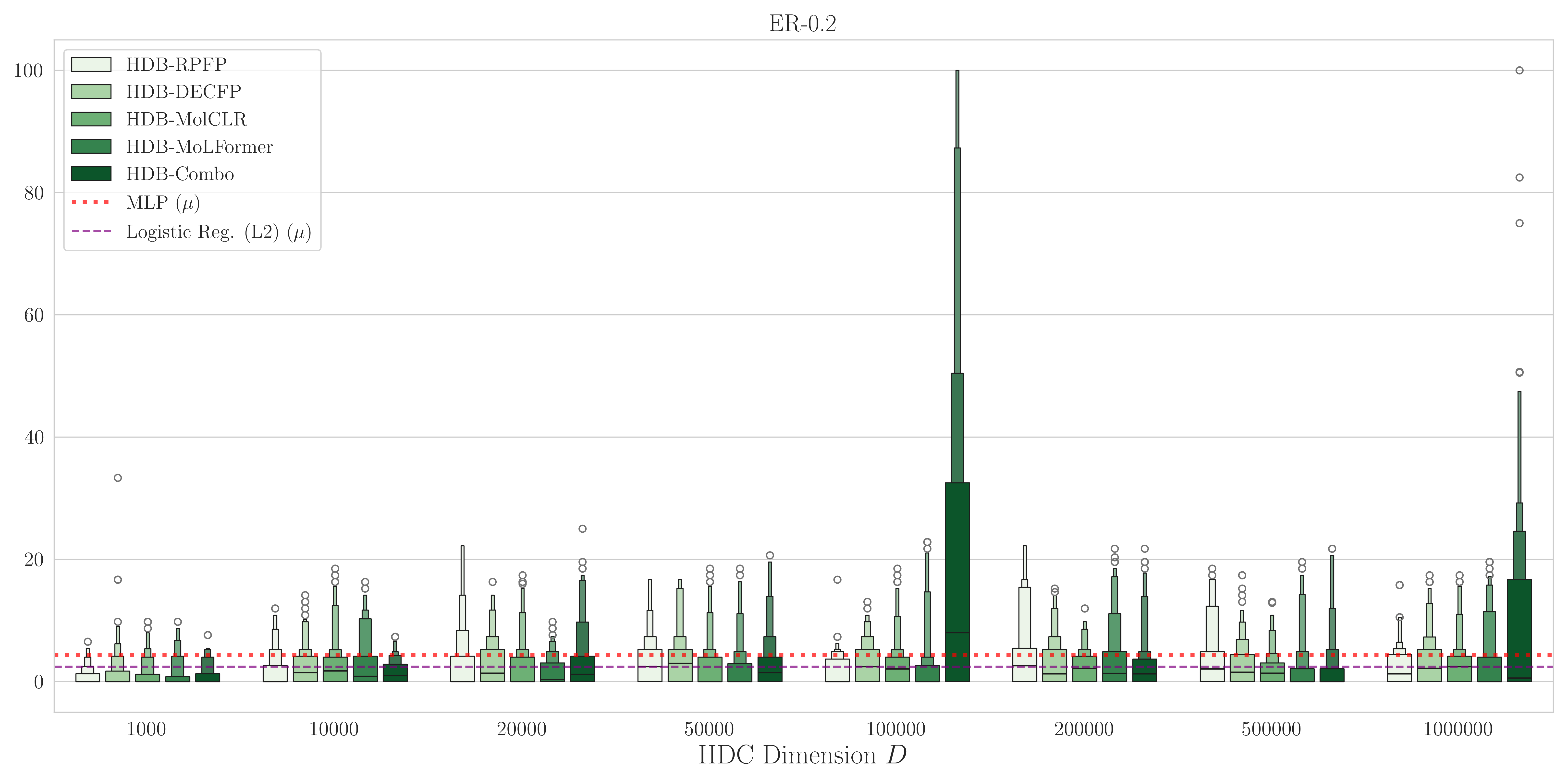}
    \caption{Distribution of the roc-enrichment with FPR=2/1000, compared to MLP and logistic regression baseline models.}
    \label{fig:er_02}
\end{figure}

\begin{figure}
    \centering
    \includegraphics[width=\linewidth]{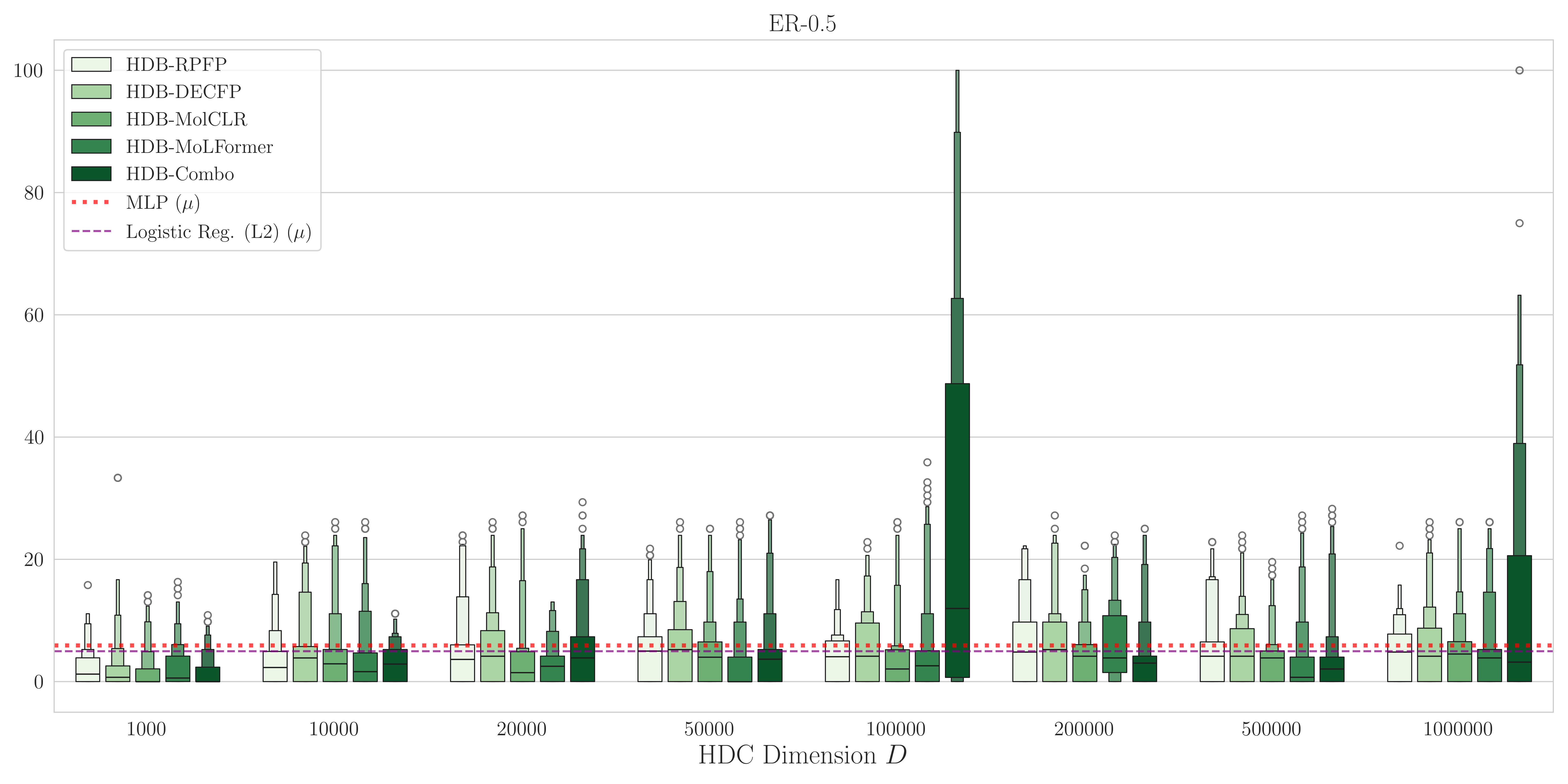}
    \caption{Distribution of the roc-enrichment with FPR=5/1000, compared to MLP and logistic regression baseline models.}
    \label{fig:er_05}
\end{figure}

\begin{figure}
    \centering
    \includegraphics[width=\linewidth]{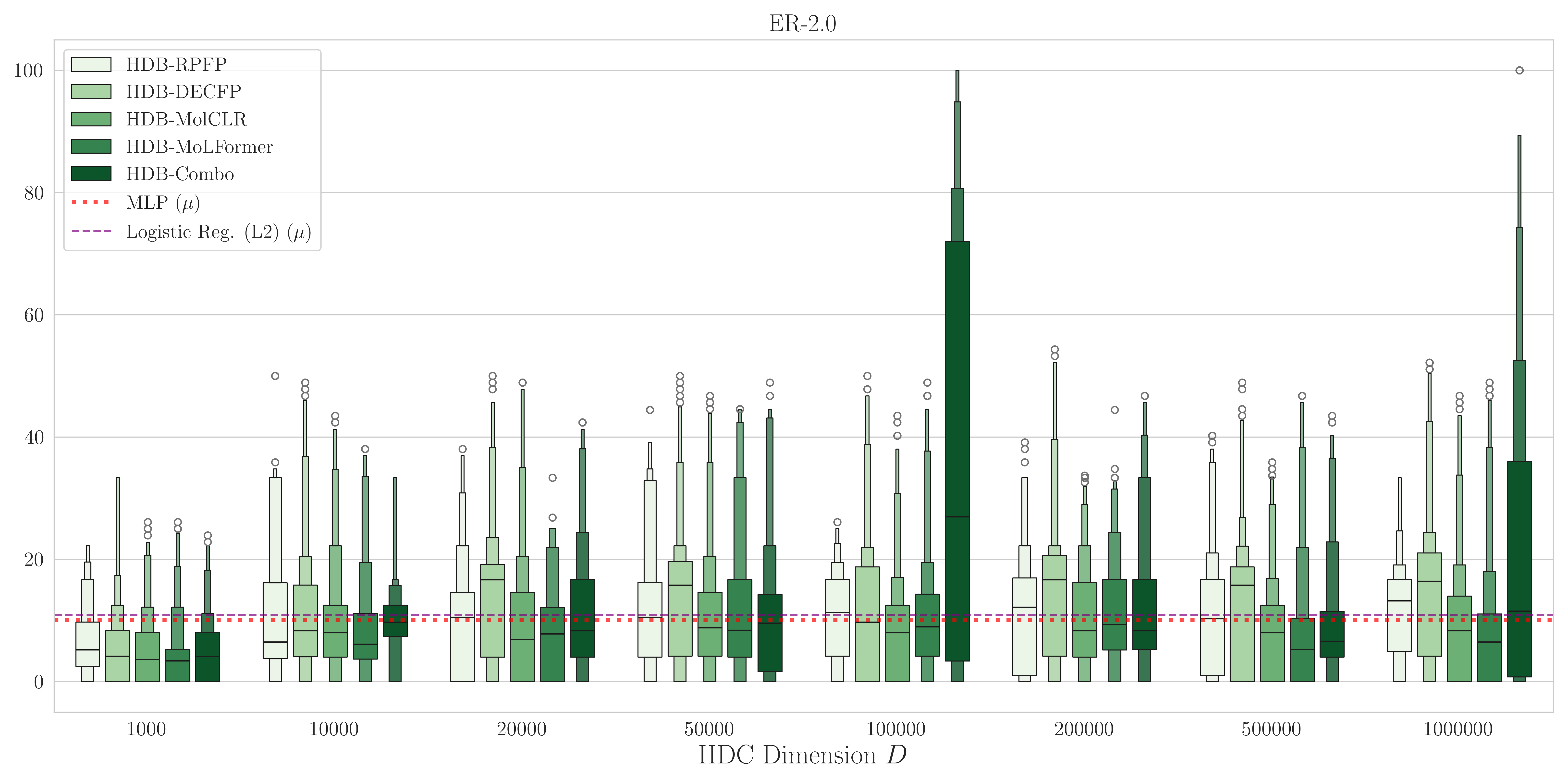}
    \caption{Distribution of the roc-enrichment with FPR=2/100, compared to MLP and logistic regression baseline models.}
    \label{fig:er_2}
\end{figure}

\begin{figure}
    \centering
    \includegraphics[width=\linewidth]{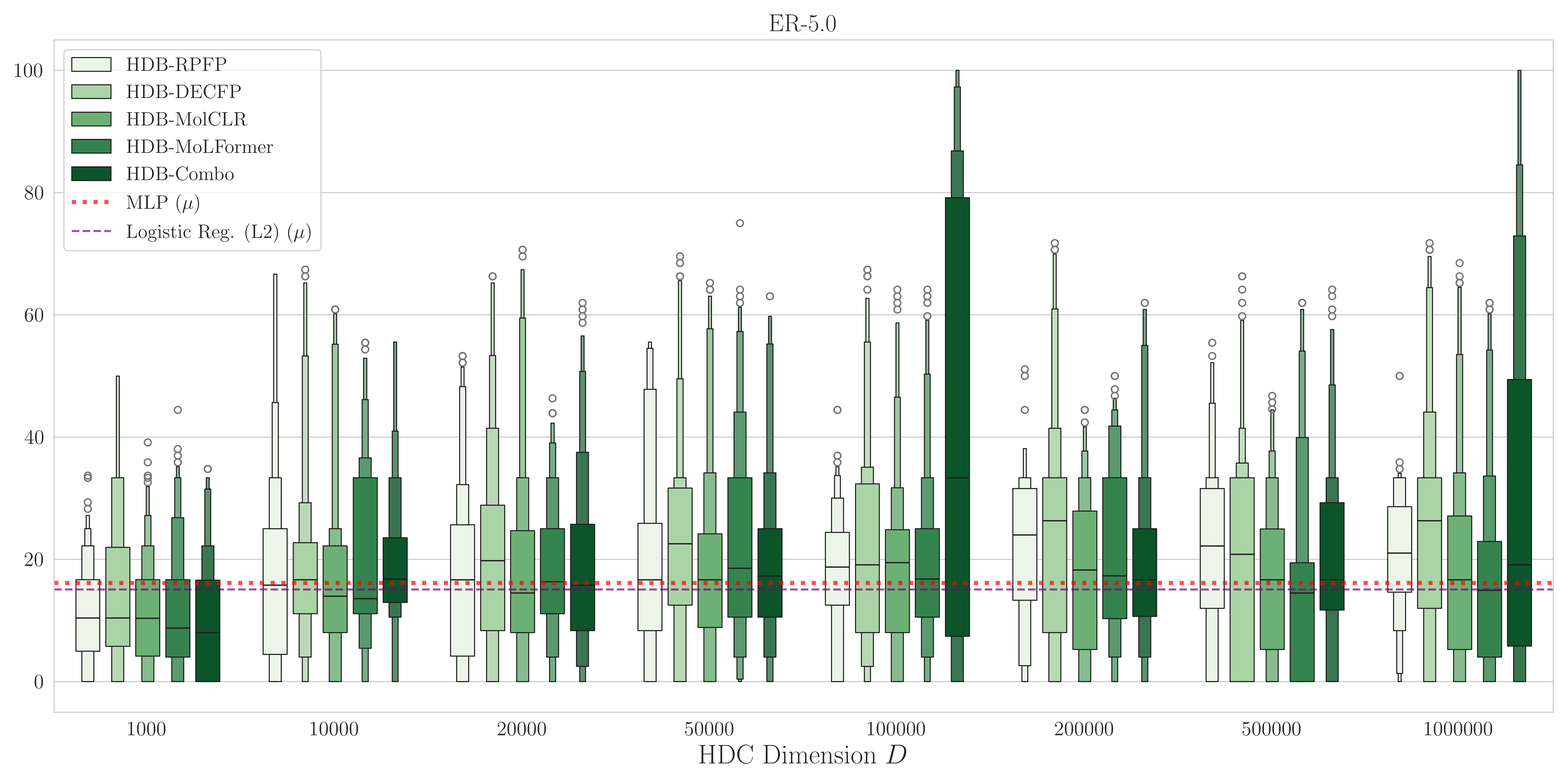}
    \caption{Distribution of the roc-enrichment with FPR=5/100, compared to MLP and logistic regression baseline models.}
    \label{fig:er_5}
\end{figure}

\begin{figure}
    \centering
    \includegraphics[width=\linewidth]{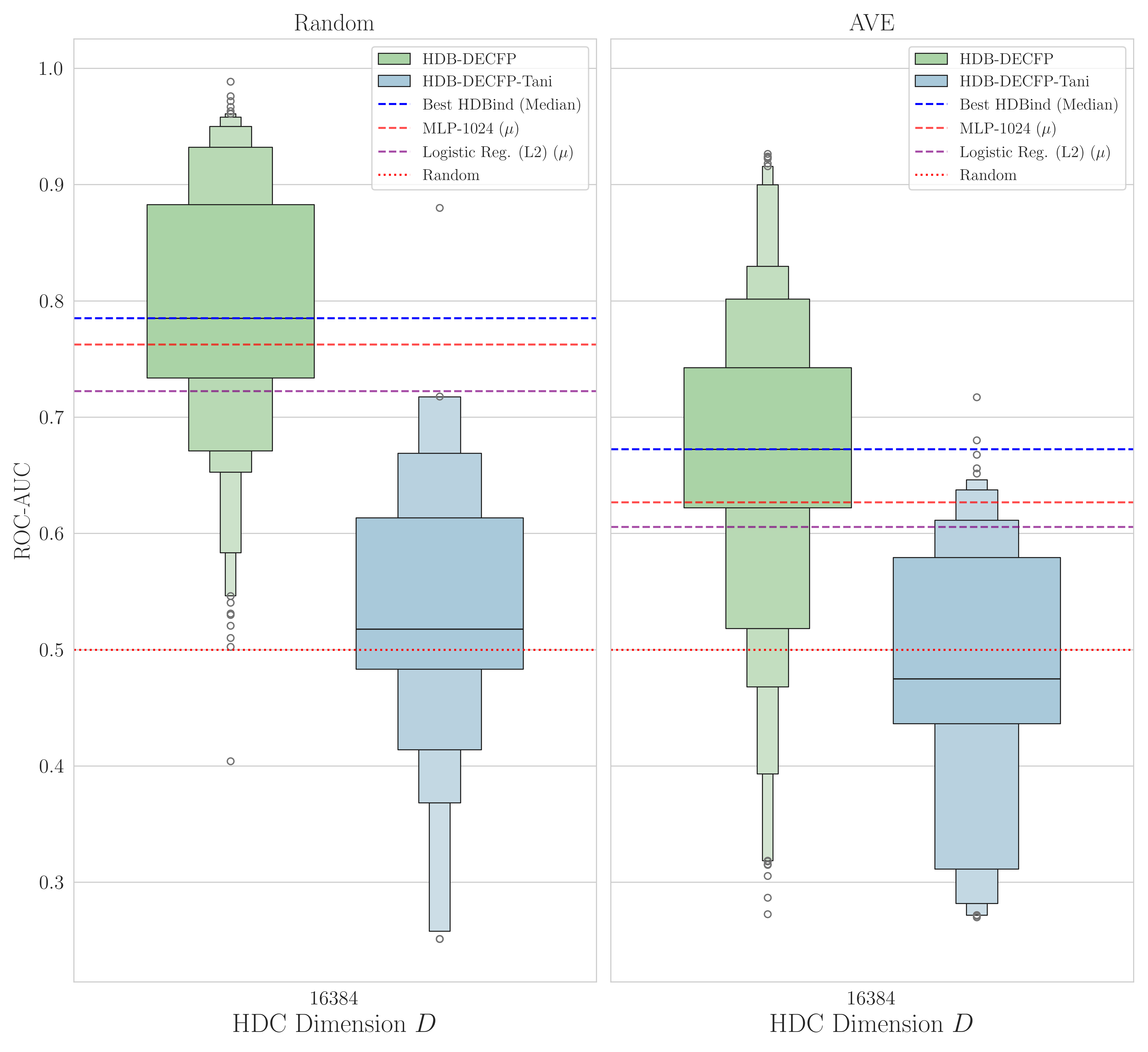}
    \caption{Distribution of roc-auc scores for HDB-DECFP models using cosine versus tanimoto similarity for training and testing with $D$=16,384 for each model. Our results suggest that the cosine similarity (green) consistently outperforms tanimoto similarity (blue) on both random and AVE bias-minimizing splits of the LIT-PCBA dataset.}
    \label{fig:16384_models_w_tanimoto}
\end{figure}

\begin{figure}
    \centering
    \includegraphics[width=\linewidth]{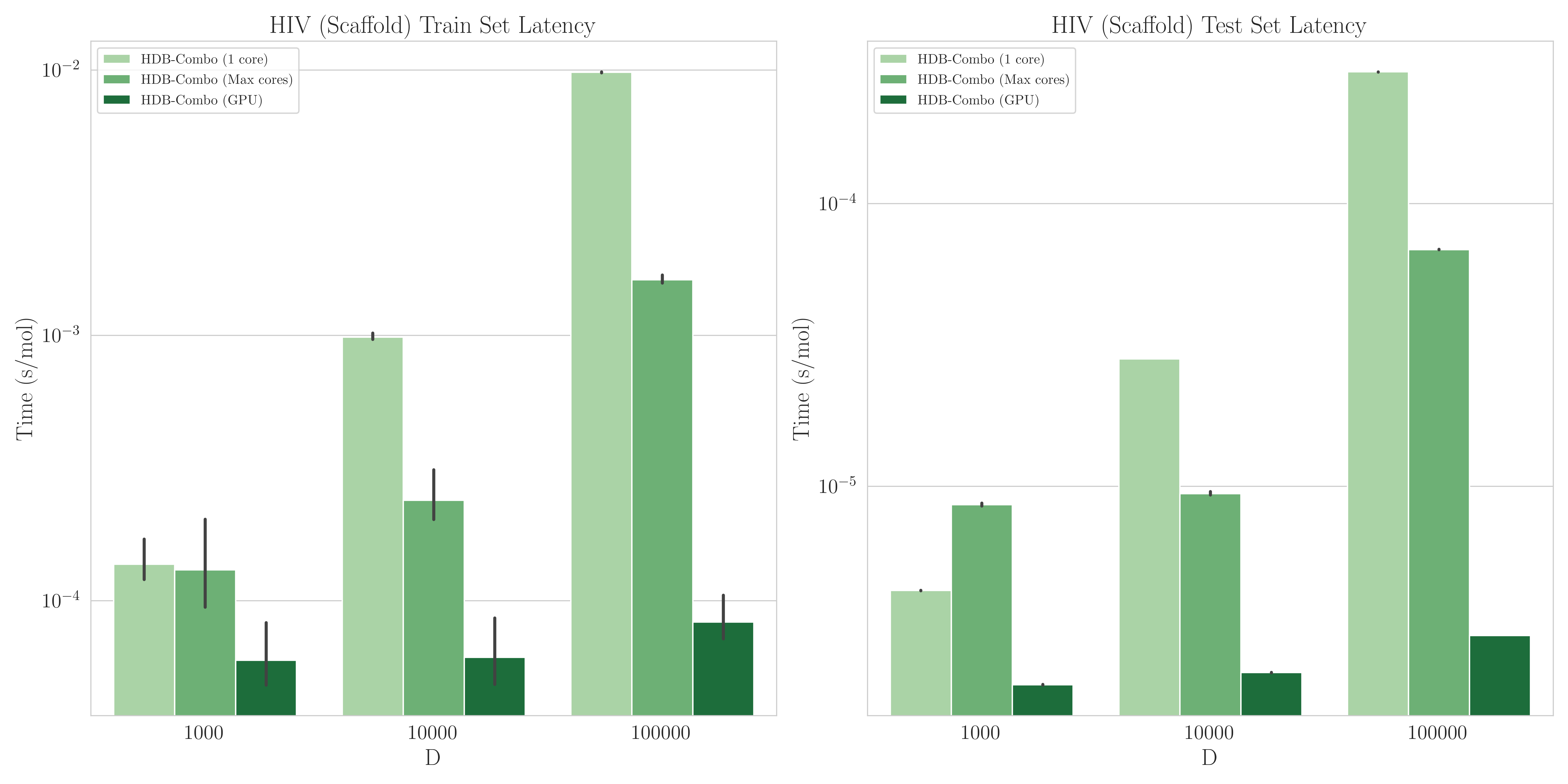}
    \caption{Comparison of the HDB-Combo training and testing latency across number of CPU cores (single or max physical cores) and GPU (with single CPU core). The HIV dataset from MoleculeNet is used. Timings are given in terms of seconds per molecule. We additionally consider the hypervector dimension size $D$ with values of 1k, 10k, and 100k. We compare these results to the MLP baseline model using a single CPU core, max physical cores, and the GPU with a single CPU core. For each column, the leftmost distribution represents single CPU core training/testing, the middle represents max physical CPU cores, and the rightmost represents the respective GPU timings. We manually set the number of threads used by the PyTorch backend by setting the environment variable \texttt{OMP\_NUM\_THREADS}.}
    \label{fig:hiv_cpu_gpu_train_test_time_comparison}
\end{figure}

\clearpage
\bibliography{bibliography}